\documentclass[preprint,3p]{elsarticle}

\usepackage{graphicx,amssymb,amsmath}
\usepackage{color}
\usepackage{rotating}

\newcommand{\be}{\begin{equation}}
\newcommand{\ee}{\end{equation}}

\newcommand{\bea}{\begin{eqnarray}}
\newcommand{\eea}{\end{eqnarray}}

\newcommand{\br}{\mathbf{r}}

\newcommand{\bt}{\mathbf{t}}

\def\eq#1{Eq.~(\ref{#1})}
\def\eqs#1#2{Eqs.~(\ref{#1},\ref{#2})}

\begin{document}

\title{Physics of base-pairing dynamics in DNA}

\author{Manoel Manghi\fnref{cor}
}
\ead{manghi@irsamc.ups-tlse.fr}
\fntext[cor]{Corresponding Author}
\author{Nicolas Destainville}
\ead{destain@irsamc.ups-tlse.fr}
\address{Laboratoire de Physique Th\'eorique, IRSAMC, Universit\'e de Toulouse, CNRS, UPS, France}
\date{\today}

\begin{abstract}
As a key molecule of Life, Deoxyribo-Nucleic Acid (DNA) is the focus of numbers of investigations with the help of biological, chemical and physical techniques. From a physical point of view, both experimental and theoretical works have brought quantitative insights into DNA base-pairing dynamics that we review in this Report, putting emphasis on theoretical developments.  We discuss the dynamics at the base-pair scale and its pivotal coupling with the polymer one, with a polymerization index running from a few nucleotides to tens of kilo-bases. This includes opening and closure of short hairpins and oligomers as well as zipping and unwinding of long macromolecules.  We review how different physical mechanisms are either used by Nature or utilized in biotechnological processes to separate the two intertwined DNA strands, by insisting on quantitative results. They go from thermally-assisted denaturation bubble nucleation to force- or torque-driven mechanisms. We show that the helical character of the molecule, possibly supercoiled, can play a key role in many denaturation and renaturation processes. We categorize the mechanisms according to the relative timescales associated with base-pairing and chain orientational degrees of freedom such as bending and torsional elastic ones. In some specific situations, these chain orientational degrees of freedom can be integrated out, and the quasi-static approximation is valid. The complex dynamics then reduces to the diffusion in a low-dimensional free-energy landscape. In contrast, some important cases of experimental interest necessarily appeal to far-from-equilibrium statistical mechanics and hydrodynamics.
\end{abstract}

\begin{keyword}
DNA \sep denaturation bubble \sep polymer dynamics \sep zipping \sep unwinding \sep free-energy barrier
\PACS 87.15.H- 87.15.A- 82.39.Pj 36.20.-r 
\end{keyword}
\maketitle

\tableofcontents


\section{Introduction}


The main reason why Nature has selected a double helical form for double-stranded DNA (dsDNA) in its more common forms is probably its stability, in order to preserve the genetic information of this key-molecule of Life~\cite{Alberts2002}, through hydrogen-bonding interactions between Watson-Crick paired bases (guanine (G)--cytosine (C) and adenosine (A)--thymine (T)). The coding sequences for genes are thus ``buried'' inside the DNA duplex and consequently poorly accessible for any kind of damage (from, e.g., OH$^{-}$ and H$_3$O$^{+}$ ions or mutagens~\cite{Frank1987a}). 

Although this DNA structure in double helix is robust enough, it is however sufficiently loose to allow the opening of the double helix. DNA hybridization and de-hybridization capabilities are in fact fundamental processes in molecular and cell biology. Spontaneous opening is rare and transient at physiological temperature but it is promoted by specialized enzymes when the genetic code has to be accessible to molecular machineries. This occurs during transcription (DNA translation into messenger RNA), replication (DNA copy), recombination (DNA ``cut-and-paste''), repair or any enzyme binding on single strands. For instance, RNA polymerases ``read'' single-stranded DNA (ssDNA), and the formation of a so-called transcriptional bubble at the transcriptional starting site is required to initiate transcription. At high enough temperature $T$ (or low enough ionic strength), thermal energy promotes partial or even complete base-pairs dissociation, a phenomenon called DNA denaturation or melting. This property is fully exploited, e.g., in Polymerase Chain Reaction (PCR) where the DNA is denatured before short sequences are hybridized which are then extended, this occurring during several dozen cycles controlled by the temperature. DNA hybridization kinetics, of evident interest in the biological processes evoked below, is also of importance for fast-developing nanotechnology designs involving DNA hybridization.

Local duplex openings, commonly called denaturation bubbles, can be observed at any temperature. They are rare at room temperature where the fraction of open base-pairs in dsDNA is experimentally found to be $10^{-6}$--$10^{-5}$ for A-T pairs~\cite{Gueron1995,vonHippel2013,FrankKamenetskii2014}, probably an order of magnitude smaller for G-C pairs~\cite{Krueger2006}. 
Their nucleation probability (and their size) increases with $T$~\cite{Krueger2006,Palmeri2008} and they proliferate when getting close to the so-called denaturation or melting temperature $T_m$ (precisely defined as the midpoint of the transition, where one half of the base-pairs are denaturated). Due to sequence heterogeneities, with AT-rich segments being less stable than GC-rich ones, the former tend to melt at lower temperatures~\cite{Nagapriya2010}. Hence, at physiological ionic strengths, $T_m$ varies between 50 and 90$^\circ$C. Therefore the DNA macromolecule manifests more thermally driven ÒbreathingÓ fluctuations at physiological temperatures in the AT-rich regions~\cite{vonHippel2013}. This property can be exploited by Nature by correlating starting sites of molecular machineries where duplex opening must be initiated and the more fragile  AT-rich regions. Duplex opening is also at play when non-linear elastic properties of DNA are involved when the molecule is strongly bent~\cite{Destainville2009} or negatively supercoiled~\cite{Adamcik2012}. This is also of biological relevance, e.g. in nucleosomes and plasmids. 

\subsection{Motivations for the study of base-pairing dynamics}

Since the discovery of DNA three-dimensional structure by Watson and Crick in 1953~\cite{Watson1953}, the DNA double helix internal dynamics have been the subject of many theoretical and experimental investigations not only from a biological or biochemical perspective but also from a physical point of view. This dynamics corresponds either to DNA zipping from the denaturated to the duplex state, or its opening from the duplex to the single stranded state, or diffusion of denaturation bubbles. Understanding the internal dynamics of DNA when in solution is indeed considered as a pivotal first step before going further and tackling the whole problem of dynamics in the nucleosome or chromatin, where proteins bound to DNA (such as histones) and higher levels of compaction make the problem even more complex~\cite{Alberts2002}. Many proteins interact with the dsDNA and some of these protein-DNA complexes are correlated to the opening of the double helix. Apart form the RNA polymerase evoked above, one can cite the DNA polymerase which plays a central role in the DNA replication, the enzymes acting in the homologous recombination~\cite{Alberts2002} such as RecA in the Escherichia coli bacteria or RAD51 in humans, and also the Single Strand Binding proteins which are essential to maintain the genetic information~\cite{Meyer1990}. Active opening of nucleic-acid double strands by helicases has also been proposed to be the result of a competition between passive zipping and helicase processivity on the single strand to which it is bound~\cite{Betterton2005}. Zipping dynamics are then directly involved in this interaction between nucleic acids and a wide class of protein at work in many molecular processes. However, we shall see that even \textit{in vitro} where the DNA is isolated, many questions remain open from an experimental point of view. 
Designing reliable theoretical models in close collaboration with experimentalists~\cite{Frank1987b,FrankKamenetskii2014} might help the latter to distinguish between different mechanisms when exploring the base-pairing kinetics in the future.

From the theoretical point of view, the richness of this quasi-one-dimensional dynamical systems comes from: 
\begin{enumerate}
	\item The cooperativity between adjacent base-pairs (related to the so-called ÒstackingÓ $\pi-\pi$ interactions between cycles), which favors a sharp crossover (within a few Kelvins) between the closed and open states when destabilizing the nucleic base interactions by raising the temperature above $T_m$ (alternatively, denaturation can be reached by decreasing the ionic strength at fixed $T$, or again by modifying the $p$H value~\cite{Lazurkin1970,Gotoh1983}).
	\item The much smaller bending and torsional moduli of a DNA bubble as compared to the double helix DNA  (given in Table~\ref{tab2}), which effectively couples base pairing (ÒinternalÓ) and chain conformation (ÒexternalÓ) degrees of freedom, increases the entropy in the denaturated state and thus destabilizes the duplex state at high enough temperatures (see e.g. Refs.\cite{Yan2004,Chakrabarti2005,Manghi2009}). DNA elastic or mechanical properties are also involved in many biological processes, notably interaction with partner proteins. 
	\item	The helical character of the molecule in its duplex state that is further stabilized through geometrical entanglement and can furthermore be supercoiled (negatively or positively). This provides an additional strain-assisted way to go from the closed to the denaturated state, notably used in Nature~\cite{FrankKamenetskii2006}, for example by actively applying a torque to the molecule through specialized enzymes~\cite{Lohman1996,Alberts2002} or by negatively supercoiling chromosomes and plasmids\footnote{Plasmids are kilobases-long \textit{closed} double-stranded DNA mini-circles, most commonly found in bacteria. As compared to the larger chromosomes containing all the essential genetic information, plasmids usually are very small and contain ``inessential''  information. Being closed, they can be passively under-twisted \textit{in vivo}.} in order to facilitate their opening~\cite{FrankKamenetskii1997,Fye1999,Jost2011}. Conservation of the Linking number Lk, a topological quantity related to the dsDNA helicity and defined below, imposes a constraint on the helical twist dynamics. Indeed, twist must be unwound at the molecule ends in order for the single strands to separate. Conversely, renaturation requires rotation of the molecule for accumulation of twist.
	\item	The heterogeneity of genetic sequences (encoded by an ÒalphabetÓ of 4 nucleotides or ÒlettersÓ, A, T, G and C, paired in tandems A-T and G-C in the double strand). The ensuing quenched disorder makes in principle thermodynamical and dynamical properties sequence-dependent, because AT rich regions are less stable than GC rich ones\footnote{In this respect, the terminology of ``random sequence'' used by physicists, which is certainly not meaningful from a \emph{biological} perspective, simply means that the sequence does not contain some specific motifs (e.g., periodic motifs, A-tracts, or palindromes) that would make the molecule have a behavior different from the average one, from a \emph{physical} point of view.}. Sequence is of notable biological importance when DNA/enzyme interactions require DNA opening at relevant sites.
\end{enumerate}

Note that even though both nucleic acids share many similarities, Watson-Crick base-pairing kinetics in DNA molecules are relatively simpler than in RNA because the paired geometry is a double helix, whereas secondary and tertiary structures of RNA can be much more complex~\cite{Alberts2002,Bundschuh2006}. DNA also possesses secondary structures different from the simple double helix (e.g. triple or quadruple helices and hairpins, cruciforms or loops~\cite{FrankKamenetskii1997,FrankKamenetskii1995,McMurray1999,Burge2006}), but they are not as often associated with biological properties as in RNAs. However, DNA base-pairing dynamics remains highly challenging because of the interplay between internal (base-pairing) and external (chain conformation) degrees of freedom in the duplex state. Zipping of DNA also shares some similitudes with the helix-to-coil transition of $\alpha$-helices~\cite{Jayaraman2007}, even though a single polymer is involved in this case.

In addition to their obvious implications in molecular biology, DNA base-pairing mechanisms have recently known a growing interest in nanotechnological applications. This started with PCR in the 1980s, more recently followed by aptamer design~\cite{Tuerk1990,Ellington1990}, DNA biochips, DNA combing~\cite{Michalet1997}, or DNA origami~\cite{Goodman2005,Rothemund2006,Zadegan2012}.

\subsection{Report outline and presentation of theoretical approaches}

\begin{figure}[h!]
\begin{tabular}{cc}
\parbox{6cm}{\includegraphics*[width=5cm]{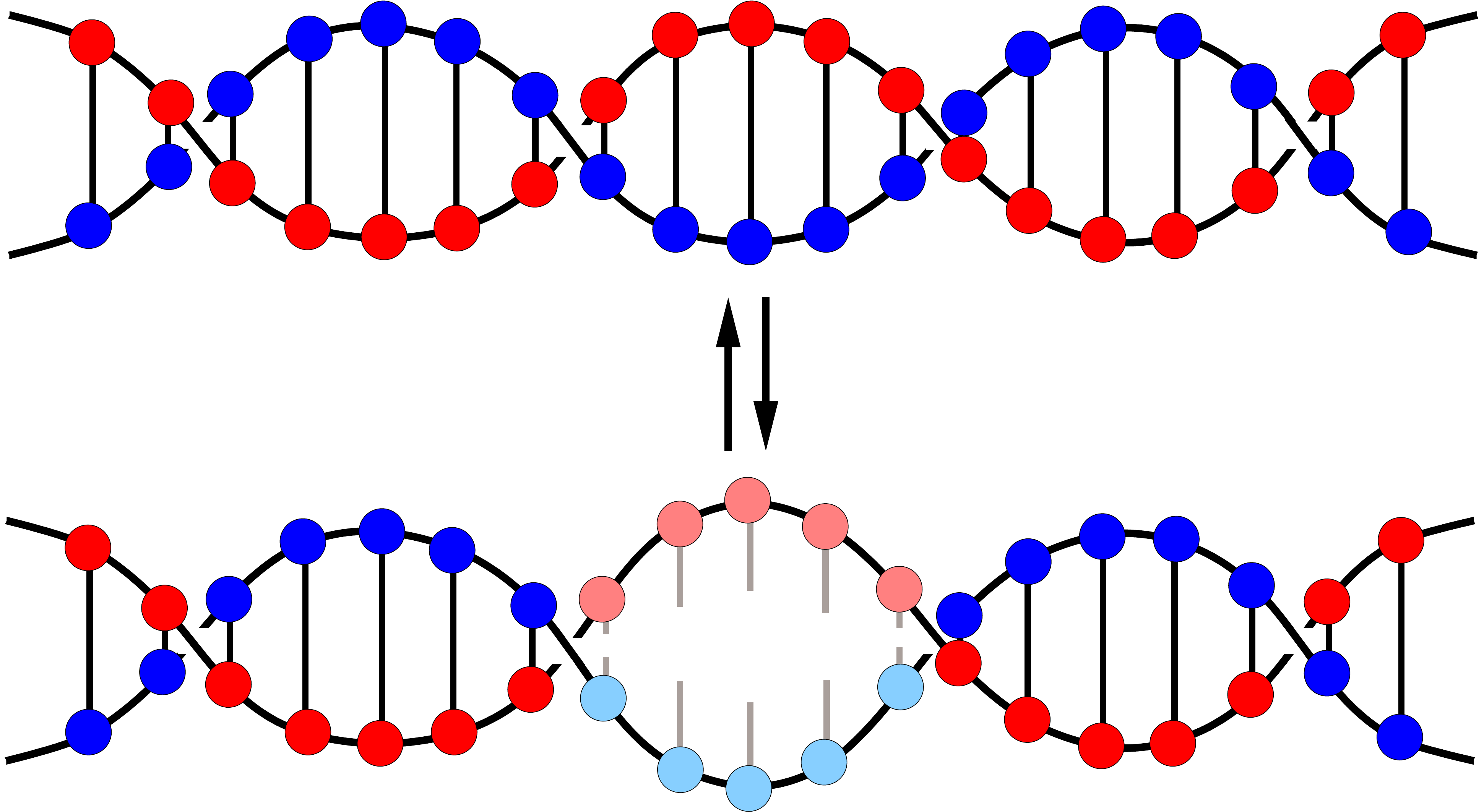}} & \parbox{6cm}{\includegraphics*[width=6cm]{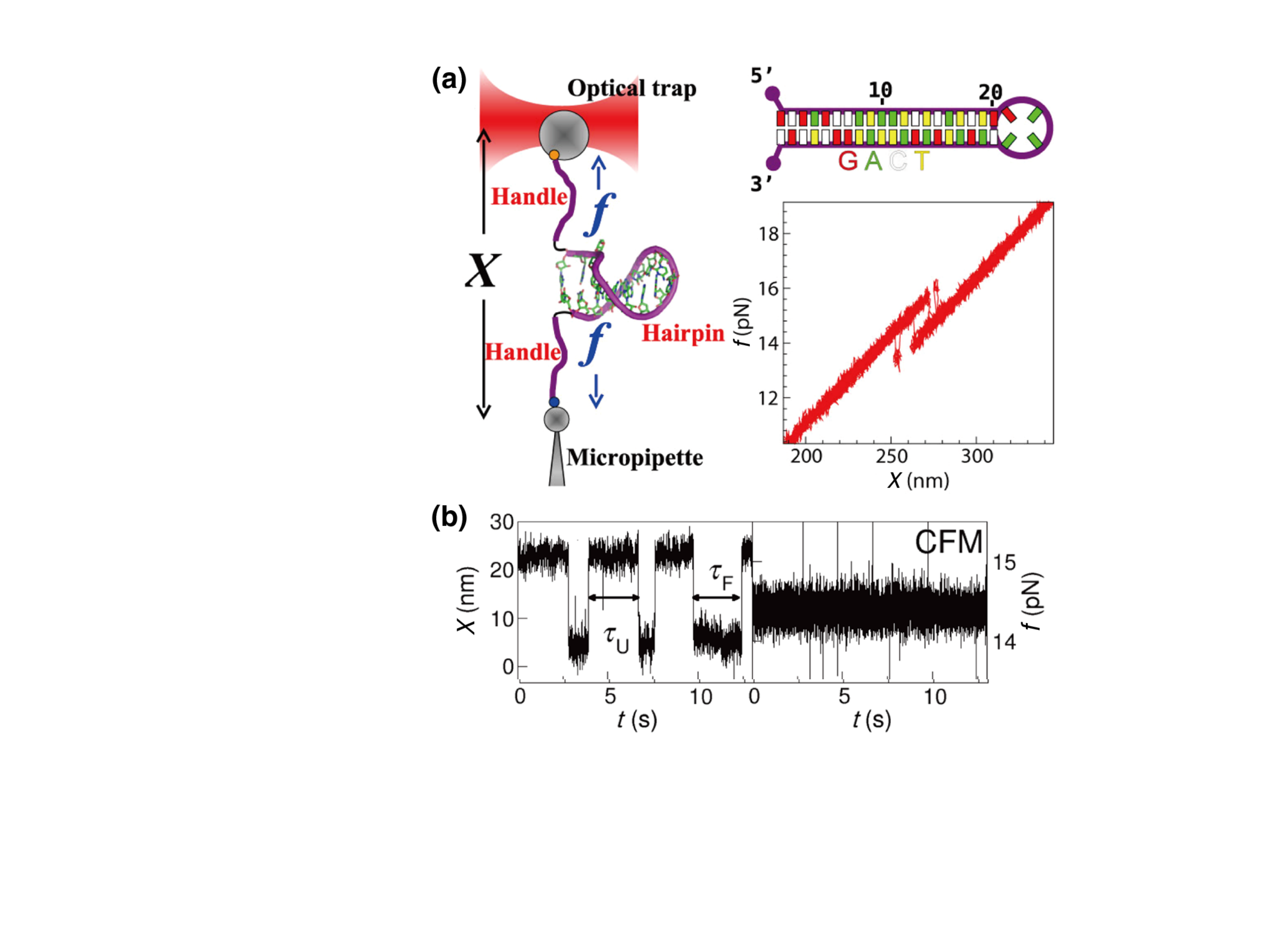}} \\
{\bf (a)} & {\bf (b)} \\
& \\
\parbox{8cm}{\includegraphics*[width=8cm]{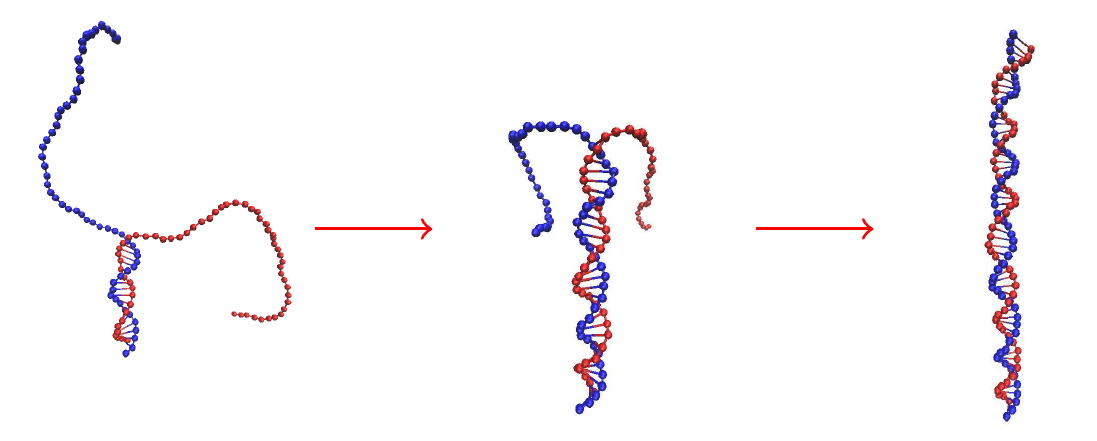}} & \parbox{6cm}{\includegraphics*[width=6cm]{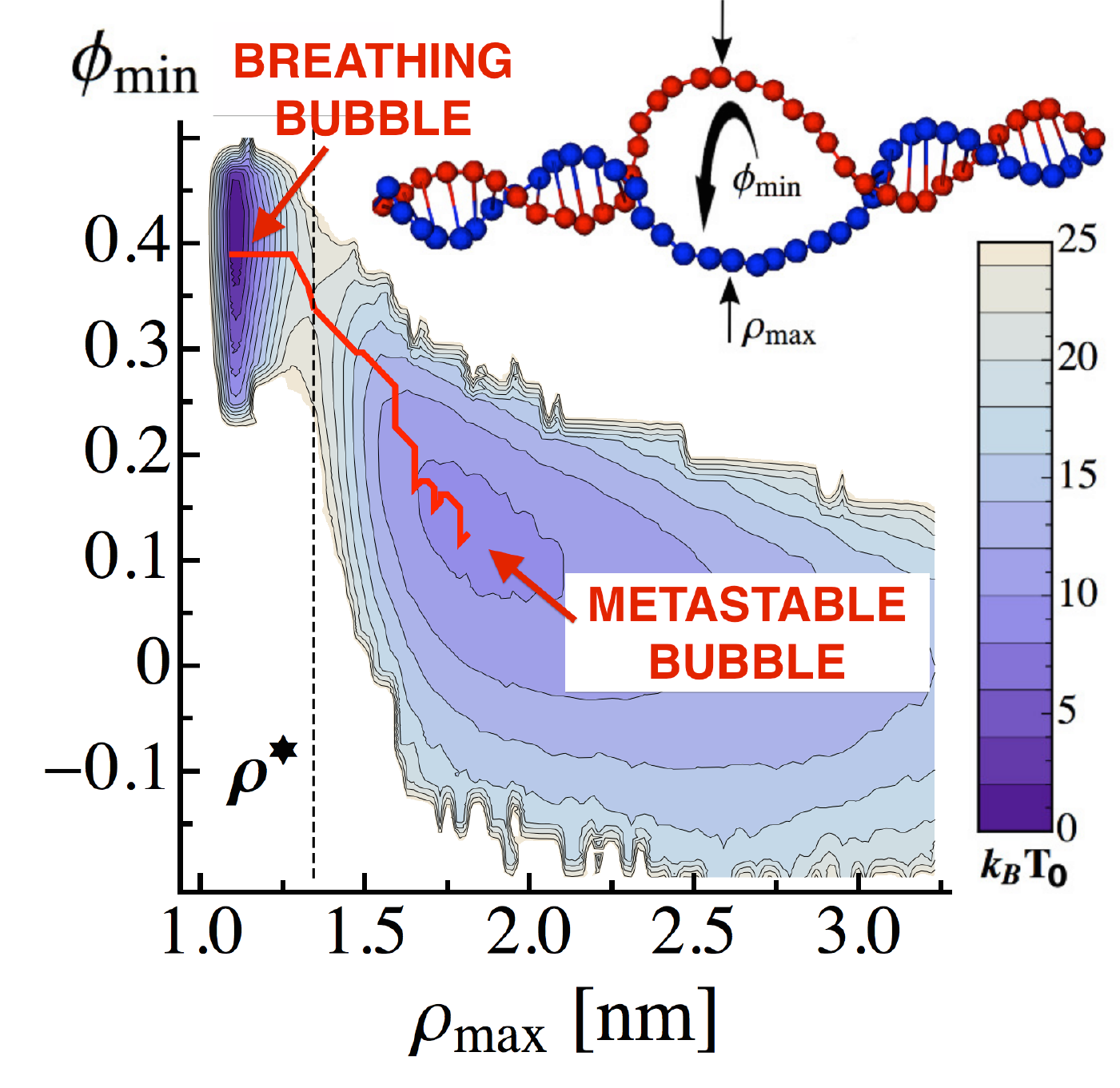}} \\
{\bf (c)} & {\bf (d)} \\
\end{tabular}
\caption{Some base-pairing mechanisms explored in this Report, where the respective roles played by internal (base-pairing) and external chain orientational degrees of freedom vary. 
(a)~Cartoon of a ``breathing bubble'': opening of few base-pairs, involving disruption of hydrogen bonds between the complementary bases, followed by their rapid renaturation. Chain orientational degrees of freedom can be considered as frozen during this fast event occurring on the nano-second time-scale, only base-pairs dynamics are at play. 
On this cartoon, we have chosen to represent breathing fluctuations as a small increase of the inter-strand distance. Alternatively, one or few base pairs might rotate around their respective strands, thus disrupting stacking with neighboring pairs. This different breathing mechanisms will be reviewed in Section~\ref{bpb}.
(b)~Principle of hairpin unfolding experiments. A force $\mathbf{f}_{\rm ext}$ is applied to the strands of a short DNA hairpin (20~bp in the figure). For a force of about 15~pN, two states are equally accessible to the hairpin, which is either open or closed. In both states, chain orientational degrees of freedom can be pre-averaged to obtain a 1D two-well free-energy landscape $F(X)$ as detailed in Section~\ref{1Damod}. Taken from~\cite{Hayashi2012}. 
(c)~Out-of-equilibrium zipping: during this processive phenomenon, chain orientational degrees of freedom cannot be considered neither as frozen nor in a quasi-equilibrium state. Both base-pairing and chain orientational degrees of freedom must be considered on an equal footing, which will be tackled in Section~\ref{ffeq}. Taken from~\cite{Dasanna2013T}.
(d)~Closure of a metastable bubble. A quasi-static approximation is valid here, and an effective 2D free-energy landscape can be constructed. In the displayed landscape, where coordinates correspond to base-pairing and torsional degrees of freedom respectively, two energy wells are distinguishable. They are related to ``breathing'' and ``metastable bubbles'' respectively. Base-pairing and chain orientational degrees of freedom must also be considered on an equal footing here (see Section~\ref{ffeq}). Taken from~\cite{Sicard2015}.
\label{MainFig}}
\end{figure}

In this Report, we chose to categorize in three major sections the numerous theoretical findings about DNA base-pairing dynamics of the past half-century. This classification, summarized in Table~\ref{tab1}, dwells on the respective roles played by internal (base-pairing) and external (chain conformation) degrees of freedom, as follows (Figure~\ref{MainFig}). 

\begin{table}
\begin{center}
\begin{tabular}{|p{6cm}|c|c|c|c|}
\hline
Addressed regime & Time-scale & Length-scale & Example & Section  \\
\hline
\hline
External (chain conformation) degrees of freedom are slow variables and are considered as  frozen; transient bubbles ``breathe'', e.g., by thermal activation. & $<1$~ns & $<10$~bp & Figure~\ref{MainFig}a & \ref{bpb} \\ 
\hline
Chain orientational degrees of freedom are fast variables and are pre-averaged. Base-pairing dynamics is explored in this pre-averaged free-energy landscape (\emph{quasi-static} approximation). & $>1~\mu$s  &  $>10$~bp & Figure~\ref{MainFig}b & \ref{1Damod} \\
\hline
Internal (base-pairing) and external (chain conformation) degrees of freedom must be considered on an equal footing.& $>10-100~\mu$s & $> \ell_p \simeq 150$~bp & Figure~\ref{MainFig}c & \ref{ffeq} \\
\hline
\end{tabular}

\caption{The different regimes with their typical time and length scales addressed in the works reviewed in this Report. The double-strand persistence length, $\ell_p$, is defined as the polymer orientational correlation length. The length-scales are given in base-pair (bp) units.
\label{tab1}}
\end{center}
\end{table}

\begin{enumerate}
	\item Section~\ref{bpb} addresses the local small distortions of the distance between strands of the DNA double helix (Figure~\ref{MainFig}a). At very short timescales (typically $\lesssim1$~ns), chain degrees of freedom remain frozen (or are very slow variables), as they are assumed to follow much slower dynamics than base-pair opening and closure. When now-and-then opening below the melting temperature $T_m$, DNA ``breathing bubbles'' (or ``breathers" for breather modes originally precisely defined in the context of non-linear physics~\cite{Dauxois1993a,Yakushevich2004}) are thus short-lived 
($\sim 1$~ns) and we shall also call them ``transient bubbles''. They are smaller than ten or so base-pairs.
	
	Note that some biophysicists and biologists also generalize the use of the word ``breather'', to refer to real openings of the dsDNA due to thermal fluctuations \textit{in vitro} (as opposed to opening induced by active proteins \textit{in vivo})~\cite{vonHippel2013}. These  long-lived denaturation bubbles are therefore hardly ever observed at physiological temperature. Close to $T_m$ they become more probable (usually close to the melting temperature of the AT sequences $T_m^{\rm AT}\simeq 50^\circ$C in physiological salt conditions). We shall return to these two different categories of bubbles below.
	\item In the opposing limit where chain orientational degrees of freedom are assumed to be fast variables and can thus be integrated out (or pre-averaged), we present in Section~\ref{1Damod} some models in which a quasi-equilibrium free-energy landscape for base-pairing (the slow variables) is constructed. These models have been historically developed to study the zipping/unzipping of long DNAs. The use of such quasi-static models is fully justified in some circumstances such as the opening/closure of small DNA hairpins (few tens of base-pairs) below $T_m$, a collective phenomenon occurring on the $\mu$s timescale which requires to cross a high free-energy barrier and is often modeled by a two-state approximation (Figure~\ref{MainFig}b). The opening induced by applied forces and torques is also evoked in this section.
	\item Section~\ref{ffeq} focuses on a wide class of phenomena where internal and external degrees of freedom must be considered on an equal footing because none of them can be considered as frozen or pre-averaged. This notably concerns DNA zipping below $T_m$ (Figure~\ref{MainFig}c) and its unwinding dynamics when denaturing above $T_m$, which occur on times {ranging from tens of micro-seconds to seconds} for long constructs. The DNA lengths studied in this regime are above the persistence length $\ell_p$, so that scaling laws can be anticipated. In some cases as the closure below $T_m$ of a ``pre-equilibrated'' bubble in the middle of a chain, the far-from equilibrium zipping stops in a metastable bubble of $\simeq 10$~bp, followed by a $\sim 50$~$\mu$s-long final bubble closure caused by a coupling between the chain bending and torsional degrees of freedom and the base-pairing ones (Figure~\ref{MainFig}d).
\end{enumerate}

The existence of two distinct categories of bubbles deserves a particular attention. An important difference is to be made between Òtransient bubblesÓ (opening of few base-pairs rapidly followed by renaturation) and Òmetastable bubblesÓ where the chain has time to equilibrate after opening the bubble and before re-closing it. By measuring imino-proton exchange with water by proton NMR in close to physiological conditions (albeit at room temperature or less; see~\cite{vonHippel2013,FrankKamenetskii2014} for detailed historical reviews), W\"{a}rml\"{a}nder and his collaborators found in 2000 two distinct timescales associated with lifetimes of open AT pairs in short duplexes~\cite{Warmlander2000}. The first ones dwell in the nanosecond range, as in earlier works~\cite{Gueron1987,Frank1987a,Gueron1995,vonHippel2013}. Quite interestingly, these new experiments also indicated the existence of longer open states, with lifetimes in the micro-second range, concerning less than 10\% of the opening events. It is tempting to speculate that the shorter lifetimes correspond to transient bubbles, whereas the longer ones should correspond to metastable bubbles. Interestingly also, long closure times ($>10$~$\mu$s) observed by Fluorescent Correlation Spectroscopy by Libchaber and its collaborators~\cite{Altan2003} presumably belong to this last category as well (these authors did not mention W\"{a}rml\"{a}nder and collaborators' work). These different experiments are based upon relatively short constructs (less than 30~bp) with CG clamping extremities and AT cores of variable lengths. The somewhat different lifetimes of metastable bubbles observed with the two different experimental techniques potentially come from the different lengths of the AT cores (8~bp in W\"{a}rml\"{a}nder and collaborators' longest construct, 18~bp in Libchaber's group experiments). We shall see below that the optimal metastable bubble length is around 10~bp.

Recently, Phelps \textit{et al.}~\cite{Phelps2013} measured DNA bubble timescales of $200~\mu$s using single-molecule F\"orster resonance energy transfer (FRET) and linear dichroism. They attributed the fact that their bubble lifetimes were $\simeq5$ times larger than those of Ref.~\cite{Altan2003} to the difference between labelling schemes used in the two studies. The proximity between the tagged base-pair and the stationnary ds-ssDNA fork (distance of 14~bp) might also be responsible for this increased lifetime. They also studied bubble lifetime closer to the fork with or without helicase (and GTP), showing that these timescales increase by a factor 3 in presence of the enzyme.

In these different contexts, DNA can be investigated theoretically at a more or less coarse-grained level. In principle, the more realistic descriptions belong to all-atom modeling with either explicit or implicit solvent. However, in the context of molecular and cell biology where complex mechanisms are often at play, one limitation of such approaches is that the key degrees of freedom associated with a given biological process are not always easily identifiable amongst the atomistic details and physical insight concerning the main mechanisms at play can be lacking. But the most critical weakness of all-atom simulations lies in the attainable timescales when considering a molecule with a size of biological interest. Real times are generally limited to a dozen of nanoseconds, they can be pushed up to a fraction of microsecond using huge numerical facilities, but timescales of interest can be in the millisecond range, possibly longer, as we have just seen. Alternative methods have been explored to reach longer timescales. A first step towards an increase of the timescale numerically accessible consists in considering small physico-chemically relevant groups of atoms (such as a water molecule, a sugar, a phosphate, or a nucleo-base) instead of single atoms. Each small group is then considered as a single effective particle, thus leading to coarse-grained models. But in most of the cases, the gain remains insufficient in regard of the needs.

Beyond this, another promising approach consists in going a step further in the coarse-graining process, thus giving up the ambition of accounting for realistic interactions at the atomic level. Effective \textit{mesoscopic} models are then at stake, where elementary ÒparticlesÓ are larger subparts of the nucleic acid (e.g. nucleotides). The relevant degrees of freedom must be identified, and then the model parameters must be finely tuned in order to correctly account for the microscopic degrees of freedom that are implicitly integrated out in the process. The gain in terms of simulation times and attainable system sizes is then of several orders of magnitude. Biologically relevant questions can be addressed in a wider class of cases, since several hundreds of base-pairs can be tackled on the milli-second timescale. The resultant increase in simple physical insight can also be substantial because each level of coarse-graining reduces the questions of interest to their more fundamental aspects. 
Analytical approaches from statistical physics relying upon mesoscopic models, eventually coupled to hydrodynamics, elasticity, or electrostatics depending on the biological issue, are also always very useful to understand the fundamental physical mechanisms at play and to span the whole range of parameters. Of course these analytical approaches are only possible for very simple models where the key degrees of freedom have been identified. Among them, continuous models, where two DNA single strands seen as Worm-Like Chains (WLC)~\cite{DoiEdwards2004} are inter-wound, can also be useful. 

The tools by which the mesoscopic models are solved are those of out-of-equilibrium statistical mechanics and non-linear physics, ranging from numerical methods (Molecular Dynamics, Brownian or Langevin dynamics, Monte Carlo simulations, biased sampling) to analytical ones (master equation, Fokker-Planck and Smoluchowski equations, mean first passage times and Kramers theory, Rouse or Zimm approximations). While they significantly increase the simulation cost or the complexity of the equations, hydrodynamic interactions can be approximately treated, for example by using the Oseen or Rotne-Prager tensors~\cite{DoiEdwards2004,Manghi2006}, in both numerical and analytical approaches.

In this Report, an important distinction will be made between \textit{quasi-static} processes and \textit{far-from-equilibrium} ones~\cite{Chaikin1995}. In the first case, most degrees of freedom are considered as fast variables that can be pre-averaged in an effective free energy depending on a few slow variables only. Such an approximation turns out to be inappropriate in some circumstances for DNA base-pairing dynamics, notably as far as chain orientational degrees of freedom are concerned. One then has to switch to far-from-equilibrium statistical mechanics.

The typical experimental values of the main quantities involved in DNA dynamics as discussed in this Report are listed in Table~\ref{tab2} in the Appendix.

\subsection{A historical survey of DNA mesoscopic modeling}

As motivated just above, following the discovery of the double-helix structure of DNA by Watson, Crick and Franklin in 1953, a succession of mesoscopic models have been proposed in the literature, with the primary objective to describe \textit{equilibrium} thermodynamical properties of DNA before addressing dynamic ones. Their respective merits will be discussed in this Report, thus we only briefly present here the most popular ones in a chronological order, without any ambition of exhaustivity. 

\begin{itemize}
	\item In the 1960's, the first physical mesoscopic models were inspired by the 1D Ising model of solid-state physics~\cite{Lazurkin1970,Wartell1972}. They simply said that a base-pair could be either open or closed, thus attributing a boolean-like variable $\sigma_i$ (in fact a classical ``spin'' $\sigma_i=\pm1$) to each base-pair. In such a model, the DNA is seen as a ladder, the rungs of which are the hydrogen bonds between base-pairs, either broken or unbroken. The helicity is therefore secondary at this level of modeling. The coupling between adjacent base-pairs accounts for their cooperativity, itself coming from the stacking $\pi-\pi$ interactions between consecutive cycles along a same strand. A rapid calculation shows that in absence of the cooperativity term, the transition occurs on more than 50~Kelvins, which is in evident contradiction with the experiments. The cooperativity narrows the transition to few Kelvins, as expected, but the transition cannot strictly speaking be a thermodynamic one because one deals with a 1D system with short-range interactions. The main contribution to the double helix stability actually does not come from the hydrogen-bonds alone, but the stacking of nearest-neighbor base-pairs contributes on an equal footing. The ``sequence-dependent stacking'' energy model was proposed later to realistically take the sequence into account by making the model parameters (on-site free-energy cost of base-pair opening and cooperativities) dependent on the position in the polymer~\cite{SantaLucia1998}. These models do not consider the chain entropy and elasticity and they give a semi-quantitative understanding of equilibrium properties and of melting profiles (see the reviews~\cite{Gotoh1983,Wartell1985,FrankKamenetskii2014}).
	\item In 1970, following the work of Zimm and Bragg in the context of the helix-coil transition in peptides~\cite{Zimm1959,Zimm1960}, Poland and Scheraga studied an improved version of the Ising model~\cite{PSbook}. As compared to its predecessors, an additional so-called ``loop-entropy'' cost $\Delta S_{\rm LE} = -k_B c \ln n$ was associated with each bubble of length denoted by $n$ ($c$ on the order of unity is the so-called loop exponent). Its origin lies in the requirement that a denaturation bubble in the middle of the polymer has its two extremities closed, thus forming a (self-avoiding) closed loop of total length $2n$. Thus this model is the first one to take some chain orientational degrees of freedom into account in an effective way. This leads (at equilibrium) to an effective long-range interaction in this originally 1D model, which in principle allows a true thermodynamic transition in the infinite polymer length limit. Several studies have focused on the exact nature of this transition, even though this is a rather academic issue since polymers are finite-sized by nature~\cite{Kafri2000}. The phenomenological numerical parameters in Poland and Scheraga's original algorithm were evaluated by Blake and Delcourt~\cite{Blake1998}. This led to the celebrated MELTSIM software~\cite{Blake1999}, now routinely used in genetics laboratories in order to predict with a correct accuracy the melting profile of any DNA sequence. In a series of papers starting in 2003, Metzler and his collaborators studied the dynamical properties of this model, thus assuming that chain orientational degrees of freedom leading to $\Delta S_{\rm LE}$ can be considered as pre-averaged. This will be discussed below.
	\item In 1979, Benham proposed a mean-field 1D model of denaturation in superhelically stressed DNA~\cite{Benham1979}, which was subsequently improved~\cite{Benham1980,Benham1990,Fye1999}. Basically, he analyzed how the supercoiling stress is  distributed among the (more flexible) denatured regions and the closed ones, by minimizing the torsional elastic energy while taking account the internal degrees of freedom in an Ising-like fashion. The model uses the topological relation between the linking number ${\rm Lk}=N/p$ for a dsDNA ($p\simeq 10.5$~bp is the pitch fo the helix and $N$ the number of base-pairs) and the twisting number ${\rm Tw}=\sum_{i=2}^N \phi_i$, the twist of the dsDNA ribbon (where $\phi_i$ is the angle between two adjacent base-pair vectors). It is given by the Fuller-White formula~\cite{Fuller1971,White1969}
\be
{\rm Lk}={\rm Tw} + {\rm Wr}
\label{Fuller}
\ee
where the writhing number Wr measures the degree of supercoiling and depends only on the shape of the ribbon axis. Neglecting the writhe, Benham related the total twist in the bubble to the residual twist in the ds region (with two different torsional moduli in ss and dsDNA regions) and the imposed linking difference $\Delta {\rm Lk}={\rm Lk}-{\rm Lk}_0$ between the imposed linking number and the dsDNA equilibrium one. It successfully predicted the location of the stress-induced duplex destabilization sites in specific genes~\cite{Benham1996,Leblanc2000}. The model can also explicitly take the sequence into account by setting two different energies of denaturation for A-T and G-C base-pairs~\cite{Jost2011}. 
	\item Starting in 1983, Yomosa, Takeno and later Yakushevich and others proposed a 1D model where the important variables are the rotational degrees of freedom of base monomers with respect to a rotational axis, $\varphi_{i,1}$ and $\varphi_{i,2}$, with one such axis for each strand~\cite{Yakushevich1989,Yakushevich2004}. This model was studied with a non-linear physics point of view. Bubble breathers are seen as solitonic excitations such as in chains of pendula coupled with torsional springs. In addition to equilibrium properties, this led to an intensive study of the dynamical properties of the denaturation processes in these systems, which will be detailed in this Report. Sequence effects have also been incorporated in this model.
	\item In 1989, Peyrard and Bishop~\cite{Peyrard1989} explored a different approach, subsequently improved in 1993 by the same authors and Dauxois (PBD model)~\cite{Dauxois1993a}. In this other non-linear 1D model, the distance between the two bases of a given base-pair is now a continuous variable $y_i$. The potential energy between both bases is modeled by a Morse potential $U_0(e^{-\alpha_1 y_i}-1)^2$. The denaturation above $T_m$ is simply due to the translational entropy gained by the $y$ coordinates, using the analogy with polymer desorption done by de Gennes in 1969~\cite{deGennes1969}. The interest of such a model is that it is fully solvable in its simplest form. Chain orientational degrees of freedom are not taken into account either but sequence effects can be included by fitting simulations of the model to ultraviolet (UV) melting curves.
	\item In the first decade of the 21st century, several coupled models were designed to explicitly take into account the coupling between internal and external degrees of freedom. This coupling arises from the lower bending ($\kappa$) and/or torsional ($C$) elastic moduli in the denatured form as compared to the duplex one. Typically, $\kappa_{\rm ds}/\kappa_{\rm bubble} \sim 25$~\cite{Yan2004,Manghi2009} and $C_{\rm ds}/C_{\rm bubble} \sim 100$~\cite{Kahn1994,Bryant2003,Lipfert2010} (see Table~\ref{tab2}). Chain orientational degrees of freedom are treated using a discrete WLC model, while internal ones are again modeled in the traditional Ising-like fashion. But the chain elastic parameters explicitly depend on the internal state, thus coupling the chain and the base-pair states. The gained entropy above the denaturation is due to the increase of the number of conformations in the denaturated state. 
\begin{itemize}
	\item Storm and Nelson mixed the features of both the WLC and the Bragg-Zimm models~\cite{Storm2003}. Their goal was to account for single-molecule experiments where a force is applied to the polymer extremities. Their so-called Ising-Discrete Persistent Chain model takes two different possible conformations of DNA into account, each with its own elastic constants as explained above, and it couples chain orientational degrees of freedom to the applied force. However, the addition of the term corresponding to the external force in the Hamiltonian prevents an exact solution of the model. An approximate variational scheme had to be implemented.
	\item When proposing their coupled model, the primary goal of Yan and Marko~\cite{Yan2004} (and others~\cite{Ranjith2005} with the same model) was to give a theoretical foundation to DNA cyclization experiments. Their solvable model was a simplified Ising-like one but without cooperativity, coupled to chain orientational degrees of freedom in the following way. A bending energy term $\frac12 \sum_i  (\delta_{n_i,0} \kappa_{\rm ds} + \delta_{n_i,1} \kappa_{\rm ss} )(\mathbf{t}_{i+1}-\mathbf{t}_{i})^2$ was added in order to assign different elastic moduli to the open ($n_i=1$) and closed ($n_i=0$)  base-pair states. The $\mathbf{t}_{i}$ are unit vectors giving the chain orientation at the level of each base-pair. 
	\item Palmeri, Manghi and Destainville~\cite{Palmeri2007,Palmeri2008,Manghi2009} went further by proposing a model were the cooperativity between adjacent base-pairs could be fully taken into account. This model was exactly solved by the transfer matrix technique. Note that a very similar approach had been proposed in 1993 by Palmeri and Leibler in 2D~\cite{Palmeri1993} and then by Chakrabarti and Levine~\cite{Chakrabarti2005,Chakrabarti2006}, but its application to the study of denaturation bubbles in dsDNA had not been explored at that time (only the helix-to-coil transition of proteins was considered, but both problems share strong similarities). When coupling chain orientational degrees of freedom to an applied force, such a model also permits to describe in an approximate but very accurate analytical way the situations where a force is applied to the DNA molecule extremities, and to provide a rationale for the rich variety of observed behaviors~\cite{Manghi2012}. 
\end{itemize}
	\item Even though the previous coupled models can in principle be apprehended analytically in order to address their out-of-equilibrium properties, this should require some important, potentially ill-controlled approximations. Consequently, in the 5 past years, several groups have adopted an alternative strategy consisting of designing somewhat more realistic models while giving up the idea of exact analytical treatments, and focusing instead on scaling arguments and/or numerical experiments. Bead-spring models have thus been developed, 
where one bead represents one nucleo-base. The polymers can be defined on a lattice or alternatively they can diffuse in a continuous medium. The model parameters are tuned in order to reliably account for equilibrium properties (such as the bending and torsional elastic moduli in both double-stranded and single-stranded forms, the melting temperature, or the helix pitch). 
\begin{itemize}
	\item In 2011, Carlon and his collaborators proposed a simplified model where the two intertwined polymers are defined on a face-centered-cubic (fcc) lattice~\cite{Ferrantini2011}. The two strands are both mutually and self-avoiding, with the exception of monomers with the same index along each strand, which are referred to as complementary monomers. They are superimposed when the duplex is closed. Such superimposed positions of complementary monomers are energetically favored in order to ensure the transition to the duplex form below $T_m$. Molecules as long as 500 base-pairs can be studied, but this model does not display the double-helix geometry in the closed state, nor the correct persistence lengths. It permitted however to study the scaling of the DNA zipping time in function of its polymerization index $N$. 
	\item Two years later, Dasanna, Destainville, Palmeri and Manghi published a more realistic model, displaying correct helix pitch and persistence lengths, as well as a state-dependent elastic twist modulus~\cite{Dasanna2013}. In this off-lattice model, the inter-strand potential is a Morse one as in the PBD approach. This model was used to propose a realistic mechanism for the observed $\mu$s-long closure times of metastable bubbles.
\end{itemize}
We shall later discuss in greater details the advances permitted by these two approaches.
	\item Beyond that, there exist a large variety of numerical coarse-grained models with more than one bead per nucleotide, developed in the two last decades. The increased degree of refinement automatically allows in principle a better account of experimental data. However, this is at the cost of increased computing cost, without necessarily providing better insight into the physically relevant mechanisms. For example, we still do not know whether, above $T_m$, open bases still have stacking interactions with adjacent base pairs in the single strands, or they are totally disordered~\cite{Krueger2006,FrankKamenetskii2014}. These models will be surveyed in the next Section.
\end{itemize}


\section{Base-pair breathing}
\label{bpb}

Since a rich literature has been devoted to the study of effective, mesoscopic models to tackle DNA base-pairing breathing dynamics, this first section primarily intends to review some of the most popular ones while discussing their intrinsic limitations. When deriving mesoscopic models, one has to give realistic values to the parameters appearing in these effective models. In this respect, experiments are of course of primary importance, but all-atom simulations are becoming everyday more efficient to give reliable insight at the base-pair scale. They likely provide very good reference points to more coarse-grained models. So we briefly discuss them in Section~\ref{all_atom} before tackling base-pairing breathing as apprehended by mesoscopic models from the more detailed ones with several beads per nucleotide (Section~\ref{3SPN}) to the models with one bead per base (Sections~\ref{mechmod} and~\ref{dissipation}).

\subsection{All-atom numerical simulations}
\label{all_atom}

As motivated above, we briefly present several types of simulations where the dsDNA base-pairs are modeled at the atomic level. Since, this Report focuses on the base pairing dynamics of relatively long DNA molecules, which involves many atoms and occurs on timescales larger than the nanosecond, we can anticipate that these numerical models with many degrees of freedom remain by nature limited to address such an issue.

With the help of current development of numerics, it is possible to simulate all-atom DNAs of few dozens of bp~\cite{Beveridge2004,Merzel2007}. Beveridge \textit{et al.} reviewed in 2004 the studies of DNA elasticity based on molecular dynamics approaches~\cite{Beveridge2004}. They focused on the effect of the sequence on the  intrinsic curvature and the flexibility of DNA for lengths up to 25~bp,  notably the well-known specificity of A-tracts, and showed that their simulations accounted well for the essential features of experimental observations. For instance, this work provided independent support for the Goodsell-Dickerson ``non-A-tract model'' of DNA intrinsic curvature. One of the major interest in these studies is to follow the effect of mobile ions (which are of course included explicitly) on the elasticity of oligomers. It was also observed that a typical time of 100~ns is needed to equilibrate ions. Other numerical simulations on 11-bp fragments computed for example the elastic constants as a function of the sequence~\cite{Lankas2000}, and found them to be in very good agreement with experimental values. The 5-ns all-atom simulations revealed a marked sequence-dependence of the stretching and torsional rigidities of DNA. In contrast, the bending moduli (though over-estimated by the force-field) appeared to depend weakly on the sequence.

Among these all-atom simulations some of them considered the binding/unbinding of a single DNA base-pair or of short oligomers. Hagan \textit{et al.} studied the characterization of kinetic pathways~\cite{Hagan2003} for a single base-pair in a 3~bp oligomer. Focusing on the flipping of a pyrimidine end base of this oligomer, they followed using transition-path sampling and umbrella sampling the four reaction coordinates (two distances and two energies) that provide a meaningful description of the reaction, associated with the inter-base interaction and the intra-strand stacking one. The simulations revealed two qualitatively different kinetic pathways for the unbinding of the base-pair: one in which the flipping base breaks its intramolecular hydrogen bonds before it unstacks and another in which it ruptures both sets of interactions simultaneously. Moreover they showed that these trajectories are not determined by a stretching of the H-bond (but by a base flipping by rotation around the backbone) nor by penetration of a water molecules and formation ot water-DNA H-bonds.

\begin{figure}[t]
\begin{center}
 \parbox{7cm}{\includegraphics[height=3.5cm]{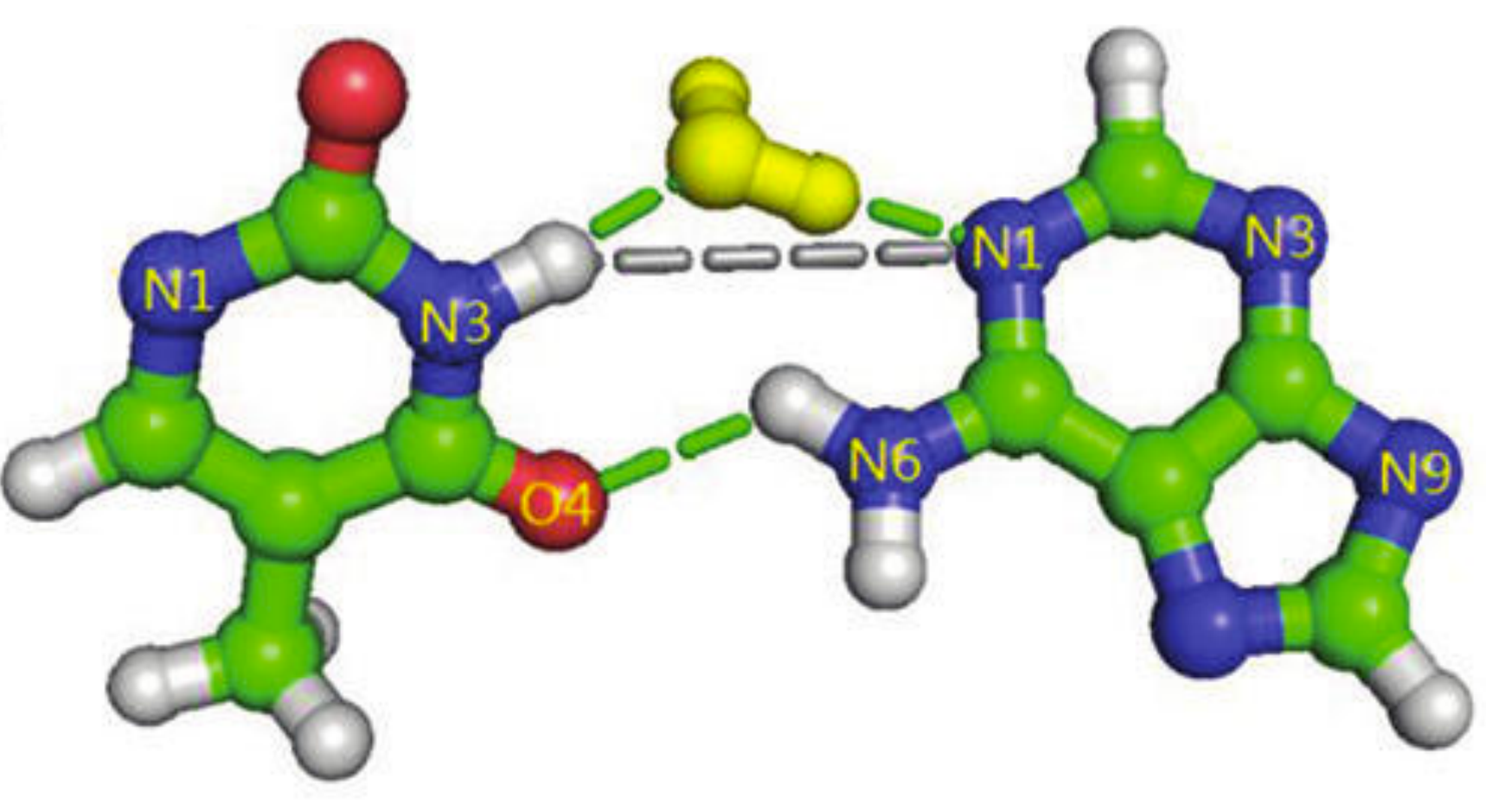}}~~\parbox{8cm}{\includegraphics*[width=8cm]{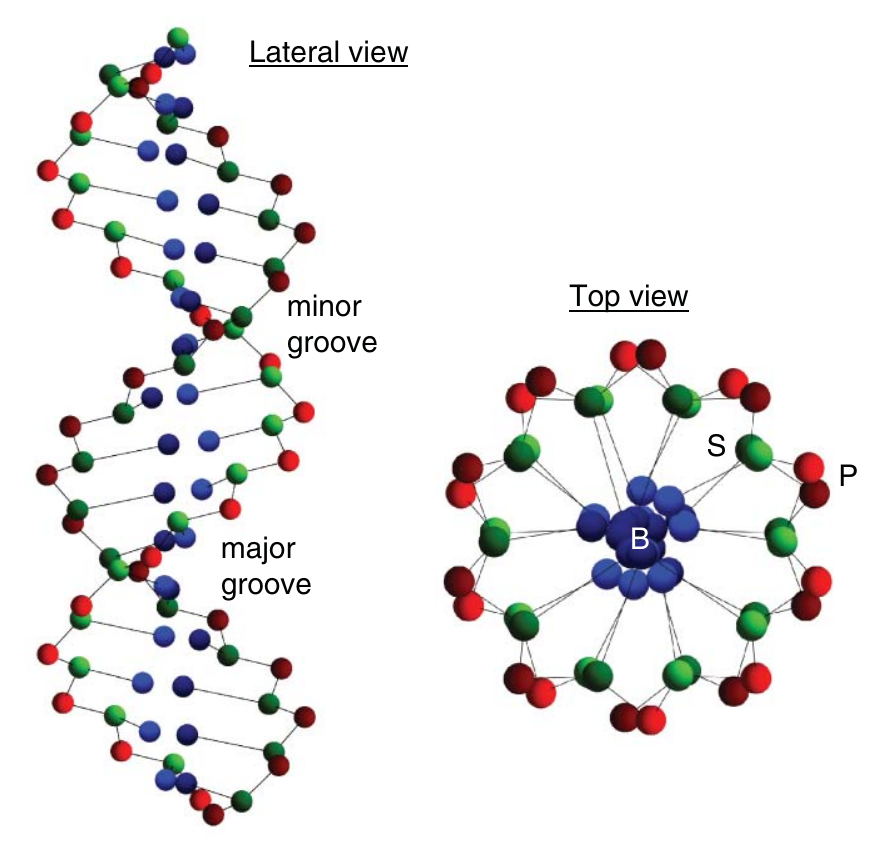}} \\
\ {\bf (a)} \hspace{6cm}  {\bf (b)} \
\end{center}
\caption{(a)~All-atom simulations showing an example of a metastable state that can hinder base-pair closure. A water molecule, in yellow, establishes two hydrogen bonds with the bases to be paired. Closing the base-pair will require the water molecule to be ejected, which represents a free-energy barrier of about $3k_{\rm B}T$. Taken from Ref.~\cite{Qi2011}. (b)~Coarse-graining of a DNA molecule (3SPN model). A model with three beads per nucleotide. Letter P indicates a phosphate group (in red), S indicates a sugar group (in green), and B indicates a base (in blue). Taken from~\cite{Florescu2011}.
\label{ThreeBeads}}
\end{figure}

One the contrary, focusing on the hybridization of a 6 A-T oligomer, some metastable structures and potential barriers have been observed in Ref.~\cite{Qi2011}, associated with the water molecules and their Hydrogen-bonding capability with base-pairs.  Using the solvent accessible surface area as an order parameter, two different metastable configurations were identified where one water molecule forms a ``bridge'' between two approaching bases (see Figure~\ref{ThreeBeads}a), impeding their immediate binding. The potential barrier to be overcome to eventually close the base-pair is then measured to be $\approx 3 k_{\rm B}T$ at room temperature. These non-trivial barriers to base-pairing likely play a role in slowing-down the hybridization dynamics (see also the discussion in Section~\ref{F:barrier}).

Using umbrella sampling simulations, free-energy pathways for base-pair opening were also computed for 13 bp oligomers with repeated GA sequence~\cite{Giudice2003}. This study showed that the larger purine (A or G) base flips favourably into the major groove and that the unstacking of the first base can be produced with small conformational changes in the DNA backbone. However in a second step once a base is removed from the interior of the helix (again by rotation around the backbone), enhanced backbone bending and untwisting appears. Here again, long-lived water bridges were observed.

Recently, the microsecond timescale (up to 44~$\mu$s) has been explored using intensive Molecular Dynamics (MD) all-atom simulations of 12 to 18~bp oligomers~\cite{Perez2007,Drsata2013,Galindo2014}. Frequent transient breathing events (breakage of few H-bonds) were observed with a duration comprised between 10~ps and 1~ns. No clear double strand opening could be observed on these $\mu$s long runs apart from the expected terminal base-pair fraying.

To finish with, Harris and coworkers have been able to investigate the effects of supercoiling on 90~bp minicircles~\cite{Harris2008}. 
Negative supercoiling can lead to partial denaturation, as we shall explore it in Section~\ref{supercoil}, and this is indeed observed in these molecular dynamics simulations. However, they are limited to simulate 4~ns only even though using a supercomputer, and it is difficult to assert that thermodynamical equilibrium has been reached.\\

\subsection{Numerical coarse-grained models with several beads per nucleotide}
\label{3SPN}

In order to tackle longer molecules and timescales, several coarse-grained numerical models have been developed in particular to study denaturation properties of short DNA at equilibriium. In these models, each nucleotide is modeled using two~\cite{Zhang1995,Drukker2000,Drukker2001,Sayar2010}, three~\cite{Knotts2007,Ouldridge2011,Linak2011,DeMille2011}, or even six~\cite{Dans2010,Zeida2012} beads. All these models involve classical effective potentials between beads, such as harmonic potentials for covalent bonds, and quadratic angle and dihedral potentials for bending and torsion. The effective non-linear potential chosen for mimicking hydrogen-bonding is either a Lennard-Jones~\cite{Florescu2011,Sayar2010,Knotts2007,Sambriski2009b} or a Morse one~\cite{Zhang1995,Drukker2000,Ouldridge2011} (see Ref.~\cite{Dasanna2013T} for an overview of these various coarse-grained models).
 
The 3SPN (for 3 sites per nucleotide) model first developed by the de Pablo group~\cite{Knotts2007,Sambriski2009b,Sambriski2009} is sketched in Figure~\ref{ThreeBeads}b~\cite{Florescu2011}. It seeks to incorporate available experimental information into the development of a coarse-grained description where the three sites are mapped onto the full atomistic representation of each DNA base. Its first evolution 3SPN.1~\cite{Sambriski2009c} includes the entropic effect of solvation on interacting strands which enables a spontaneous rehybridization of a bubble. The water solvent is treated implicitly and the electrostatic interactions are taken at the Debye-H\"uckel level. In the last evolution 3SPN.1-I~\cite{Freeman2011}, monovalent and divalent ions are included explicitly which allows one to reproduce the experimental melting temperature for oligonucleotides. 
A similar model developed just after in Oxford, the oxDNA model~\cite{Ouldridge2011,Sulc2012} also considers the cross-stacking interactions and ensures a physical representation of the ssDNA. Linak \textit{et al.}~\cite{Linak2011} which also develop a model similar to the 3SPN one introduces cross-stacking (Hoogsteen bonds) as well. 
In these three models, the effective potential shapes are fitted on potentials of mean force obtained by Boltzmann inversion of the relevant probabilities distributions coming from all-atom simulations. Hence, the effective interaction parameters, e.g. elastic moduli, can be inferred from more realistic atomistic molecular simulations. Moreover since the solvent is implicit, Langevin dynamics simulations are performed with time-step on the order of 5 to 30~fs~\cite{Sambriski2009c}.
 
In these works, melting profiles are computed by equilibrating the small oligomers on nanoseconds for various temperatures, and then compared to experimental ones, which is assumed to validate the choice of the parameter values.

In all these coarse-grained models, the denaturation and hybridization kinetics is followed on a maximum of a few hundreds of nanoseconds. In Ref.~\cite{Zhang1995}, the  hydrogen-bond displacement of base-pairs are simulated on a few ns, which corresponds to the base-pair breathing timescale as defined in the Introduction, where chain orientational degrees of freedom are not equilibrated. 

Using the 3SPN model, Knotts \textit{et al.}~\cite{Knotts2007} followed the bubble formation at 360~K (i.e. close to the melting temperature where dynamics is faster) and rehybridization after quenching it to 300~K in a 60~bp DNA sequence. They measured opening and closure times of 15~ns and 44~ns respectively, but as explicitly said in~\cite{Knotts2007}, these figures should be viewed with caution since friction is not included is this coarse-grained model. Using the evolution 3SPN-1 and replica exchange simulations (hence varying the temperature), only the average fraction of open base-pairs is dynamically followed on 200~ns but no kinetic information is given~\cite{Sambriski2009c}. Using transition path sampling~\cite{Sambriski2009}, the nucleation pathways in the ssDNA-dsDNA renaturation for oligomers are analyzed which show the significant role played by the sequence. For repetitive sequences the reassociation is non-specific with a molecular ``slithering'' mechanism, whereas for random sequences favors a restrictive pathway involving the formation of key base-pairs.

To study the early stages of a 14~bp DNA hybridization kinetics, Ouldridge \textit{et al.}~\cite{Ouldridge2013} (oxDNA model) had to accelerate dynamics by choosing a higher (by a factor 16) diffusion coefficient than physical DNA and to implement the so-called Forward Flux Sampling technique. They found that strand association proceeds through a complex set of partially hybridized intermediate states in a temperature-dependent way. More formed base-pairs are required in the intermediate states to render duplex closure probable when $T$ gets closer to $T_m$.

Hence, due to current CPU time limitations, most of these numerical coarse-grained models with several beads per nucleotides only follow the melting and renaturation of short oligomers on times smaller than 1~$\mu$s. The bubble nucleation and life times are therefore not accessible. One exception is the work by Zeida \textit{et al.}~\cite{Dans2010,Zeida2012} where a Langevin dynamics with a time step of 20~ps (3 order of magnitude faster than in 3SPN model) was used. Simulations up to 50~$\mu$s where done and the results are discussed at the end of Section~\ref{ffeq}.

\subsection{Mechanical models}
\label{mechmod}

To go a step further in the physical modeling and coarse-graining process, many mechanical mesoscopic models~\cite{Yakushevich2004} have been developed where the DNA chain is viewed as a straight chain of masses (points, spheres or disks) coupled by linear or torsional springs, in the line of the old concept of coupled atomistic chains for solids. Among these models one should distinguish between \textit{linear} physics models (the potentials are quadratic in the position and angular variables which implies linear equation of motions), which describe the DNA dynamics in terms of phonons~\cite{Barkley1979}, and \textit{nonlinear} physics ones. The latter follow the original ideas of Englander and coworkers in 1980~\cite{Englander1980}, which enable one to study larger excursions of molecules, such as base-pair unpairing by using the appealing concepts of nonlinear physics such as solitons or breather modes. However, as pointed out by Yakushevich himself in Ref.~\cite{Yakushevich2004}, processes of dissipation are completely neglected in these mechanical models (this is why we call them ``mechanical''). This major approximation is discussed in Section~\ref{dissipation}.

Although \textit{linear models} do not permit to study the base pairing/unpairing dynamics, they are able to yield the typical time scale of these mechanical models. In the continuum approximation ($a\to 0$ where $a$ is the distance between neighbouring disks or masses), three acoustic branches appear associated with torsional, longitudinal, and transverse modes. The frequencies are generically given by the ratios of the associated elastic modulus $k_l$ along the DNA axis and the mass $M$ (or moment of inertia $I$),
\be
\omega_{\rm a}^2(q)=\frac{k_l a^2}M q^2
\label{acoustic_freq}
\ee
When the double-stranded structure is taken into account, usually with two coupled chains, optical branches appear with a second eigenvalue
\be
\omega_{\rm o}^2(q)=\omega_{\rm a}^2(q)+ \frac{2k_t}M
\label{optical_freq}
\ee
where $k_t$ is the coupling elastic constant between masses of the same base-pair. Adjusting these formula to typical values of sound velocities in DNA, $v=\partial \omega/\partial q$, (both for longitudinal and torsional waves) which are measured around $v=100-2000$~m/s~\cite{Yakushevich2004,Yakushevich2002}, and using $a$ equal to a few Angstr\" oms, and $M\simeq 100\ m_{\rm p}$ (where $m_{\rm p}=1.67\times 10^{-27}$~kg is the proton mass) yields $k_l$ equal to a few N/m. Moreover several works have measured by microwave and infrared absorption/transmission experiments $\omega(0)=\sqrt{2k_t/M}$ equal to $35$~cm$^{-1}$ to 80~cm$^{-1}$~\cite{Kim1987,Yakushevich2004,Yakushevich2002} which yields $k_t$ smaller than 1~N/m. This estimation will be useful below when investigating the non-linear model predictions.

\textit{Non-linear models} have indeed been extensively studied in the context of DNA. The first works dealt with single rod-like models where the potential between successive monomers were chosen to be non-quadratic both for longitudinal~\cite{Muto1989}, transversal~\cite{Christiansen1990,Christiansen1993}  or bending modes~\cite{Ichikawa1981}. However, the main interest of non-linear models is that large excursions of bases, and therefore base-pair unpairing, can be accessible with a proper modeling of the interactions between the bases of the two single strands. Assuming simple two-body potentials, analytical solutions emerge as \textit{solitons} by analogy with mechanical solitons in pendula chains as first shown by Englander~\cite{Englander1980} and Scott~\cite{Scott1969}. Two main models can be distinguished.\\

\begin{figure}[t]
\centerline{\includegraphics*[width=8cm]{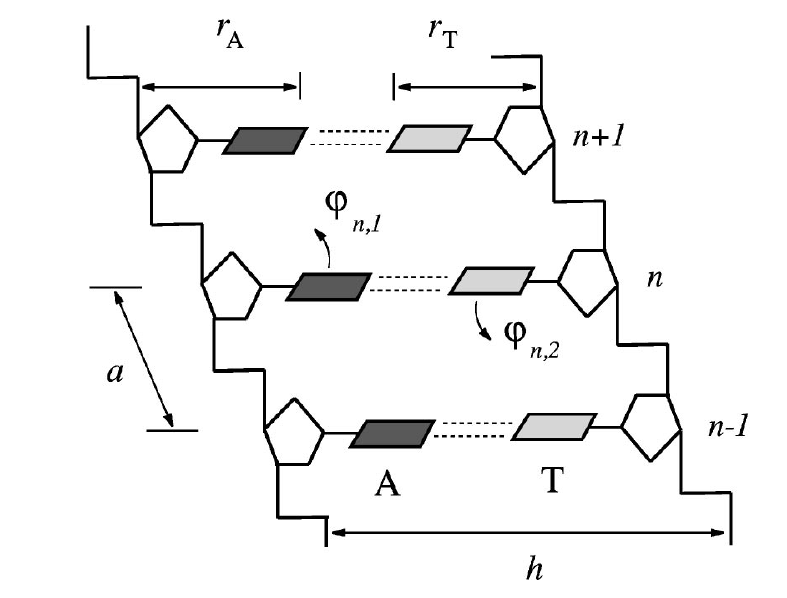}}
\caption{Sketch of the discrete Yakushevich model. The two single strands of the macromolecule form two parallel straight lines, placed at a distance $h=1.6$~nm from each other. The nucleo-bases, represented by rectangles, can make only rotational motions around their backbone chain, being all the time perpendicular to it. The angular displacement of the $n$-th base of the first (resp. second) strand is denoted by $\varphi_{n,1}$ (resp. $\varphi_{n,2}$). Taken from~\cite{Yakushevich2002}.
\label{Yaku}}
\end{figure}

One of the most famous double rod-like model that leads to solitonic solutions is due to Yomosa~\cite{Yomosa1983,Yomosa1984} further developed by Takeno~\cite{Takeno1983}, Yakushevich~\cite{Yakushevich1989,Yakushevich2001,Yakushevich2002,Yakushevich2004} and others~\cite{Salerno1991,Gaeta1992}, where the contribution to base-pair opening comes from rotational motions of the bases around the backbone. The important variables are thus the rotational degrees of freedom of base monomers with respect to a rotational axis, chosen as (O$z$), parallel to the duplex axis (see figure~\ref{Yaku}). In its continuous version, the Hamiltonian reads
\be
\mathcal H_{\rm Y}=\frac12\int dz\left\{\sum_{i=1,2} \left[I_i \left(\frac{\partial\varphi_i}{\partial t}\right)^2+k_l a^2 \ell^2 \left(\frac{\partial^2\varphi_i}{\partial z^2}\right)^2 -4 k_t \ell^2\cos\varphi_i \right]+ 4k_t \ell^2\cos(\varphi_1+\varphi_2) \right\}
\label{Hamiltonian_Y}
\ee
where $\varphi_i(z,t)$ are the angles defining the bases orientations (on the two strands labelled by $i=1,2$, see Figure~\ref{Yaku}). The bases have radius $\ell$ and are assumed to be in contact, $I_i$ is their moment of inertia along $(Oz)$, $a$ is the distance between base-pairs, and $k_l$ and $k_t$ are the longitudinal and transversal spring constants. The terms in the square brackets correspond to the classical sine-Gordon Hamiltonian for a chain of torsion springs~\cite{Scott1969,Englander1980}, whereas the last term models the interaction potential  between both strands. The linearization of \eq{Hamiltonian_Y} leads to the same frequencies as \eqs{acoustic_freq}{optical_freq} where the mass $M$ is replaced by $I/\ell^2$ in the symmetric case. The full mechanical problem associated with \eq{Hamiltonian_Y} is difficult to solve exactly but it has been shown within  some approximations that it leads to at least 4 soliton-like solutions where $\varphi_i(z-v t)$ (where $v$ is the wave velocity) increases by $\pm2\pi$, corresponding to local breaking of base-pair bonds, i.e., in the present context, to small denaturation bubbles. Using the method of Hereman \textit{et al.}~\cite{Hereman1986}, it can be shown analytically that an approximate solution is
\be
\varphi_1(Z,T)=-\varphi_2(Z,T)= 4 \arctan\left[\exp\left(\frac{Z-VT-Z_0}{\sqrt{(1-V^2)/2}}\right)\right]
\ee
where $Z=z/a\sqrt{k_t/k_l}$, $T=t \ell\sqrt{k_t/I}$, and thus $V=v/(a\sqrt{k_l\ell^2/I})$. The coordinate $Z_0+VT$ is the center of the soliton moving at constant velocity $V$. Typical values of $V$ are between 0.5 to 1~\cite{Yakushevich2002}. Hence denaturation bubbles of typical size $L\simeq a\sqrt{k_l/k_t}/\sqrt{1-V^2}\simeq a\sqrt{k_l/k_t}$ travel at a velocity slightly lower than the sound velocity $v_0=a\sqrt{k_l\ell^2/I}$. Using the values of $k_t$ and $k_l$ cited above (respectively 1 and a few N/m) leads to bubble sizes between 2 to 10 bps. More details about these types of models can be found in the book by Yakushevich~\cite{Yakushevich2004}. Several extensions have been developed in recent years~\cite{Gaeta2006,Cadoni2007,Gaeta2008,Tabi2008,Vasumathi2009,Vanitha2012}. One drawback of these models is that the large fluctuations of the inter-strand separations are not allowed (the distance between the strand backbones is maintained fixed), which makes difficult a proper description of all denaturation bubbles. In addition, the number of effective parameters can become very large when the model is refined further~\cite{Yakushevich2001}, which makes their estimation very difficult in practice.\\

In the alternative model proposed by Peyrard and Bishop~\cite{Peyrard1989} and later extended by Dauxois \textit{et al.}~\cite{Dauxois1993a,Dauxois1993b,Peyrard2004} and others~\cite{Cule1997,Campa1998,Barbi1999}, rotational motion of the bases is ignored, and base-pair opening is now modeled by the stretching of hydrogen-bonds, where the essential variable, $y_i$, is the distance between the two bases of the base-pair of index $i$. The two strands are also supposed to remain parallel. In addition to a non-linear Morse potential that models the interaction between bases of the same pair (a Lennard-Jones potential is also used in Ref.~\cite{Muto1990}), another potential describes the stacking interaction between the adjacent bases $i$ and $i-1$ of the same strand. The part of the Hamiltonian which depends on $y_i$ is therefore 
\be
\mathcal H_{\rm PBD} = \sum_i \left[ \frac{M}2 \left(\frac{d y_i}{d t}\right)^2+ U_0\left(e^{-\alpha_1 y_i}-1\right)^2 + \frac{k_s}2(y_i-y_{i-1})^2 \right]
\label{Hamiltonian_PB}
\ee
where $U_0\simeq 1.1\,k_{\rm B}T$, $\alpha_1\simeq 45$~nm$^{-1}$, and the stretching modulus is $k_s\simeq 1$~N/m. 

The dynamics of such a system has been extensively studied numerically using for instance molecular dynamics simulations coupled to a Nos\'e thermostat~\cite{Dauxois1993a} or Langevin simulations~\cite{Joyeux2005,Kalosakas2006,Alexandrov2009b}.  They show small regions of a few base pairs where $y_i$ reaches a large amplitude which have been named as DNA \textit{breathers}. It has been suggested by the authors that these breathers could be the precursors of the denaturation bubbles observed in experiments.
Again \eq{Hamiltonian_PB} leads to so-called non-linear waves (breathers), and using a multiple time expansion of the low amplitude expansion of the continuous dynamical equation, solitonic solutions appear. The order of magnitude of the frequency~\cite{Peyrard2004} is the same as in the Yakushevich model, and is given by the usual dispersion relation for planar waves
\be
\omega^2=2\frac{U_0\alpha_1^2}m+4\frac{k_s}m\sin^2\left(\frac{q}{2\alpha_1}\right)
\ee
corresponding to the optical frequencies [\eq{optical_freq}] in the limit $q\to0$. Interestingly, this approach also describes, in the continuous limit, the  domain wall between a large denaturated region and a closed one as the linear growth of a kink, the typical size of which is given by $\sqrt{2/S}\simeq 4-5$~bp. Indeed, the dimensionless parameter $S=k_s/(U_0\alpha_1^2)\simeq 0.1$, with the parameter values given above.

This PBD model has subsequently been improved in two directions. 
When comparing experimental DNA denaturation curves, the PBD model yields a too smooth transition. They thus added a non-linear coupling $k_s(y_{i-1},y_i)=k_s\left[1+\rho \exp(-\alpha_2(y_i+y_{i-1}))\right]$ ($\rho=0.5$, $\alpha_2=3.5$~nm$^{-1}$) in the spirit of the Poland-Sheraga model~\cite{PSbook} in which the stacking interaction (or cooperativity) parameter narrows the transition. The solution of this new potential can only be obtained numerically. Other types of non-linear couplings have been investigated~\cite{Joyeux2005,Buyukdagli2006,Buyukdagli2007}, taking advantage of the experimental knowledge of stacking interactions from thermodynamics calculations. Moreover, Buyukdagli~\textit{et al.} inserted a finite stacking by allowing the bases to move in the plane perpendicular to the strand axis~\cite{Buyukdagli2006} hence extending the number of degrees of freedom.

Barbi and collaborators~\cite{Barbi1999} extended further the PBD model by taking the helicoidal geometry into account, but still assuming that the DNA is infinitely stiff. The Hamiltonian has thus two variables per base-pair which are coupled through a geometrical constraint. The usual distance between the bases, noted $r_i$, has an equilibrium value $R_0$, and the twist angle $\phi_i$ is such that the difference $\phi_i-\phi_{i-1}$ is forced to be close to its equilibrium value $\phi_{\rm eq}=2\pi/p$ ($p \simeq 10.5$~bp is the B-DNA pitch) by a new term in the Hamiltonian
	\be
	\mathcal{H}_{\rm hel.}=K\left[(h^2+r_{i-1}^2+r_i^2-2r_{i+1}r_i\cos(\phi_i-\phi_{i-1}))^{1/2}-L_0\right]^2
	\ee
where $L_0=\sqrt{h^2+4R_0^2\sin(\phi_{\rm eq}/2)^2}$ is the equilibrium distance along each strand between bases $i$ and $i-1$ ($h$ is the distance along the DNA axis). Non-linear localized breather-like solutions are obtained using the same approximate methods as above, showing a local untwisting (kink) of 0.001~rad. Molecular Dynamics simulations were performed very close to the melting temperature on the 10~ns timescale~\cite{Barbi2003}, and showed breathers of a 5 to 10~bp. A central peak was also observed in the dynamical structure factor at $\omega\simeq 0.3$~ps$^{-1}$, attributed to the dynamics of bubble boundaries. Campa~\cite{Campa2001} observed numerically that a breather, created by locally unwinding the DNA of $\simeq 1$~rad, travels on long chains made of $N=2500$~bp both for homogeneous and heterogeneous DNA sequences. Gaeta and Venier have shown that traveling solitary wave solutions cannot exist for physical values of the parameters and must be associated with a global over-twisting of the helix~\cite{Gaeta2008}. Note that another type of interaction has been proposed to take into account the helicity in Refs.~\cite{Dauxois1991,Zdravkovic2011}.

\subsection{Mechanical models are not adapted to all denaturation bubbles dynamics}
\label{dissipation}

Both Yakushevich and PBD-like models have first been developed to study DNA dynamics, analytically~\cite{Peyrard1989,Barbi1999,Peyrard2004,Sulaiman2012} or numerically~\cite{Peyrard1992,Dauxois1993a,Barbi2003,Alexandrov2009b,Kalosakas2006}.
Both viscous and stochastic forces induced by the surrounding aqueous solvent were not included in the first versions of the models. It is clear that for a nucleo-base in water, the ``coasting time'' 
\be
\tau_{\rm c}\simeq \frac{M}{6\pi\eta d}
\ee 
which corresponds to the time spent for a mass $M$ of typical size $d$ to coast in a fluid of viscosity $\eta$ thanks to inertia~\cite{Purcell1977,Lauga2009,Manghi2006} is 0.1~ps (the base plus backbone mass is $M=500~m_p=8\times10^{-25}$~kg, the maximal backbone-base distance is $d=0.95$~nm, and $\eta=8.9\times10^{-4}$~Pa.s). Hence for times longer than 0.1~ps, all inertial effects are completely damped~\cite{DoiEdwards2004,Chaikin1995}. This is an important point, since the solitonic solutions of these mechanical models come from wave-like equations where inertial terms in $\partial^2 y/\partial t^2$ are central in the theory.
This point has been recently raised  by Frank-Kamenetskii and Prakash in their critical review~\cite{FrankKamenetskii2014} and admitted by Peyrard and Bishop in their comment~\cite{Peyrard2014} and even earlier in Ref.~\cite{Peyrard2009}.\\

Several attempts to introduce a friction force in the mechanical models have been proposed~\cite{Yakushevich2002,Zdravkovic2011,Sulaiman2012,vanErp2005,Das2008,Deng2008}. For instance, in the geometry of the PBD model, this friction force is written as $-\zeta_0 \partial y/\partial t$ where $\zeta_0\simeq 6\pi \eta a \sim 10^{-11}$kg/s is the friction coefficient of a base-pair. By accounting for the interactions forces $-{\rm d}V/{\rm d}y$ where $V(y)$ is the base-pair potential energy, and the Langevin stochastic force $f_{\rm L}(t)$, one obtains the Langevin equation:
\be
M\frac{\partial^2 y}{\partial t^2}= -\frac{{\rm d} V}{{\rm d}y} - \zeta_0 \frac{\partial y}{\partial t} + f_{\rm L}(t)
\ee
where $\langle f_{\rm L}(t)\rangle=0$ and $\langle f_{\rm L}(t)f_{\rm L}(t') \rangle= 2 \zeta k_{\rm B}T \delta(t-t')$. The damping coefficient $\gamma=\zeta/M=\tau_c^{-1}$ corresponds to the inverse of the coasting time defined above. For a nucleotide base it is thus around 10~ps$^{-1}$.

Since the success of mechanical models is based on the emergence of solitons, inertial terms where kept and several values of the damping coefficient $\gamma$ where introduced. To simulate properly inertial effects, one needs $\tau_c\gg\Delta t$, the simulation time step, usually on the order of the femtosecond (hence a simulation of typically $10^6-10^7$ time steps corresponds to a real time of a few nanoseconds). Commonly $\Delta t\simeq10^{-4}$ to $10^{-2}\tau_c$ in the numerical resolution of the Langevin equations. Therefore the damping coefficient values were chosen very small so that $\Delta t$ is large enough and simulation durations in real time are of physical interest. They range  from $\gamma=0.005$~ps$^{-1}$~\cite{Joyeux2005,Buyukdagli2006,Buyukdagli2007,Sanrey2009} to $\lesssim1$~ps$^{-1}$~\cite{Yakushevich2002,Vasumathi2009,Hien2007,Das2008}, values which are much smaller than the value of 10~ps$^{-1}$ estimated above\footnote{Note that, using the PBD model, some thermodynamic quantities such as the time-averaged fraction of open base-pairs are underestimated when the damping coefficient $\gamma$ is too small, smaller than 0.05~ps$^{-1}$~\cite{Das2008,vanErp2009Comment,Das2009Reply}. This is presumably due to the fact that the system did not reach equilibrium.}.

These damped mechanical models lead to the same conclusions: viscous damping induces a quick stoppage of the soliton which thus travels less than 10~bp~\cite{Yakushevich2002,Vasumathi2009} and decreases drastically the soliton velocity~\cite{Hien2007,Sulaiman2012}, and the ``bubble'' lifetimes which becomes on the order of a few picoseconds only~\cite{Alexandrov2006,Alexandrov2009,Alexandrov2010}.
Deng~\cite{Deng2008}, using  a stochastic differential PBD equation, found base-pair opening times of about 10--400~ps. They propose to obtain opening times on the order of microseconds by choosing an extremely small damping coefficient of $\gamma= 5 \times 10^{-9}$~ps$^{-1}$ corresponding to a non-physical base-pair radius $a=10^{-7}$~nm.

In order to compensate the slowing down effect due to friction, some studies have been done at temperatures close to $T_m$. Joyeux and collaborators~\cite{Joyeux2005,Buyukdagli2006,Buyukdagli2007,Sanrey2009} focused on the melting transition such that large bp excursions are observed.
Das and Chakraborty~\cite{Das2008} simulated numerically the nucleation of denaturation bubbles larger than 4 to 5 bps close to the denaturation transition (only $T>40^\circ$C was explored) such that the opening was possible in $10^6$ time steps.\\

All these results are coherent with the simple estimate of a soliton (breather) which survives during $\tau_c$ and travels on a distance $d\simeq v \tau_c$ where $v$ is the soliton velocity (on the order of the sound velocity). If one uses the real values measured experimentally (see Section~\ref{mechmod}) and includes the proper friction coefficient values, one finds that solitons travels on a distance $d\simeq 0.01$ to $1$~nm during less than 1~ps.
These short-living opening events therefore clearly do not correspond to a complete opening of a bubble but to breathers defined by small distortions of the distance $y(t)$ between bases less than 1~nm  (indeed the threshold for opening are taken in these works to lie between 0.05~\cite{Ares2005} and 0.3~nm~\cite{Alexandrov2009c}), as discussed in the Introduction.

Since in a highly viscous medium such as water, the motion is diffusive for times larger than $\tau_c\simeq0.1$~ps, larger bubble-breather lifetimes of a few ps at room temperature are coherent with the rough picture of a particle diffusing in an harmonic potential with spring constant $k_t$, the correlation function of which is~\cite{DoiEdwards2004}
\be
\langle y(t)y(0) \rangle =\frac{k_{\rm B}T}{k_t}\exp\left(-\frac{t}\tau\right)
\label{diffusion_potential}
\ee
where $\tau=\zeta/k_t$. Inserting the values given above, $k_t=1$~N/m, $\zeta=2\times 10^{-11}$~kg/s at room temperature, one finds $\tau \simeq20$~ps and a mean-squared deviation of the intra-base pair distance of $\sqrt{\langle y^2\rangle}= \sqrt{k_{\rm B}T/k_t}\simeq 0.6$~nm. Note that the entropy associated with the chain orientational degrees of freedom and the forces exerted by the adjacent base pairs do not modify quantitatively this rough estimate.\\

In all these models, base-pair breathing in dsDNA appears at a typical timescale of a few ps, in any case $<1$~ns. It thus raises a crucial question: what is missing in the model to reach the microsecond life time observed in some experiments~\cite{Warmlander2000,Altan2003,Phelps2013}? As it will be discussed in the next Sections, many chain degrees of freedom are not taken into account in the previous models, which can induce a free-energy barrier between the closed and open states. 

To reach accessible opening times with friction included in the model, Duduiala \textit{et al.}~\cite{Duduiala2009} defined a non-linear stochastic differential model to study the breathing of a structural defect, a thymine base being replaced by a difluorotaluene (F) with only one hydrogen-bond for the A-F base-pair. By fitting the parameters using MD simulation data, they deduced the potential of mean-force that should be introduced in the mesoscopic model to reproduce the MD results. The damping coefficient was found to be around 3~ps$^{-1}$, and importantly, a barrier from the closed state to the breathing one was found around $16~k_{\rm B}T$ for under-twisted DNAs.

Peyrard and co-workers proposed in 2008 to introduce an \textit{ad hoc} barrier of $6 k_{\rm B}T$ which enforces the times scales to be larger by 2 or 3 orders of magnitude~\cite{Peyrard2009}. The time scale is then obtained by the Kramers process escaping controlled by the base-pair diffusion. The origin of this barrier is not made explicit in Ref.~\cite{Peyrard2009} and several unknown parameters are introduced. Some of them are determined through the melting curves which turn to be much sharper. Indeed entropic effects sharpen the melting transition as explained previously by Cule and Hwa~\cite{Cule1997} who showed that a $y$-dependent stretching coefficient, $k_s(y_{i-1},y_i)$, introduces an entropic barrier. They obtained opening times of about 7~ns but no systematic study as a function of the barrier height or the friction coefficient (not given) was performed. Moreover, it thus shows that at these timescales the velocity distribution function has relaxed towards a Maxwellian as expected at timescales $\gg \tau_c$. Local thermodynamical equilibrium has been reached, confirming the fact that the dynamics is controlled by diffusion.

\subsection{Chain dynamics timescales}
\label{chain_dynamics}

In the preceding section, it was argued that to be able to catch the base-pairing dynamics at the microsecond scale, as observed in the experiments, some additional degrees of freedom must be included in the modeling. This makes the connection with polymer physics, where it is well known that the polymer entropy controls the typical time scales of polymer dynamics~\cite{PGGbook,DoiEdwards2004}. The Rouse model for the dynamics of a polymer chain of $N'=Na/\ell_{\rm K}$ physical monomers ($\ell_{\rm K}=2\ell_p$ is the Kuhn length) leads to the Rouse relaxation time of the mode $p$ (i.e. corresponding to a chain section of $N'/p$ physical monomers)
\be
\tau_{R}(p)\simeq \tau'_0 \left(\frac{N'}{p}\right)^2\qquad \mathrm{with}\qquad \tau'_0\simeq \frac{\zeta \ell_{\rm K}^3}{k_{\rm B}T}
\label{Rouse1}
\ee
where $\zeta\simeq4\pi\eta$ is the perpendicular component of the friction coefficient (per unit length) of a rod of length $\ell_{\rm K}$. Note that $\tau'_0$ is also the characteristic relaxation time of bending fluctuations. Indeed, the bending energy of a polymer parametrized by $\br(s)$ where $s$ is the curvilinear index is
\be
E_{\rm bend}=\frac12 \int_0^L \kappa_b \left(\frac{\partial^2 \br}{\partial s^2}\right)^2 ds
\ee
The equation of motion of $\br(s)$ is given by balancing the friction and bending forces, which leads to
\be
\zeta\frac{\partial \br}{\partial t}=-\kappa_b\frac{\partial^4 \br}{\partial s^4}
\ee
in the directions perpendicular to $\bt (s)=\frac{\partial\br}{\partial s}$. Hence the relaxation time for a segment of length $\ell$ is $\tau_{\rm bend}=\ell^4\zeta/\kappa_b$, which, for the Kuhn length $\ell_{\rm K}=2\beta\kappa_b a$ where $a=0.34$~nm is again the chemical monomer (base-pair) length, leads to $\tau_{\rm bend}=2\tau'_0$. Putting numbers yields $\tau'_0\simeq 80~\mu$s for dsDNA ($\ell_p\simeq 50$~nm) and $\tau'_0\simeq 0.6$~ns for ssDNA ($\ell_p\simeq 1$~nm).

Note that when excluded volume is taken into account the Rouse time (\eq{Rouse1} for the slowest mode $p=1$) of a chain of $N$ monomers and end-to-end distance $R\equiv \sqrt{\langle {\bf R}^2 \rangle}\simeq \ell_{\rm K} N'^\nu$ where $\nu\simeq 3/5$ is the Flory exponent, can be generalized using a simple scaling argument
\be
\tau_{\rm R}\simeq \frac{R^2}{D_{\rm chain}}\simeq R^2\frac{ N' \zeta}{k_{\rm B}T}\propto N^{1+2\nu}
\label{Rouse2}
\ee
because it is assumed that the total chain friction coefficient  is simply $\simeq N'\zeta$ in the free-draining regime ($\zeta$ is the physical monomer friction coefficient). When hydrodynamics interactions are included and using the pre-averaging Kirkwood approximation~\cite{PGGbook,DoiEdwards2004} one obtains the Zimm time
\be
\tau_{\rm Z} \simeq \frac{R^2}{D_{\rm chain}}\simeq R^2 \frac{\eta R}{k_{\rm B}T}\propto N^{3\nu}
\label{Zimm}
\ee
where the friction coefficient is the one of a sphere of radius $R$ (non-draining regime). It has been shown experimentally (Fluorescence correlation spectroscopy experiments)~\cite{Petrov2006} that hydrodynamic interactions control the segmental dynamics of dsDNA. Hence, in principle, the dsDNA relaxation time follows the Zimm scaling \eq{Zimm}.

The observation that the largest measured lifetimes of DNA denaturation bubbles (as described in the Introduction) turn to be on the order of timescales of bending fluctuations led some authors to suggest that bending fluctuations play a major role in DNA bubble dynamics~\cite{Jeon2006,Dasanna2013}. Indeed, it has been shown previously that the huge difference in bending moduli of ss- and ds-DNA plays a major role in the DNA equilibrium statistical physics~\cite{Palmeri2007,Palmeri2008,Manghi2009}, and it presumably influences the bubble dynamics too. These works will be discussed in Section~\ref{wooed}. 

One attempt in coupling bending fluctuations to base-pairing ones has been done by Jeon \textit{et al.}~\cite{Jeon2006,Kim2008} who developed a breathing DNA model by explicitly considering two different bending rigidities for denaturated ss segments and ds ones. Using a ladder model with two interacting single WLC strands through a Morse potential, they studied the interplay between bending motions and bubbles dynamics for DNAs of 300~bp. Starting with a railroad-track configuration, they simulated the DNA on hundreds of ns and examined the bubble lifetime as a function of their size. At room temperature they observed lifetimes on the order of 5~ns for bubbles of 10~bp, which slightly increase close to the melting temperature $T_m$. These measured times are much smaller than the relaxation time of the chain, which is between the Rouse time of one ssDNA and the one for a dsDNA, i.e. between 1 to hundreds of $\mu$s (note that 300~bp corresponds to $\ell_{\rm K}$ for a dsDNA). Hence the chain is far from being equilibrated and essentially keeps its initial railroad-track conformation during the whole simulation. Furthermore, they cannot have a significant sampling of chain degrees of freedom with such a model. To properly study the interplay between the bending motions and the denaturation bubble dynamics, longer simulations are thus necessary.

\medskip

To conclude this section devoted to all-atom, coarse-grained Molecular Dynamics and mechanical mesoscopic models, the explored timescale with these models is generally less than $1~\mu$s and therefore much smaller than the  relaxation time of a dsDNA segment of size equal to the Kuhn length, $\tau'_0\simeq 80~\mu$s. Hence these models implicitly assume that conformational chain degrees of freedom are frozen when base-pairing evolves. 
The base-pair dynamics which is followed with these models is therefore the base-pair breathing, i.e. the local opening of few base-pairs rapidly followed by their immediate renaturation. These transient bubbles have lifetimes $<1$~ns, during which chain degrees of freedom cannot relax. We have discussed why the transient bubble dynamics should essentially be governed by a diffusive process in the potential well, close to the dsDNA equilibrium structure.

A whole category of base-pairing processes cannot be addressed reliably in this mesoscopic context. Additional models have thus been developed to tackle DNA dynamics on longer timescales, e.g., the milli-second one as studied experimentally by Altan-Bonnet, Libchaber and Krichevsky (ALK)~\cite{Altan2003}. Reviewing them is the goal of the next Sections. In particular, we shall see that far-from-equilibrium processes can then be at play at the chain level, which definitely precludes the use of effective models where chain degrees of freedom are considered as frozen, but underlines the need of models that consider the coupling between chain orientational degrees of freedom and the internal ones related to base-pair dynamics.


\section{One-dimensional pre-averaged dynamical models}
\label{1Damod}

As discussed in the Introduction, after the discovery of the double-helix structure of DNA, the first physical mesoscopic models to be proposed in the early 1960s were inspired by the 1D Ising model of solid-state physics~\cite{Lazurkin1970,Wartell1972}. As it is common in statistical physics, when trying to account for experimental observations, one starts from the simplest reasonable model where the maximum number of degrees of freedom are pre-averaged. The simplest models assume that a base-pair is simply either open or closed. The DNA molecule is seen as a ladder, the helicity being secondary at this level of modeling, the rungs of which are the hydrogen bonds between base-pairs, either broken or unbroken. The conformational degrees of freedom of the semi-flexible polymers, whether ds- or ss-DNAs, involved in the physical description of the system are thus considered to be pre-averaged as if they were in equilibrium. This is equivalent to consider the chains in a quasi-static approximation, and is the essence of the zipper model to be presented now, before discussing its more elaborate successors in various biophysical experimental contexts.

\subsection{Zipper model}
\label{zipper:model}

Zipping (sometimes also called zippering) is in fact the second stage of renaturation (or (re-)hybridization) of DNA, which is the formation of duplexed strands from two complementary single strands after annealing from a high-temperature, fully denaturated state. Indeed, renaturation occurs in two stages, first a generally limiting, complex-forming or nucleation step (see Section~\ref{nucleation}), followed by a faster, processive and complete closure of the double helix, usually known as zipping~\cite{Wetmur1968,Cantor1980,Sikorav2009}. 

The question of the kinetics of the renaturation of DNA was addressed shortly after the discovery of its double-helix structure in 1953, as detailed in the early review~\cite{Marmur1963}. These  experimental studies essentially focused on the establishment of the second-order character of the $2\,\mathrm{ssDNA}\rightleftarrows \mathrm{dsDNA}$ reaction, and possible deviations from it, without any experimental access on the faster intramolecular zippering stage, because second-order kinetics only take into account complementary strand recognition and not the following irreversible zipping stage, once recognition has been ensured. Furthermore, severe complications inherently arose from the insufficiently controlled quality of the DNA molecules due to the extraction processes, most nucleation events leading only to partial Watson-Crick base-pairing of the fragments, and single strand regions eventually remaining unpaired (``dangling tails'')~\cite{Britten1976}. It has not been possible to work with large amounts of well-controlled, strictly complementary strands before the discovery of molecular biology techniques (PCR) in the 1980s~\cite{Sikorav2009}. It was then possible to measure finely melting curves, as well as association and dissociation rates which both display Arrhenius behaviors (see e.g.~\cite{Paner1992,Morrison1993,Yamashita2008}).

\begin{figure}[ht]
\begin{center}
\includegraphics*[width=9cm]{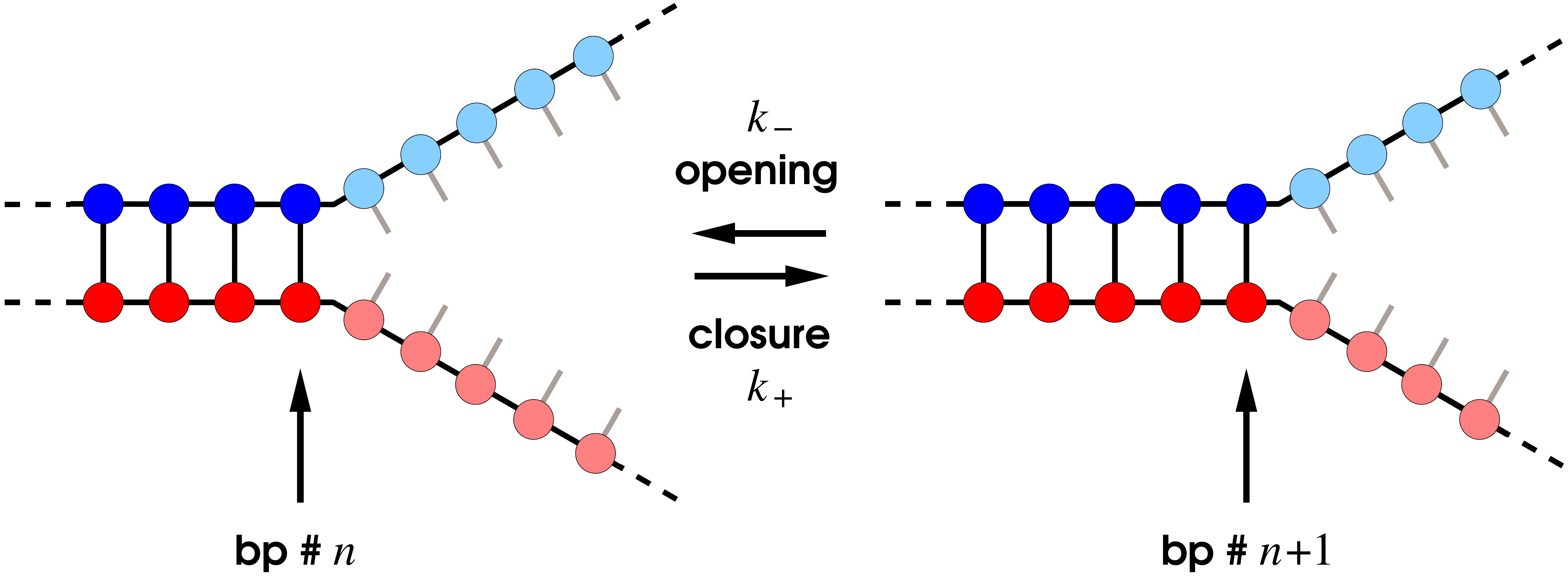}
\end{center}
\caption{Sketch of the zipper model. A number $n$ of base pairs are closed on the left part of the complex, $N-n$ remain unclosed on the right. Closure and opening rates are denoted by $k_+$ and $k_-$ respectively. Closure (resp. opening) increases (resp. decreases) $n$ by 1 unity. The zipper model is thus formally equivalent to a biased 1D random walk. 
\label{zipper:fig}}
\end{figure}

From a theoretical perspective, first questioning was raised by Flory as soon as 1961~\cite{Flory1961}, together with helix-coil transition of $\alpha$-helices, even though adopting a quite basic approach, essentially amounting to a biased random walk in one dimension (the bias of which depends on temperature in order to match Boltzmann distributions at equilibrium). Indeed, the zipper model (Figure~\ref{zipper:fig}), one of the dynamical counterparts of the 1D Ising model, has been developed as an effective model of DNA double helix formation. It supposes for simplicity sake~-- and assumes this approximation to be correct for sufficiently short constructs~-- that the partially double-stranded molecule is allowed to have only one unbroken sequence of successive paired bases growing from the initial nucleus because successful nucleation events are rare~\cite{Porschke1971,Porschke1974}. 

Once a successful nucleation event has occurred, processive, base-by-base zipping can proceed. The zipper model fundamentally assumes that the renaturation process can be described by a one-dimensional reaction coordinate (the number $n$ of successive closed base-pairs) and thus it implicitly supposes that chain orientational degrees of freedom can be pre-averaged so that the quasi-static motion is diffusive in a one-dimensional free-energy landscape. It can easily be analytically solved as a biased 1D random walk~\cite{Zimm1960,Applequist1963,Bicout2004}. If one denotes by $k_+$ and $k_-$ the closure and opening rates (Figure~\ref{zipper:fig}) and by $\Delta F_0$ the average free-energy gained when closing one base-pair (including the chain conformational entropy)\footnote{$\Delta F_0 \approx 2.8k_{\rm B}T$ at room temperature (25$^\circ$C $= 298$~K) when averaging over the sequence and $\Delta F_0 \approx 2.3 k_{\rm B}T$ at physiological temperature (37$^\circ$C $= 310$~K) where entropic effects get stronger. The loss of entropy when closing a base-pair is $\Delta S_0\simeq10 k_B$ at these temperatures~\cite{SantaLucia1998}.}, then the detailed balance leads to
\be
\frac{k_-}{k_+}=e^{-\beta \Delta F_0}
\label{balance}
\ee
As expected, $k_-<k_+$ if the molecule unzips ($\Delta F_0>0$) and $k_->k_+$ if it zips ($\Delta F_0<0$). Close to $T_m$, $k_-\simeq k_+$ and $\Delta F_0\propto T-T_m$~\cite{Wartell1972}.

The averaged velocity of the junction is given by $v=a(k_+-k_-)$. The velocity is also given by $v=f/\zeta_0$ where $f\approx \Delta F_0/a\sim 10$~pN is the driving force exerted on the DNA at the fork junction associated to the variation of the DNA free-energy when one base-pair opens or closes. It is thus implicitly assumed that $f$ is essentially uniform along the chain included at the sub-base-pair length-scale. The friction coefficient is $\zeta_0 \approx 6 \pi \eta a_{\rm H} \sim 10^{-11}$~kg/s by assuming that only one base-pair moves, with $a_{\rm H}\approx 1$~nm the base-pair hydrodynamic radius. 
The characteristic time needed to (un)zip a single base-pair is then for $\Delta F_0\neq 0$, i.e. $T\neq T_m$
\be
\tau_f \equiv \frac{a}{v} = \frac{a\zeta_0}{f} \simeq \frac{a^2\zeta_0}{ |\Delta F_0|},
\label{tau00}
\ee
and the average zipping time of a chain of $N$ base-pairs is  
\be
\tau_{\rm zip}= \tau_f N \propto N.
\label{ziptime}
\ee
This model anticipates $\tau_f \sim 0.1$~ns. We shall discuss this order of magnitude below when comparing it to experimental values. The underlying issue is whether the single base-pair closure or opening is controlled by $\Delta F_0$ as supposed above, or by a (yet unknown) microscopic barrier. This issue will be discussed in Section~\ref{F:barrier}

The diffusion coefficient of the 1D random walk is equal to $D=a^2 (k_++k_-)/2$. A short calculation leads to 
\be 
D = \frac{\Delta F_0}{2\zeta_0} \frac{e^{\beta \Delta F_0} +1}{e^{\beta \Delta F_0} -1}.
\label{D}
\ee
For the special case $T\simeq T_m$, i.e. in the limit $\beta |\Delta F_0|\ll1$, one recovers from \eq{D} the Einstein relation $D=k_{\rm B}T/\zeta_0$, as expected close to equilibrium. The relaxation time $\tau_m\equiv\tau(T_m)$ is then given by the 1D diffusion law
\be
\tau_m\simeq \frac{(aN)^2}{D}\simeq\tau_0 N^2 \propto N^2
\ee
with $\tau_0=6\pi\eta a^2a_{\rm H}/(k_{\rm B}T)\simeq 0.5$~ns is the characteristic diffusion time of a base-pair to be discussed in Section~\ref{F:barrier} as well.

As compared to early studies, the zipper model can be refined further, for example by taking some form of cooperativity between base-pairs into account in an empirical way~\cite{Murugan2002}. In general, it is assumed for simplicity sake that the binding energies of all base-pairs are identical, as a first approximation, thus ignoring possible sequence effects. Jarayaman and coauthors~\cite{Jayaraman2007} (see also~\cite{Szu1979}) went further into the analysis of this one-dimensional effective model by assuming that a base-pair can be closed even though its two neighbors are open, but that this is less probable than the closure of a base-pair a neighbor of which is already closed (the basic assumption of the original zipper model). The authors propose an analytic expression for the time-correlation function between nearest neighbors and next-nearest neighbors through a perturbation approach. However, this calculation is limited to the special case where rate constants for reverse and forward reactions are identical in order to ensure the Hermitian character of a transition matrix. It limits the applicability of this approach to the strict vicinity of the melting temperature. 

Applying the zipper model or its further developments notably accounts to ignore staggering, slithering or transitory hairpin formation, that has been shown both experimentally and numerically to play a role by slowing down the hybridization process because of the presence of either complementary sub-sequences in single strands~\cite{Rief1999,Montanari2001,Gao2006,Schreck2015} or repetitive, periodic motifs~\cite{Sambriski2009,Sambriski2009b,Hinckley2014}. Pushing further the  one-dimensional biased random walk point of view in the continuous limit and solving convection-diffusion equations, it can for example be shown~\cite{Redner_private} that allowing the formation of one or several defects during zipping (such as secondary mismatched partial hybridizations) can significantly delay final closure because the system gets stuck in a dynamical trap: it can be long before the erroneously wound segment unwinds, all the more reason when the temperature is low and the random walk is biased. This effect can in principle be quantified. A reduction of hybridization rates by a factor 10 is given in Ref.~\cite{Schreck2015}. Consequently, zipping time predicted by the zipper above can only give the lower bounds of real closure times.

Furthermore, independently of the quality of these different approaches, we shall see in Section~\ref{ffeq} that because of time-dependent, non-local frictional forces that depend on the chain conformation, zipping of long molecules is better described by far-from-equilibrium approaches. It cannot be fully understood through this over-simplified zipper model because it relies upon the strong assumption that the chain is close to equilibrium (differently said in a quasi-static regime). For the relevant regimes of driving forces, the polymer is likely far from equilibrium because it cannot respond to the driving force all at once.

\subsection{Bubble closure in the Poland-Scheraga landscape}

In the zipper model, the chain thermodynamical contribution is completely included in the local parameter $\Delta F_0$. An improvement of this simple model has been proposed  by Poland and Scheraga~\cite{PSbook} where the entropy of the denaturated strands (seen as a ssDNA loop of size $2n$) is taken into account in a non-local term depending on $\ln n$ in the free-energy.
Given the predictive character of the Poland-Scheraga model for DNA thermodynamics, and especially the DNA denaturation, many works on DNA base-pairing dynamics also explored the diffusive dynamics in the Poland-Scheraga free-energy landscape (see for instance the review by Metzler \textit{et al.}~\cite{Metzler2009}). In this framework the mean life time of an initial bubble of size $n$ has been the focus of several studies which are surveyed here. It is important to note that this model only applies for the zipping of a bubble located in the middle of the DNA. For DNA zipping in the Y-shape, it reduces to the zipper model. 

The Poland-Scheraga model is an extension of the zipper model by Zimm and Bragg~\cite{Zimm1959,Zimm1960} which can be reformulated as an Ising model~\cite{Wartell1972}. At each base-pair $i$ is associated an Ising variable $\sigma_i$ which can take two values, $+1$ for a closed base-pair and $-1$ for an open base-pair. The Ising Hamiltonian is therefore
\be
\mathcal{H}_{\rm Ising}[\sigma] = - \sum_{i =1}^{N - 1} \, \left[J \sigma _{i + 1} \sigma _i +\frac{K}2 (\sigma_{i+1} + \sigma _i )\right] - \mu \sum_{i = 1}^N\,\sigma_i 
\label{H}
\ee
where $2\mu$ corresponds to the free energy to pay to break one base-pair (whatever the state of the neighbouring base-pairs), $2J$ the energetic cost to create a domain wall between two adjacent segments in different states and $2K$ the free energy difference between two adjacent base-pairs closed or open\footnote{In the limit $N\to\infty$ or with periodic boundary conditions, $\mu$ and $K$ play the same role in the model and can be embedded in a single parameter, $L=\mu+K$.}. They are related to the melting temperature $T_m$ (for a homopolymer) by $L\equiv \mu+K\propto(T_m-T)$. Note that $L$ and the previously defined $\Delta F_0$ are simply related, $\Delta F_0=2L$. The cooperativity parameter  $\sigma=e^{-4\beta J} $ controls the width of the denaturation transition. Poland and Scheraga propose to add a \textit{loop entropy factor} to this parameter,
\be
\sigma_{\rm LE}=\frac{e^{-4\beta J}}{(n_0+2+2n)^c}
\label{LE}
\ee
where the denaturation bubble of size $n$ is viewed as a closed flexible ssDNA polymer loop of size $2+2n$. This is therefore an entropic term which favors the closed state. The empirical ``stiffness'' parameter $n_0$, first introduced by Fixman and Freire~\cite{Fixman1977}, takes implicitly the rigidity into account at small $n$ values\footnote{A reasonable value of $n_0$ would be $\simeq3-10$, the number of base-pairs in a ssDNA Kuhn length. However, when comparing the Poland-Scheraga model to experimental denaturation profiles, the fitting value for $n_0$ may be as large as 100, which has been correlated to the fraction of stacked residues in the loop}~\cite{Blake1987,Palmeri2008}. and the loop exponent $c=\nu d$ (where $d$ is the dimension and $\nu$ the Flory exponent). Hence, in 3D, $c=3/2$  for a phantom chain and $c \approx1.764$ for a self-avoiding chain. If excluded volume interactions are taken into account between loops, $c\approx 2.115$~\cite{Kafri2000}. The free energy of a bubble of size $n$ reads
\be
\Delta F (n) = 4 J + n\Delta F_0 + k_{\rm B}T c \ln(n_0+2+2n)
\label{G_PS}
\ee

Hanke and Metzler simplified \eq{G_PS} by neglecting $n_0$ and studied the Smoluchowski equation (in the approximation of the continuous limit) for the probability density $P(n,t)$ to have a bubble of size $n$ at time $t$~\cite{Hanke2003,Banik2005,Chaudhury2009}
\be
\frac{\partial P(n,t)}{\partial t} =\hat D \frac{\partial}{\partial n}\left[ \frac{\partial(\beta \Delta F)}{\partial n}+ \frac{\partial}{\partial n}\right] P(n,t)
\label{smol}
\ee
where as in the zipper model, $\hat D=D/a^2=(k_++k_-)/2$ is the diffusion coefficient, $k_\pm$ being the rate constants to close/open a base-pair, taken as adjustable parameters in the model. The drift term is $\partial (\beta \Delta F)/\partial n = \beta \Delta F_0 + c/n$. 

Although appealing, the Smoluchowski equation \eq{smol} in a Poland-Scheraga energy landscape suffers from two major assumptions: (1)~As observed by Metzler \textit{et al.}~\cite{Metzler2009}, it is again assumed that $n$ is the slow variable of the system compared to the chain degrees of freedom. Especially, the loop is assumed to be in a local thermal equilibrium at any given time during its evolution~\cite{Bar2007}. We have previously discussed in Section~\ref{bpb} that this assumption is questionable. (2)~The single strand rigidity is completely ignored, in particular for small bubbles $n\sim10$, it is known that the loop entropy factor in \eq{LE} is wrong since it has been calculated in the $n\to\infty$ limit~\cite{PGGbook}.

Thanks to a mapping to the quantum Coulomb problem (the potential of which is in $1/r$ and \eq{smol} is equivalent to an imaginary time Schr\"odinger equation), bubble lifetime distributions as well as auto-correlation functions have been computed analytically~\cite{Fogedby2007a,Fogedby2007b,Bar2007,Bar2009}. The dynamics depends on the sign of $\Delta F_0$:
\begin{itemize}
	\item For $T<T_m$ ($\Delta F_0>0$) the loop entropy factor plays a negligible role as soon as $n$ is large enough. In particular, the mean bubble lifetime, for an initial bubble size $n$, is
\be
\tau=n\tau_f\frac{K_{(c-1)/2}(n\beta L)}{K_{(c+1)/2}(n\beta L)}
\ee
where $\tau_f=a^2/(D\beta \Delta F_0)$ is the same characteristic time as in the zipper model (see \eq{tau00}) for $\beta\Delta F_0$ not too large, and $K_i(x)$ is the Bessel function of order $i$. The mean bubble lifetime scales as $n$ as for the zipper model, the factor which depends on the $c$ value tends to 1 for large bubbles, $n\beta L\gg1$. 
	\item At the melting temperature $T=T_m$, since $\Delta F_0=0$, the dynamics is uniquely controlled by the loop exponent $c$. The mean bubble lifetime is $\tau=(an)^2/[D(c-1)]$, and scales as $n^2$ as a signature of any free diffusion in 1D. 
	\item For $T>T_m$, the picture of one single bubble is no more valid since the dynamics evolves towards the chain denaturation.
\end{itemize}

Sequence effects have been studied following essentially the same approach but in the discrete form and with Ising parameters depending on the sequence. Hence the problem has been solved numerically for the sequence of the ALK experiment~\cite{Altan2003} or various biological sequences~\cite{Ambjornsson2006,Ambjornsson2007a,Ambjornsson2007b,Talukder2011}.
The coalescence of two DNA bubbles initially located at weak (AT rich) domains and separated by a more stable (GC rich) barrier region in a designed construct of dsDNA has also been studied in Ref.~\cite{Pedersen2009}. 
In these works, the time averaged fluorescence autocorrelation function has been fitted, $k_+$ and $k_-$ being the adjustable parameters.
Probability distribution functions connected to the autocorrelation function have different forms for small and large bubbles~\cite{Bandyopadhyay2011}. For small bubbles the dynamics depends essentially on the exponent $c$, whereas for large ones the key parameter is $\Delta F_0$.

Changes of temperature have also been studied using this model, either under conditions of rapid heating from a subcritical to the critical temperature $T_m$~\cite{Murthy2011} where a bubble of size $n$ grows in a time $\propto n^{c+1}$, or with a temperature ramp which induces some hysteresis effects~\cite{Allahverdyan2009}.

Other Langevin dynamics, based on the Poland-Scheraga free-energy landscape, were studied. Kunz \textit{et al.}~\cite{Kunz2007} describes the dynamics in terms of the order parameter $\rho=\lim_{N\to\infty} n/N$, again controlled by $c$. If $1<c<3$, $\rho(t)$ decreases exponentially with typical time in $(T_m-T)^{c-3}$ for $2<c<3$ and  $(T_m-T)^{1-c}$ for $1<c<2$ close to $T_m$ (and a logarithmic dependence in the case $c=2$).

To sum up, the important underlying assumption in describing the closure of DNA bubbles using the free energy of the Poland-Scheraga model of \eq{G_PS}, where all the degrees of freedom except the number of open base-pairs have been integrated out, is that these degrees of freedom equilibrate much faster than $n$. However we have seen that some conformational degrees of freedom of the DNA chain such as bending, but also torsion and entropic stretching, have long relaxation times. This is the reason why the simple scaling laws found within this approach, $\tau\sim n$ at $T<T_m$ and $\tau\sim n^2$ at $T=T_m$ are the same as the ones of the zipper model and do not display anomalous exponents, as measured in numerical experiments. We shall return to this issue in Section~\ref{ffeq} with more elaborate arguments.

\subsection{Mechanical unzipping below the melting temperature by applied force}
\label{force_unzip}

\begin{figure}[ht]
\begin{center}
\includegraphics*[height=4.5cm]{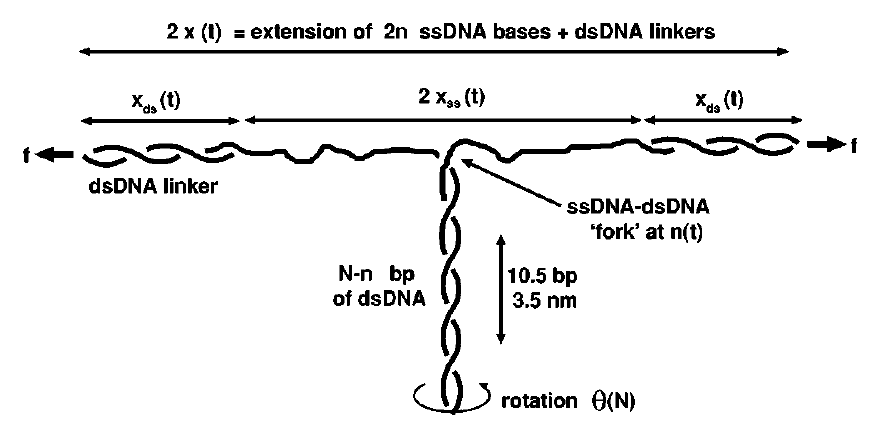}\includegraphics*[height=5cm]{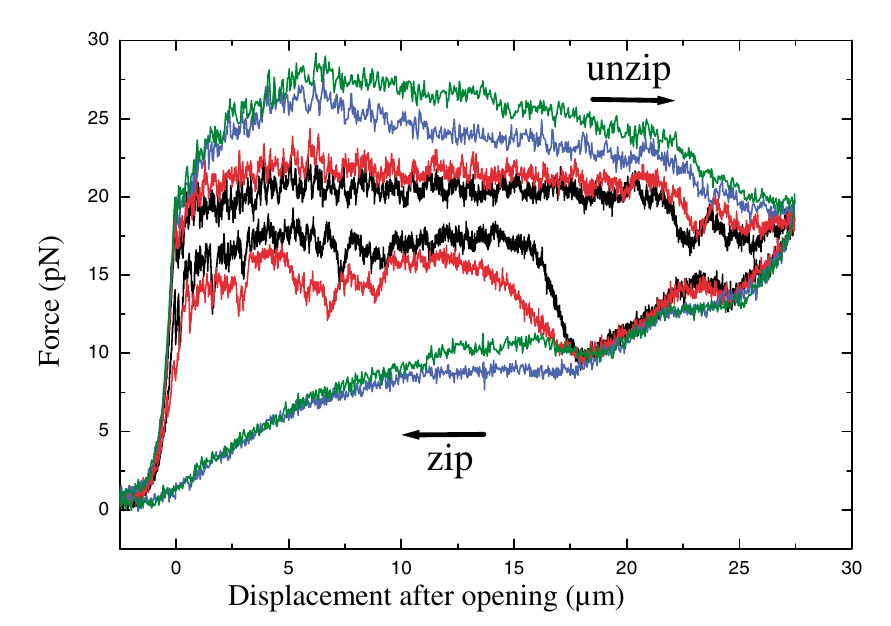}\\
\ \hspace{1cm} {\bf (a)}~\hspace{7.2cm}~{\bf (b)} \
\end{center}
\caption{(a)~Principle of force-induced duplex unwinding experiments. Two dsDNA linkers are attached to the ssDNA ends, and are used to pull ssDNA ends apart with a force denoted by $\mathbf{f}$. The linkers are themselves attached to two solid supports, e.g., beads trapped in optical tweezers (not represented). When the dsDNA denatures ($n(t)$ base-pairs are unzipped in the figure), the twist of the dsDNA must be evacuated, forcing the remaining dsDNA of length $N-n$ to rotate at angular velocity $\omega={\rm d \phi}/{\rm d} t$ ($\phi$ is denoted by $\theta$ in the figure). A rotational drag ensues. Taken from~\cite{Cocco2002b}. (b)~Extension-force experimental plots ($f$ vs $2x(t)$ with the left figure notations) of a $\lambda$-phage DNA of 48.5~kbp, recorded upon opening and closing of the same molecule at different unzipping/zipping velocities $v=4$~$\mu$m/s (black), 8~$\mu$m/s (red), 16~$\mu$m/s (blue), and 20~$\mu$m/s (green). Hysteresis gets stronger with increasing velocity. Taken from~\cite{Thomen2002}.
\label{force:unzip}}
\end{figure}

We now address force-induced unzipping well below the melting temperature, which is of experimental relevance. The fundamental motivation is to give solid foundations to more complex \textit{in vivo} situations where active, force-induced unwinding is performed by helicases or RNA polymerases~\cite{Alberts2002}, which exert a force on the order of piconewtons. In the single-molecule experiments, in general close to the physiological conditions, two dsDNA linkers are attached to the ssDNA ends, and are used to pull ssDNA ends apart (Figure~\ref{force:unzip}a)~\cite{Essevaz1997,Bockelmann1997,Bockelmann1998,Bockelmann2002,Bockelmann2004,Danilowicz2004}. The Y-shape assumption is exact in this case because temperature-activated bubbles are extremely rare in dsDNA at physiological temperature. Note that this unzipping process where the applied force is transverse to the molecule should not be confused with DNA structural changes induced by a force applied longitudinally, as studied for example in Ref.~\cite{Manghi2012} (and references therein).

In practice, the velocity $v$ of the linker extremities is held constant while the applied force $f_{\rm ext}(t)$ is monitored from the measurement with a sub-nanometer precision of the relative positions of the solid supports to which the linkers are attached. The dsDNA begins to open when the force is larger than $10$ to $15$~pN (at low velocity), the lower value corresponding to the pure AT-limit and the higher one to the pure GC-limit~\cite{Bockelmann1997}. Rief et \textit{al.} measured by AFM the forces at equilibrium $9\pm3$~pN (resp. $20\pm3$~pN) for pure AT (resp. pure GC) constructs~\cite{Rief1999}. More generally, the unzipping force vs displacement plots (Figure~\ref{force:unzip}b) present reproducible features which are governed by the sequence being opened. The molecule to be opened is several kbp long, e.g. a $\lambda$-phage DNA of $N \simeq 48.5$~kbp. Since twist must be evacuated at one of the molecule extremities, a part of the molecule will rotate and one expects that the ensuing rotational friction torque should contribute to oppose the torque $\mathcal{T}_{\rm ext}$ associated with the applied force $f_{\rm ext}$ needed to unzip DNA.

In the low velocity regime, $v < 1$~$\mu$m/s, i.e. typically $v < 1$~kbp/s, one observes that the force intensity $f_{\rm ext}$ does not depend significantly on the velocity $v$ (Figure~\ref{force:unzip}b). At higher velocities, a significant increase of unzipping force is observed~\cite{Thomen2002}, indicating that the rotational friction torque $\mathcal{T}$ on double-stranded DNA leads to an additional contribution to the opening force, this effect increasing with the length of the molecule. This torque is estimated to be  $\mathcal{T}\approx 5 k_{\rm B}T$ on a $\lambda$-phage DNA at $v=20$~$\mu$m/s.

The whole process can be reversed ($v<0$), thus leading to DNA rehybridization, also sometimes called re-annealing. The signal presents hysteresis, an effect that increases with velocity. At velocities $|v|> 10$~kbp/s, the force measured during zipping decreases significantly, indicating that the ``velocity of re-annealing of the molecule is less than the velocity imposed'' by the experimental device~\cite{Thomen2002}. This provides an estimate of the zipping velocity of long duplexes, $v \approx 10^{-2}$~bp/$\mu$s which will be useful in Section~\ref{ffeq}.

As far as reversibility is concerned, the process is quasi-static at low velocities typically slower than 1~kbp/s for a 50-kbp dsDNA, and the zipper model is fully relevant provided that an additional force is applied at the fork level, the external force $f_{\rm ext}$ transmitted by the linkers. The ssDNA polymers being held under a tension of $\sim10$~pN, their fluctuations are significantly suppressed. Chain fluctuations that were integrated out in the zipper model in a disputable way become in fact unessential. Note that the entropy gain when going from the ds to the ss form is consequently lower, which increases the value of $\Delta F_0$ that should in principle be used. At higher velocities $v$, the equilibrium approach ceases to apply and out-of-equilibrium effects cannot be ignored anymore~\cite{Cocco2002b}: (a)~In the unzipping regime, the rotational viscous drag on the dsDNA becomes important, as discussed above. (b)~In the zipping regime, the rehybridization is slower than the imposed velocity and resembles far-from-equilibrium zipping in absence of force, where the viscous torque delays the rewinding of the single strands. This situation will be discussed in detail in Section~\ref{far:from:eq:zipping} below.

From a theoretical point of view, Cocco, Monasson and Marko proposed a model where the different contributions to the free energy are carefully taken into account, notably the elastic energies of the different involved DNAs displayed in Figure~\ref{force:unzip}a~\cite{Cocco2002b,Cocco2001}. The precise relation between the total extension $2x(t)$ and the number $n(t)$ of open base-pairs can be inferred at equilibrium, when $v \rightarrow 0$. By carefully comparing equilibration times and (un)zipping timescales, they demonstrated that dsDNA and ssDNA stretching degrees of freedom \textit{are} in equilibrium on experimental timescale for the $\lambda$-phage DNA discussed above. 

The relaxational dynamics of the fork position $n(t)$ is governed by the equation
\be
\frac{2w_{\rm ss}(f_{\rm ext})-\Delta F_0 -\mathcal{T} \phi_{\rm eq}}{a} -\zeta_0a\frac{{\rm d} n}{{\rm d}t}=0.
\label{edoCMM}
\ee
In this equation, the fraction in the left-hand-side is the total force acting on the fork base-pair: $w_{\rm ss}(f_{\rm ext})$ is the variation of the elastic energy stored in each ssDNA when closing a base-pair, at fixed tension $f_{\rm ext}$ and fixed single-strand extension $x-x_{\rm ds}$. It accounts for the extensibility of ssDNAs, described, e.g. as freely-jointed chains~\cite{PGGbook}. Its form is not central at this stage but it satisfies  $w_{\rm ss}(f_u)=\Delta F_0$ at zero torque and at equilibrium, where $f_u$ is the unzipping force at zero torque and at equilibrium. Furthermore $\zeta_0 \approx 6 \pi \eta a$ is again the friction coefficient of a base-pair (here for sake of simplicity, the base-pair hydrodynamic radius $a_{\rm H}$ and the dsDNA monomer length $a$ are assumed to be equal). The angle $\phi_{\rm eq}=2\pi/p \simeq 0.6$~rad is the equilibrium twist angle between adjacent base-pairs ($p\simeq 10.5$~bp is the B-DNA pitch) and relates the number $n(t)$ of denaturated base-pairs and the total rotation angle of the dsDNA through $\phi(t)=\phi_{\rm eq} n(t)$ (see Figure~\ref{force:unzip}a). Thus $\omega={\rm d \phi}/{\rm d} t = \phi_{\rm eq} {\rm d}n/{\rm d} t$. Finally, 
\be
\mathcal{T} = 4 \pi \eta r^2 a \left[N -n(t)\right] \omega
\label{torque}
\ee
is the rotational friction torque exerted on the dsDNA	 assumed to be linear, in the Rouse regime~\cite{DoiEdwards2004}, where $r\simeq 1$~nm is the dsDNA radius. Combining \eqs{edoCMM}{torque} leads to a first-order ordinary differential equation on the fork position $n(t)$:
\be
\frac{{\rm d} n}{{\rm d}t} = \frac1{k_{\rm B} T} \frac{2w_{\rm ss}(f)-\Delta F_0}{\tau(n)}
\label{edo2CMM}
\ee
where the typical time is linear with $n$:
\be 
\tau(n)=\frac{6 \pi \eta a^3}{k_{\rm B} T} +  \frac{4 \pi \eta a r^2 \phi_{\rm eq}^2}{k_{\rm B} T} \left[N -n(t)\right] \approx 10 + 2  \left[N -n(t)\right] \mbox{(ns)}.
\ee
The so-obtained evolution equation also applies to rezipping, at least at low velocities where ssDNAs remain under tension. At fixed $v$ as in experiments\footnote{Note that the definition of $v$ differs by a factor 2 between the experiments and this theoretical work because here the total molecule extension is denoted by $2x$, as illustrated in Figure~\ref{force:unzip}a. The conversion rule is $2v_{\rm th}=v_{\rm exp}$.}, $x(t)=vt$. Changing the variable from $t$ to $x$ in Eq.~(\ref{edo2CMM}), expending the force $f$ and the fork position $n$ at order 1 in $v$, and plugging them in the equation, one eventually gets~\cite{Cocco2002b}
\be
f(x)=f_{u}\left[1+\frac{v}{v^*(x)} + \mathcal{O}(v^2)\right].
\ee
Here $v^*(x)$ grows slowly with the displacement $x$, and is always larger than 20~$\mu$m/s in the present context. One recovers that the force is essentially independent of $v$ while $v \lesssim 1~\mu$m/s, and that the quasi-static approach is then fully justified. 

Above this threshold, we can already anticipate the discussion of Section~\ref{ffeq} by examining how the rotational viscous drag prevents instantaneous equilibration. The force $f_{\rm ext}$ required to unzip the double helix then grows significantly (if $v>0$ then $f_{\rm ext}>f_u$), while it decreases during rezipping (if $v<0$ then $f_{\rm ext}<f_u$). This explains the hysteresis observed experimentally. The order of magnitudes are in agreement with experimental values even though this agreement can be refined further by appealing to more elaborate far-from-equilibrium arguments in order  to take into account the relatively slow twist propagation \textit{inside} the rotating dsDNA~\cite{Cocco2002b}. Writhing might also complicate the issue of twist propagation through Eq.~(\ref{Fuller}). 

More recently, the hysteresis has also been investigated on a more fundamental basis by Kapri, with the help of either Monte Carlo simulations on a simplified ladder-like 2D lattice model~\cite{Kapri2012,Kapri2014}, and by Kumar and collaborators using 3D Brownian Dynamics~\cite{Kumar2013,Mishra2013}, with a number $N$ of base-pairs up to 512. In the case where the system is periodically driven with pulsation $\hat\omega$ (not to be confused with $\omega = {\rm d \phi}/{\rm d} t$ above) and force amplitude $G$, original scaling laws are found. Let us define the area of the hysteresis loop as $A_{\rm loop} = \oint \langle x(f)\rangle {\rm d}f$ on a cycle, where $\langle x(f)\rangle$ is the separation between ssDNAs extremities at a given force $f$, averaged over realizations. This dynamical order parameter measures the intensity of the hysteresis and vanishes in the quasi-static regime $\hat \omega \rightarrow 0$. Then a clear scaling law $A_{\rm loop} = N^\delta \mathcal{G}(G,\hat \omega N^z)$ is observed in simulations, where $\delta \simeq 1$, and $z \simeq 1$~\cite{Kapri2014,Mishra2013}. 

The experimental situation discussed above, where rezipping is too fast to keep the single-strands under tension because of rotational drag, can also be tackled in the framework developed by Cocco, Monasson and Marko~\cite{Cocco2002b} by setting $f_{\rm ext}=0$, i.e. $w_{\rm ss}(f_{\rm ext})=0$ in Eq.~(\ref{edo2CMM}). The so-obtained differential equation can be integrated exactly. If one chooses $n(t=0)=N$, that is to say if one starts from a completely unzipped configuration, then $N-n(t) \propto t^{1/2}$. As an example, the closure times in the $\lambda$-phage DNA case appear to be on the order of 1~s, which eventually sets the lower zipping velocity limit where the $f_{\rm ext}=0$ assumption is correct, $v>25$~$\mu$m/s. The effect gets stronger when zipping progresses, as it is also visible in Figure~\ref{force:unzip}b (lower blue and green curves).

Sequence effects which have a clear experimental signature, especially at low velocity~\cite{Essevaz1997,Thomen2002}, can also be included in the theory. Bockelmann, Essevaz-Roulet, and Heslot~\cite{Bockelmann1997,Bockelmann1998}, and later Cocco, Monasson and Marko~\cite{Cocco2002b} have extended the previous ideas to the case where $\Delta F_0(i)$ is a function of the base-pair $i$ being unzipped/zipped. More sophisticated approaches have also been proposed by Lubensky and Nelson~\cite{Lubensky2002}. The results of these calculations are in good agreement with experimental data on the $\lambda$-phage sequence. As in the experiments, the more stable GC-rich parts of the sequence generate a ``stick-slip''-like motion~\cite{Bockelmann1997,Bockelmann1998}, giving a sawtooth-like pattern in the extension-force plots~\cite{Lubensky2002}. However using this approach~\cite{Baldazzi2006} for the fast sequencing DNA at the base-pair level, even though feasible in principle, does not seem realistic at this stage.  Only sequence features appearing at a 10~bp scale have been observed so far~\cite{Bockelmann2002}. Theoretical considerations led to converging conclusions~\cite{Thompson2002}.

\subsection{Unfolding and folding of hairpins: Two-state model approximation}
\label{hairpins}

\begin{figure}[ht]
\begin{center}
\includegraphics[width=11cm]{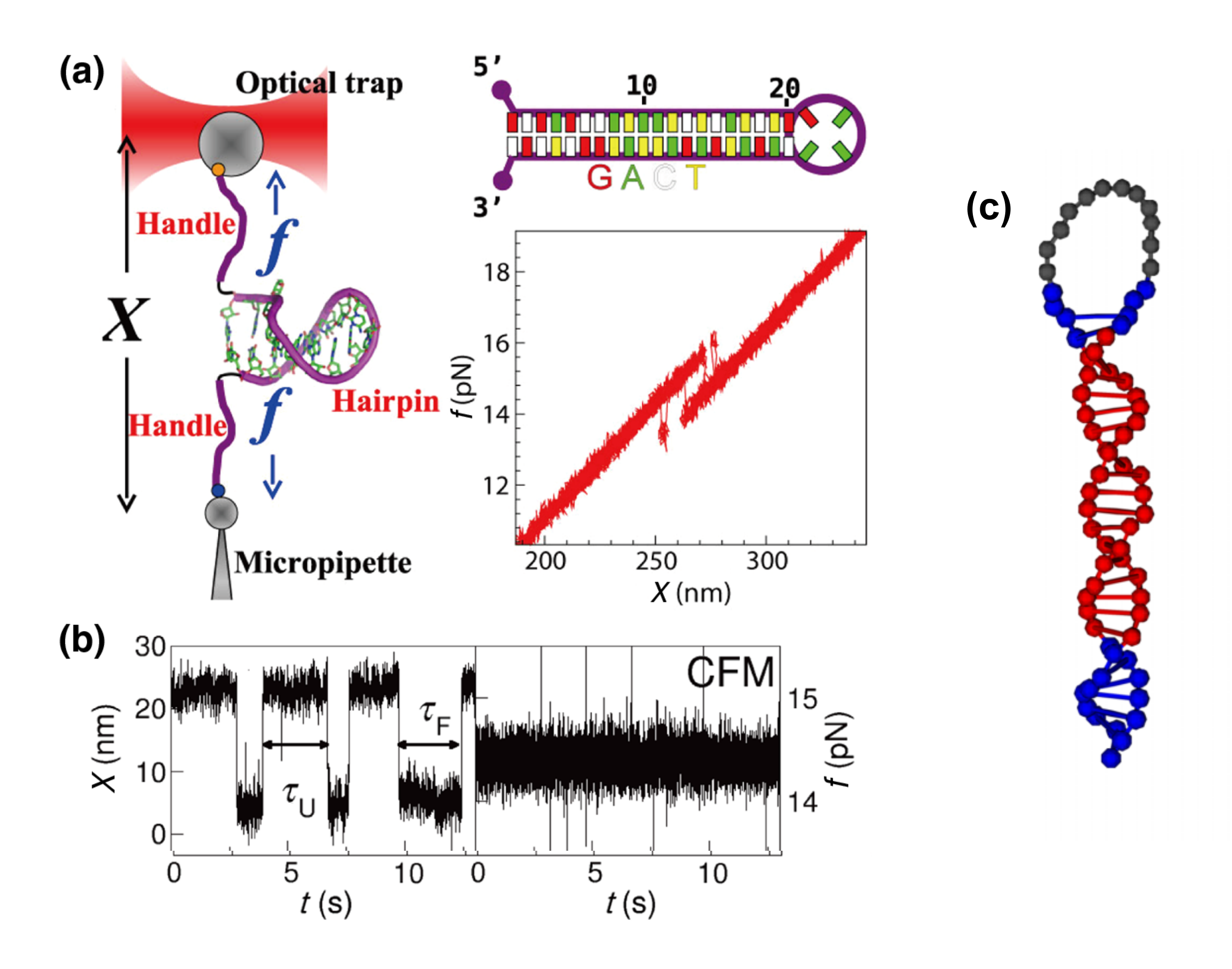}
\end{center}
\caption{(a)~Sketch of the unfolding/folding experiment.  The hairpin (upper right: DNA sequence) is stretched by a mechanical force $f_{\rm ext}$ (denoted by $f$ in the figure) applied with optical tweezers. An experimental force-distance curve obtained from a pulling experiment is shown (lower right). A first-order transition between the open and closed states occurs at $f_{\rm ext}\approx 15$~pN. (b)~Typical extension ($X$) and force traces obtained in the controlled force mode showing the hopping between the two states open and closed; (a) and (b) taken from~\cite{Hayashi2012}. (c)~Snapshot of a DNA beacon/hairpin with a sequence analogous to the one considered in Altan-Bonnet experiments~\cite{Altan2003} (GC rich regions in blue, AT sequence in red and T-rich loop of length $N_{\rm loop}$ in gray). Illustration from~\cite{Sicard2015}.
\label{hairpin:fig}}
\end{figure}

When the nucleic acid length is reduced to a few dozens of base-pairs with a short ssDNA loop at one of its ends to avoid a full separation in two strands, the DNA construct is usually called an ``hairpin''. The same experimental protocol as above can then be applied to pull on the single strands with a given force $f_{\rm ext}$. For an applied force suitably chosen around 15~pN, the open and closed configurations become roughly equiprobable, as illustrated in Figure~\ref{hairpin:fig}. 

Dynamics of folding/unfolding of DNA and RNA hairpins, with or without external force, can be studied by measuring fluorescence correlation spectroscopy~\cite{Altan2003,Bonnet1998,Jung2006}, absorbance vs. time after laser temperature jumps~\cite{Ansari2005}, or using mechanical force spectroscopy~\cite{Hayashi2012,Liphardt2001,Hanne2007,Neupane2012}. Very precise equilibrium energy landscapes $F(X)$ can be inferred from experiments ($X$ or $x$ measures the hairpin extension)~\cite{Woodside2006} which indeed display two marked potential wells. The ensuing two-state model has been the object of several experimental and theoretical investigations, together with the associated unfolding and folding dynamics. The central assumption of this two-state approximation is that the short sequence closure and opening occur in a cooperative manner.

\begin{figure}[ht]
\begin{center}
\includegraphics[width=7cm]{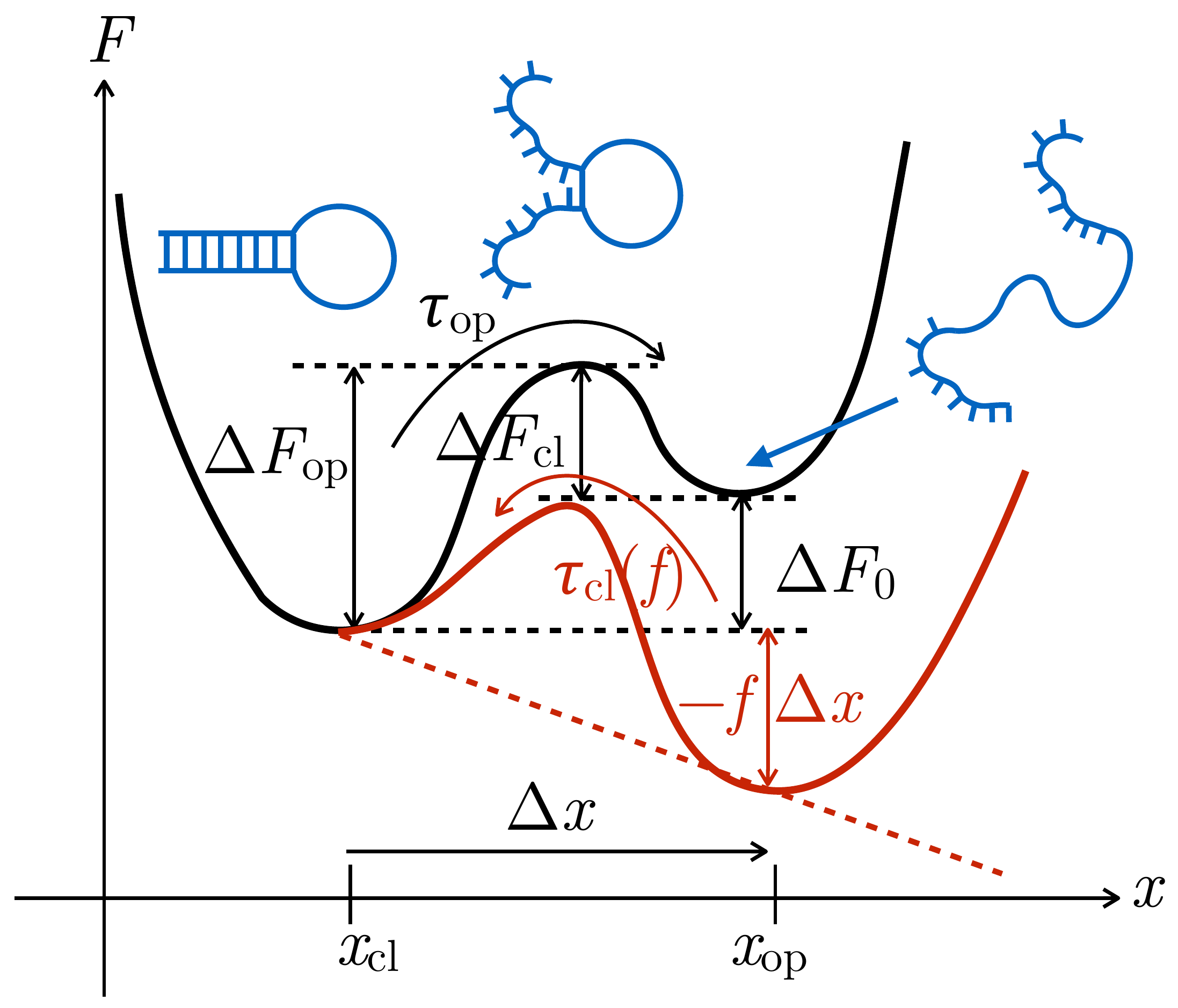}
\end{center}
\caption{Sketch of the 1D landscape in the pre-averaged two-state model, valid for instance for the hairpin opening and closure. A marked free-energy barrier separates the closed (left) and the open (right) states (equilibrium and transition hairpin states are sketched in blue).  When an external force, $f$, is applied to pull the hairpin, the free-energy profile is tilted in favor of the open state with a smaller opening barrier (red curve).
\label{2states:fig}}
\end{figure}

Kramers theory is particularly adapted to this situation (see for instance Ref.~\cite{Hanggi1990}). Adopting a quasi-static viewpoint supposed to be valid here~\cite{Hummer2012}, it states that in the overdamped regime of interest here, the closure time is given by
\be
\tau_{\rm cl}=\frac{2\pi\zeta}{\omega_{\rm met}\omega_{\rm TS}} \exp\left(\frac{\Delta F_{\rm cl}}{k_{\rm B}T}\right)
\label{Kramers}
\ee
where $\omega_{\rm met}$ and $\omega_{\rm TS}$ are the spring constants associated to the free-energy in the metastable basin (located on the right of the black curve in Fig.~\ref{2states:fig}) and at the transition state ($\zeta$ is the friction coefficient of the hairpin). For the reverse case, i.e. the hairpin opening, \eq{Kramers} also gives the opening time $\tau_{\rm op}$ once $\Delta F_{\rm cl}$ is replaced by $\Delta F_{\rm op}=\Delta F_0+\Delta F_{\rm cl}$ and $\omega_{\rm met}$ is replaced by $\omega_{\rm eq}$ the spring constant in the equilibrium well\footnote{In this section, $\Delta F_0$ is the \textit{total} ``reaction'' free-energy, i.e. $\Delta F_0=F(x_{\rm open})-F(x_{\rm closed})$ for the whole hairpin, not for a single base-pair.}.

To our knowledge, the first study of the hairpin kinetics at zero external force is due to Bonnet \textit{et al.}~\cite{Bonnet1998} (see also the review~\cite{Gurunathan2008}). The measured opening (or unfolding) times were around $\tau_{\rm op}\simeq100~\mu$s for a duplex stem of $n=5$~bp at room temperature. The closure (or folding) times were shown to depend strongly on the hairpin loop size $N_{\rm loop}$ (in bp units; see Figure~\ref{hairpin:fig}c) as $\tau_{\rm cl}\sim N_{\rm loop}^{2.6}$. Indeed as $N_{\rm loop}$ increases, the loop becomes more and more flexible, thus decreasing the probability that the two strands encounter back. According to the sequence, the mechanism might also be a three-state one with an intermediate state where only the distal CG base-pairs are paired~\cite{Jung2006}. The relaxation time between the open and intermediate states was also measured to be around $50~\mu$s in this case.

Ansari and Kuznetsov~\cite{Ansari2005} focused on the dependence of the hairpin dynamics upon changes in solvent viscosity. They found an Arrhenius behaviour following \eq{Kramers} for the opening ($\tau_{\rm op}$) and closure times ($\tau_{\rm cl}$) which scale almost linearly with the solvent viscosity. These long relaxation times, $\tau_{\rm r}^{-1}=\tau_{\rm cl}^{-1}+\tau_{\rm op}^{-1}$, are measured between $10~\mu$s (at water viscosity) to 1~ms (with 70\% of glycerol). The Arrhenius behaviour is supported by Monte-Carlo kinetic simulations~\cite{SalesPardo2005} with an activation barrier $\Delta F_{\rm op} \simeq 6~k_{\rm B}T$ for an hairpin stem of $n=7$~bp. Brownian dynamics simulations also show the Arrhenius behavior for short hairpins~\cite{Kenward2009}.

Starting with the pioneering work by Liphardt \textit{et al.} in 2001~\cite{Liphardt2001}, the hairpin folding/unfolding dynamics was later explored by applying a mechanical force of several pN that helps the unfolding. Within the two-state model, the unfolding dwell times depend on the applied force, $f_{\rm ext}$: 
\be
\tau(f_{\rm ext})=\tau_{\rm op} \exp\left(-\frac{f_{\rm ext} \Delta x}{k_{\rm B}T}\right)
\label{Kramers_force}
\ee
where the opening time at zero force $\tau_{\rm op}$ is related to free-energy barrier following \eq{Kramers}, and $\Delta x$ measures the characteristic distance between the open and the closed states (see Figure~\ref{2states:fig} and Ref.~\cite{Woodside2006}). The major interest in force spectroscopy is to dramatically increase the dwell times in the open state and thus to get enhanced sampling of the energetic landscape~\cite{Woodside2006}.

Liphardt \textit{et al.} made force-extension measurements for three different hairpin constructs by laser tweezers~\cite{Liphardt2001}. Using the two-state model they measured a barrier of $\Delta F_{\rm op}=59~k_{\rm B}T$ for an hairpin of $n=22$~bp with $\Delta x=5$~nm. The barrier was related to the experimentally measurable work as the area under the rip observed in the force-extension curves~\cite{Manosas2005} (see Figure~\ref{hairpin:fig}). Their experiment was carefully studied in Ref.~\cite{Manosas2006} (see also Ref.~\cite{Cocco2003}) by applying Kramers theory and taking into account the entropy associated with the single strands. The one-dimensional experimental effective barrier was reconstructed as a function of the number of open base-pairs for various force values. The extrapolation at zero force yields $\tau_{\rm op}=10~\mu$s, and a barrier $\Delta F_{\rm op}$ linear in $n$, showing the collective character of the opening, with a value $\Delta F_{\rm op}\simeq 60~k_{\rm B}T$ for the full opening ($n=19$). These different experiments indicate a barrier height roughly proportional to the number $n$ of base-pairs in the duplex stem, $\Delta F_{\rm op}/(k_{\rm B}T) \simeq n$ to 3$n$.

Other experiments were done on different DNA hairpins with a duplex length of $n=10$~bp~\cite{Hanne2007}. The zero force opening time was measured to be around 7~s, much larger than the value presented above. Moreover they found a much smaller value of $\Delta x\simeq 0.5-1$~nm as defined in \eq{Kramers_force}. These huge differences were assumed to be due to the different experimental conditions ($p$H and salt concentration, sequence)~\cite{Hanne2007}. As experimentally shown by Bonnet \textit{et al.} the loop size $N_{\rm loop}$ also has a big influence on the folding rates~\cite{Bonnet1998,Woodside2006b}.

Finally, a step further was done by Neupane and co-workers~\cite{Neupane2012}, who were able to measure experimentally the so-called transition time between the two states of hairpins with $n=10$ to 30~bp in the duplex. This transition time is generally shorter than the opening and closure times, because it does not include the contribution of the waiting time spent in the potential wells. It was found to vary linearly with the duplex length, $\tau_{\rm TS}\simeq 6-30~\mu$s for several unfolding rates. They were measured by reconstructing the force-dependent one-dimensional energy landscape from the observed distributions, thus again assuming the validity of the quasi-static approximation. An inherent complication arises from the presence of the optical trap and DNA handles, which have been corrected. The transition time is given by the formula~\cite{Hummer2004,Roland2015}
\be
\tau_{\rm TS}\simeq \frac{\ln(2 e^\gamma \beta \Delta F_{\rm op})}{D\beta\omega_{\rm TS}}
\ee
where $\gamma\simeq0.577$ is the Euler constant and $\omega_{\rm TS}$ is defined in \eq{Kramers}. Moreover they experimentally determined the diffusion coefficient $D\simeq 0.5~\mu$m$^2$/s. 

To explain the two-state shape of the free-energy landscapes, they were shown to be modeled with a good accuracy by a simple zipper-like model taking into account both the sequence-dependent $\Delta F_0$ (from nearest-neighbor free energy parameters) and the elastic energy stored in the stretched and curved polymers~\cite{Woodside2006,Cocco2003,Woodside2006b}. The release of the last base-pair reduces the free-energy by increasing the entropy of the ssDNA segment of length $N_{\rm loop}>\ell_p^{\rm ss}$ forming the initial loop. This is the origin of the potential well in the fully open hairpin state (Figure~\ref{2states:fig}).

However some experimental works questioned the two-state approximation for modeling the hairpin kinetics at zero force~\cite{Jung2006,Ma2007,Jung2008} because of the existence of intermediate states (possibly mis-folded) between the fully zipped and the fully open ones (see also~\cite{Ouldridge2013}). By finely analyzing FCS experimental data on an hairpin of size $n=20$~bp to investigate its terminal fluctuations, Chen and coworkers established that the zipper model was equally unable to account for the short-time ($<0.1$~ms) dynamics and to consistently fit the experimental FCS~\cite{Chen2008}. They had to appeal to stretched exponentials $\exp[-(\omega t)^b]$, $b \neq 1$ (instead of simple ones, $b=1$, in the original zipper model, as in ordinary chemical reaction kinetics) to model elementary processes, and the so-obtained fits were then very satisfying. The ensuing average zipping rates were found to be on the order of a fraction to a few tens of bp/$\mu$s (with significant error bars), in correct agreement with earlier estimates based on more rudimentary techniques~\cite{Craig1971,Porschke1974}. The occurrence of stretched exponentials was attributed to ``static disorder'', without proposing at this stage any microscopic mechanism that could be responsible for it.

Altan-Bonnet, Libchaber, and Krichevsky have also done FCS experiments at zero force~\cite{Altan2003} on DNA hairpins to follow the opening/closure dynamics of a AT sequence of 18~bp clamped by GC-rich regions at its extremities (see Figure~\ref{hairpin:fig}c). In contrast to the previous hairpin constructs, the internal AT bubble can open by thermal activation and then close again without the molecule extremities opening. In these experiments, the quencher and fluorophore were located in the middle of the AT sequence and not at the hairpin extremity, thus measuring the lifetime of these partially open state. They showed that bubble closure times follow an Arrhenius law, which indicates the presence of an energy barrier impeding closure of $\simeq11~k_{\rm B}T$ (for 18 AT bp) and the relaxation time $\tau_{\rm r}$, dominated by the closure time, was $\simeq 10$ to $100~\mu$s. One of the first idea to model this hairpin kinetics was naturally to use the zipper model~\cite{Altan2003,Bicout2004}. It was essentially used to fit the auto-correlation function with the bubble closure and opening times as fitting parameters. For instance, in an analytical work Bicout and Kats~\cite{Bicout2004} deduced a bubble lifetime $\tau\simeq \tau_1 n^2$. However $\tau_1$ was an adjustable parameter only (fitted using $n=18$ as in~\cite{Altan2003}). Other works assumed a quadratic confining potential (modeling the GC base-pairs) to fit the auto-correlation function~\cite{Chatterjee2007,Chakrabarti2011}, but again without estimating the relaxation time. We shall see below in Section~\ref{wooed} that the two-state approximation is not fully adapted to obtain the bubble closure lifetime. The opening of smaller bubbles limited to the AT-core has to be considered as a specific third state. In this case, we shall see that the chain dynamics is then coupled to the base-pair kinetics in a very specific way, which explains why neither the zipper model nor the two-state approximation are fully adapted. It is worth mentioning that the two-state approximation for hairpins already couples base-pair and chain orientational degrees of freedom at play in the open-state free-energy well.

\subsection{Bubble dynamics in torsionally constrained DNA}
\label{supercoil}

It has been known for a while that a torsionally constrained DNA, either by applying an external torque on it or by closure of the molecule into a superhelical ring (such as in plasmids) or by the formation of loops through regular attachments to a network of proteins in the nucleosome, can be locally denaturated through the nucleation of one or several denaturation bubbles under certain conditions~\cite{Adamcik2012,FrankKamenetskii1997,Dean1971,Benham1996,Strick1998,Takahashi2015}. Still with the goal of deciphering basic biological mechanisms on physical grounds, the statistical physics of this phenomenon in equilibrium has been extensively studied by Benham and others~\cite{Fye1999,Benham1979,Benham1980,Benham1990,Jeon2010}, including sequence effects and their connection with biologically relevant sites~\cite{Jost2011}.  

Less is known about the underlying dynamics of the nucleation of bubbles in superhelical DNAs. Lankas \textit{et al.} observed kinking (but not bubble nucleation) of duration between 10 and 50~ns on MD simulations of  DNA minicircles of 94~bp~\cite{Lankas2006}. Note that different force fields led to the different conclusion that a bubble can also be nucleated~\cite{Mitchell2011}. Very recently, Oberstrass \textit{et al.} used a high-resolution approach for measuring the torsional response of short DNA sequences (20 to 100 bp)~\cite{Oberstrass2013}. In particular, they observe the reversible hopping between two states of AAT-repeats, related to DNA breathing. The opening and closed mean dwell times (between 1 and 10~s for 5 to 8 negative rotations) were measured as a function of the applied negative torque, an increase in negative twist favoring the open state.

Hwa \textit{et al.}~\cite{Hwa2003} studied the dynamics of twist induced denaturation bubbles using the Poland-Scheraga model in the single bubble approximation. According to the applied negative torque value $-\mathcal{T}$, they focus on the transition between a delocalized small breather which freely diffuses along the chain for small torques, $\mathcal{T}<\mathcal{T}_{\rm loc}$, and a localized bubble due to the sequence heterogeneity for larger torques,  $\mathcal{T}_{\rm loc}<\mathcal{T}<\mathcal{T}_{\rm d}$ (for $\mathcal{T}>\mathcal{T}_{\rm d}$ the DNA is fully denaturated). The critical torque, $\mathcal{T}_{\rm loc}$, was estimated to be around 8~pN.nm $\approx 2 k_{\rm B}T$. An analogy can be drawn between glassy dynamics and the bubble dynamics in the localized state. In particular the escape time is shown to be sub-diffusive $\tau\sim N^z$ with $z>2$ ($z=2$ in the delocalized state), with an average bubble length which grows logarithmically with time, and an aging phenomenon is observed in the MD simulations.

Benham and co-workers interested in the DNA transcription dynamics, studied the dynamics of DNA minicircles~\cite{Mielke2005} (of length $L\simeq \ell_p$) with a dynamically imposed superhelicity, to model the torque generated by the RNA polymerase. The system being out-of-equilibrium, they used the Brownian dynamics simulations developed in Ref.~\cite{Mielke2004}. They focused on the sequence-dependent location, essentially in the AT sequences, in agreement with the equilibrium Benham model and found that initial duplex opening occurs around 100~$\mu$s after the imposed torque (for a imposed angular velocity of about $4\times 10^4$~s$^{-1}$, i.e. at least 2 orders of magnitude greater than that of transcription, to speed up the simulation). Jost and collaborators also explored the connection between sequence and the probability of bubble opening, while insisting on the difference between metastable unwinding of a large bubble and small scale breathing~\cite{Jost2011}, as already mentioned in the Introduction.

A full comprehension of the bubble dynamics under applied torque is still lacking. One major difficulty arises from the Fuller-White theorem, \eq{Fuller}. Indeed, increasing the torque or the superhelicity on a plasmid for instance imposes a 3D writhe (supercoiling) to the circle even before a bubble nucleation occurs. Supercoiling being a non-local effect, it is challenging to include it in an effective Hamiltonian. Hence, a complete theoretical model should consider the total bending and torsional energies of the whole plasmid, which is a formidable task.

\medskip

To conclude this section, we have seen that pre-averaged models are able to describe correctly and in a simplified way number of experimentally relevant situations (force-induced zipping/unzipping at low velocity, folding/unfolding of hairpins) but that their use is more questionable in other circumstances because chain orientational degrees of freedom have relaxation times on the same order of magnitude as or longer than base-pairing. The next section is devoted to the survey of these far-from-equilibrium phenomena where base-pairing and chain orientational degrees of freedom must be tackled on an equal footing.


\section{Interplay between base-pairing and chain dynamics}
\label{ffeq}


In this section, we explore the coupling between base-pairing and chain orientational degrees of freedom, with a special emphasis on far-from-equlibrium chain dynamics, as motivated at several places in the previous sections. We explore the different regimes of interest, below, close to, and above the melting temperature $T_m$ with an emphasis on the scaling laws in the limit of long molecules. We are especially interested in rehybridization below $T_m$. It is the result of several distinct stages: a nucleation step where diffusing complementary single strands encounter, followed by a zipping step and, in some circumstances, by a last metastable state before complete closure can occur. We first review some experimental results that motivate the theoretical investigations reviewed in this section.

\subsection{Experimental motivation}
\label{exp:motiv}

One of the best candidates for far-from-equilibrium dynamics in the context of DNA base-pairing are related to single strands rehybridization at $T<T_m$, as illustrated in Figure~\ref{MainFig}c, 
of pivotal importance, e.g., in PCR. Processing zipping velocities can conveniently be measured on short molecules of few base-pairs. Early measurements estimated them to be around 1 to 10~bp/$\mu$s~\cite{Sikorav2009,Porschke1974,Craig1971} at physiological salt concentration and room temperature, by use of a temperature-jump apparatus and dynamical measurements of UV absorbance. The authors of Ref.~\cite{Chen2008} were led to similar conclusions using FCS experiments on hairpins. All-atom simulations predicted somewhat slower velocities of about 0.2~bp/$\mu$s on very short 6-bp constructs, in similar conditions~\cite{Qi2011}. However, for long molecules, it has been understood that closure times grew faster than the number of base-pairs $N$. For example, as already indicated in Section~\ref{force_unzip}, it can be concluded from Thomen \textit{et al.}'s more recent experimental data on single molecules~\cite{Thomen2002} that the zipping velocity is $v_{\rm zip} \simeq 10^{-2}$~bp/$\mu$s for a $\lambda$-phage 
DNA of 48.5~kbp at comparable temperature. 
Therefore it is tempting to assume that zipping times follow a simple scaling law  
\be
\tau_{\rm zip}\propto N^{\alpha}\qquad\mathrm{with}\qquad\alpha>1.
\label{def:alpha}
\ee
The non-trivial dynamical exponent $\alpha$ is the object of Section~\ref{far:from:eq:zipping}. The zipping velocity is $v_{\rm zip} \simeq aN/\tau_{\rm zip}  \propto N^{1-\alpha}$. The previous data indicate a reduction of $v_{\rm zip}$ by a factor 100 to 1000 when $N$ increases by a factor $\sim 10^4$. The rough estimate $\alpha\approx 1.5-1.75$ ensues. To our knowledge, no systematic experimental study of $v_{\rm zip}$ as a function of $N$ has been performed so far. This issue is addressed in Section~\ref{poq}.

In the same way, unzipping at $T>T_m$ has been experimentally explored by Record and Zimm for long DNAs and shown to display apparent reaction rates $k(t)$ decreasing with time~\cite{Szu1979,Record1972}. This suggests that unzipping velocities decrease when denaturation progresses and thus depend on the DNA molecular size, as it will be made explicit below. No scaling law or exponent can be inferred from these bulk measurements, however denaturation rates can be estimated, e.g., $\simeq 10^{-3}$~bp/$\mu$s for the T2-phage DNA\footnote{In Ref.~\cite{Record1972}, T2-phage DNAs of size $N\simeq 2 \times 10^5$~bp have been submitted to temperature jumps from which an unzipping step rate has been estimated to be $\simeq 5 \times 10^4$~bp/min $\simeq 10^{-3}$bp/$\mu$s. A 3-fold lower value was later inferred by Szu and Jernigan through a more elaborate analysis~\cite{Szu1979}. Note however that these experiments were performed in a HCONH$_2$-rich solvent in order to lower $T_m$ and that HCONH$_2$ is 3 times more viscous than water.}.

Finally, by controlling the $p$H of the DNA solution, Crothers measured the relaxation base-pair dynamics (on a T2-phage DNA) close to $T_m$~\cite{Crothers1964}, followed by others~\cite{Davison1966,Davison1967,Spatz1969}. The scaling law for the relaxation time $\tau_m$ follows the classical 1D diffusion law  $\tau_m\propto N^{\alpha_m}$ with $\alpha_m \simeq 2$ below a threshold above which it saturates, presumably because of multiple bubble nucleations. However the experimental diffusion coefficient found by Crothers is $\sim10^3$ smaller than expected. This discrepancy will be examined in Section~\ref{F:barrier}.

This section examines how coupling base-pairing and chain orientational degrees of freedom casts light on these different experimental facts and scaling laws.

\subsection{The nucleation step preceding the zipping of long molecules}
\label{nucleation}

As already pointed out above, when complementary single strands can freely diffuse (i.e. when the geometry is not that of an hairpin and complementary strands have been totally separated at sufficiently high temperature), zipping at $T<T_m$ follows a limiting nucleation step. A few-bp stable nucleus must be formed after complementary ssDNA successful encounter, following aborted mismatched partial hybridizations~\cite{Jayaraman2007,Sambriski2009,Sambriski2009b,Ouldridge2013,Marmur1963,Porschke1974,Craig1971}. This limiting character of the nucleation step was understood in the first pioneering works on DNA renaturation, where zipping following the nucleation step was even considered as ``instantaneous'', given the time resolution of experimental techniques at this time~\cite{Marmur1963,Rau1978} (except at high salt and low temperature conditions where zipping was supposed to be slowed due to the formation of secondary structures in the single strands~-- above $T^\star \simeq(T_m-30)^\circ$C, secondary structures become unstable~\cite{Sikorav2009}).

The succession of events preceding nucleation in the dilute regime has been controversial. The question, first raised by Wetmur and Davidson in 1968~\cite{Wetmur1968}, can be stated as follows: is the nucleation event diffusion- or reaction-limited? Recently, Sikorav, Orland and Breslau brought a fresh and decisive eye on this issue in a theoretical work~\cite{Sikorav2009}, where using available experimental facts (namely the dependence of reaction rates on DNA length, solvent viscosity, ionic strength and temperature) and modern concepts from polymer physics, they demonstrated that the only realistic encounter mechanism leading to the nucleation event is a Kramers' process and is thus reaction-limited~\cite{Hanggi1990}. This work has been later confirmed by Ferrantini, Baiesi and Carlon~\cite{Ferrantini2010} by numerical simulations performed in the same regime of parameters 
as experiments. Both works conclude that the main obstacle to nucleation is related to the interplay between base-pairing and chain dynamics. Nucleation requires interpenetration of the mutually self-avoiding complementary single strands in their random-coil configuration below $T_m$. The free-energy cost is therefore of entropic nature, $\Delta F\simeq \theta_2 k_{\rm B}T\ln (N^\nu)$ where $\nu$ is again the Flory exponent and $\theta_2\simeq 0.82$ is the contact exponent~\cite{desCloiseaux_book}. The bimolecular nucleation rate grows like $k_2\propto k_{\rm nu} N$ where $N$ is the DNA molecule length in base-pair units and $k_{\rm nu}$ is the nucleation rate constant. Applying the Kramers law in the high friction regime (see \eq{Kramers} above) for $k_{\rm nu}$ yields $k_2\propto \exp(-\beta \Delta F)\propto N^{1-\nu\theta_2}$ with $1-\nu\theta_2\simeq 0.51$.

\subsection{Far-from-equilibrium zipping of long molecules}
\label{far:from:eq:zipping}

Once a successful nucleation event has occurred, the zipping dynamical exponent $\alpha$ defined in Eq.~(\ref{def:alpha}) characterizes the anomalous dynamics and is discussed now from a theoretical perspective. Note that $\alpha=1$ in the zipper model, see \eq{ziptime}. Given the analogy that we shall discuss below (Section~\ref{transloc}) between DNA zipping and field-driven polymer translocation across a nano-pore, and given that the value of the exponent $\alpha$ in this latter context remains controversial in spite of two decades of intensive numerical and theoretical investigations~\cite{Vocks2008,Palyulin2014}, we shall see that the value of the dynamical exponent in the present situation also remains a source of debate. Nevertheless, a common emerging feature is that  the anomalous exponent comes from the fact that during zipping, a significant part of the unpaired single strands are strongly out of equilibrium, between the already zipped part of the molecule and ``quiescent'', still at equilibrium single-strand extremities, as illustrated in figure~\ref{stem:flower} below. Vocks and collaborators use the terminology of ``memory effect'' to describe the time-correlations related to this phenomenon~\cite{Vocks2008}. A similar description holds for tethered polymers in moderate or strong flow. It was proposed by Brochard-Wyart more than 20 years ago~\cite{BrochardWyart1993,BrochardWyart1995} and will be useful below.

Unless explicitly mentioned, all works cited until the end of this section, either numerical of analytical, are performed in the free-draining (or Rouse) approximation scheme, thus ignoring hydrodynamic interactions~\cite{DoiEdwards2004,Manghi2006}. Note that the Kinetic Monte Carlo simulations discussed below implicitly take into account the Rouse dynamics~\cite{Newman1999}. In general, hydrodynamic interactions likely speed up the polymer dynamics, and their effect will also be discussed when required.

\begin{figure}[ht]
\begin{center}
\includegraphics*[width=12cm]{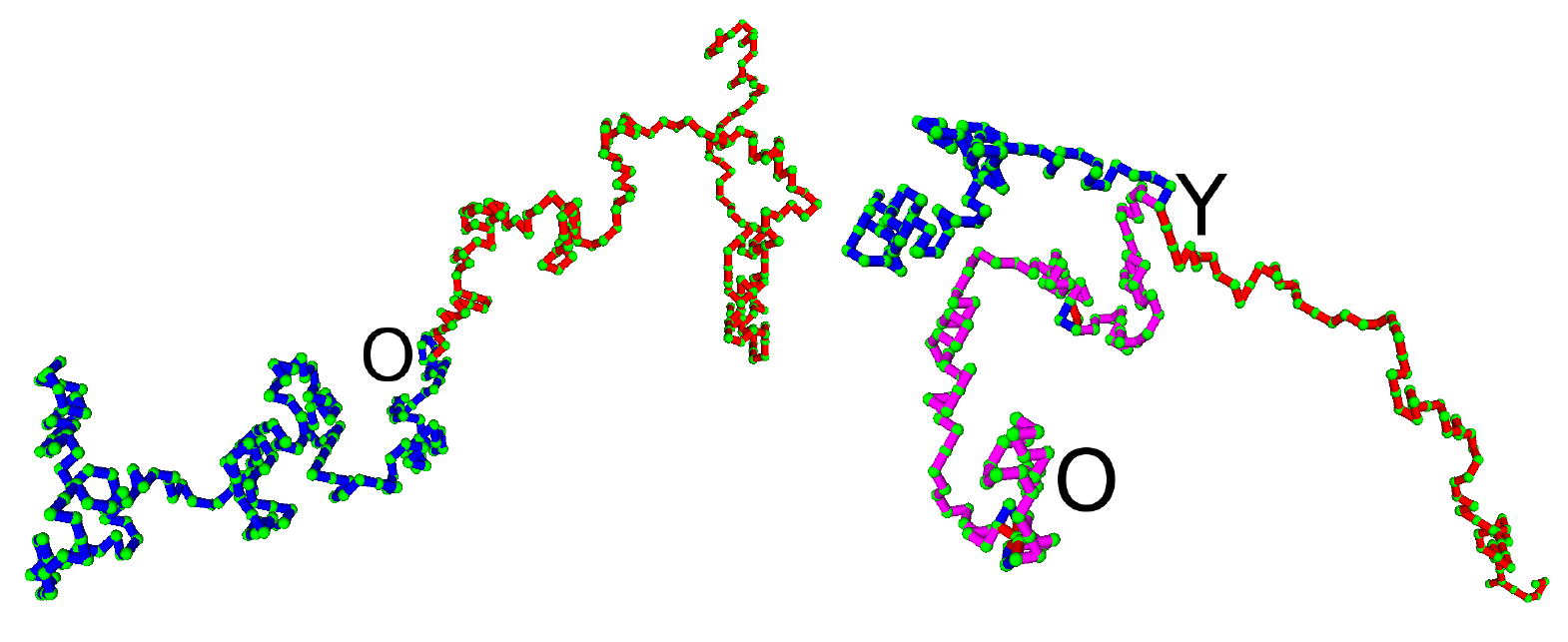}
\end{center}
\caption{Zipping in the case of the lattice model of Ref.~\cite{Ferrantini2011}, discussed in the text. Left: At time $t=0$, both ssDNA molecules (in blue and red) share a single closed base-pair. At $T<T_m$, the duplex starts closing. Right: the typical Y-shape adopted by the DNA molecule. Between points $O$ and $Y$, the duplex (in pink) is closed and both polymers are superimposed,  whereas their free extremities remain single-stranded. Note that in this illustration, some blue and red small denaturation bubbles, where both strands are not superimposed are visible in the pink duplex region. The zipper model studied in Section~\ref{zipper:model} forbids this possibility.
\label{Ferrantini:fig}}
\end{figure}

In Ref.~\cite{Ferrantini2011}, in 2011, Ferrantini and Carlon studied DNA zipping in a simplified numerical approach, where neither the helical character of the duplex state nor its large persistence length were explicitly taken into account (see Figure~\ref{Ferrantini:fig}). The polymers are simply defined on a face-centered-cubic (fcc) lattice. The two strands are both mutually- and self-avoiding, with the exception of monomers with the same index along each strand, which are referred to as complementary monomers. Two complementary monomers bind when overlapping on the same lattice site. Molecules as long as $N=500$~bp can be studied, at temperatures ranging from well below to slightly below the melting transition. All so-obtained scaling laws are compatible with an exponent value $\alpha \simeq 1.37$. The authors remark that this value is itself remarkably compatible with the predictions of Ref.~\cite{Vocks2008} in the context of field-driven polymer translocation, $\alpha=(1+2\nu)/(1+\nu)=1.37$, where $\nu \simeq 0.59$ (see \eq{Rouse2}) is again the Flory exponent in 3D ($\nu=0.75$ in 2D). However, this value $\alpha=(1+2\nu)/(1+\nu)$ has been contested in the context of driven translocation, as detailed in the next section, and Ferrantini and Carlon's numerical value is more likely due to finite-size effects~\cite{Carlon_private}. Note that in this work, the zipper-model hypothesis~-- namely the fact that the partially two-stranded molecule is allowed to have only one unbroken sequence of successive paired bases growing from the initial nucleus, and thus adopts a Y-shape as in Figure~\ref{Ferrantini:fig} (right)~-- was also tested numerically. As far as scaling properties are concerned, conclusions remain unchanged when relaxing this constraint and allowing secondary renaturation nuclei away from the original Y-shape junction. 

In a more recent publication~\cite{Frederickx2014}, Frederickx, in't Veld and Carlon propose an alternative argument based on the ``stem-and-flower'' description by Brochard-Wyart, because the driving force is strong enough~\cite{BrochardWyart1995}. Indeed, the average free-energy $\Delta F_0\simeq 2.5 \; k_{\rm B}T$~\cite{Manghi2009,SantaLucia1998}  gained when closing one base-pair at physiological temperature and salt concentration for a ``random'' sequence is larger than $k_{\rm B}T$. The closest part to the two-stranded region of each single strand (the ``stem'') is under tension and into motion, whereas the extremities of the same single strands (the ``flowers'') are still unaffected by the zipping process and are close to their equilibrium initial state, as illustrated in Figure~\ref{stem:flower}. Thus only the stems move at a given time and undergo friction. If $f\simeq \Delta F_0/a$ is again the typical driving force (in the piconewton range), averaged over CG and AT base-pairs for simplicity sake, the average number $n(t)$ of bound bases in the double strand obeys the relation (neglecting the Langevin random force)
\begin{equation}
f-\zeta_0\gamma(n)a\frac{{\rm d} n}{{\rm d}t}=0
\label{friction}
\end{equation}
where $\zeta_0 \gamma(n)$ is the stem-length dependent friction (the average stem length is a function of $t$ and thus of $n(t)$; for a similar approach, see also Ref.~\cite{Dasanna2012}). The velocity of the junction is $v=a\ {\rm d} n/{\rm d}t$. If the flower does not move significantly during zipping, it can be easily argued~\cite{Frederickx2014} that the typical distance between the zipped region and the flower is proportional to the stem length $an_s$ (supposedly stretched). The $n+n_s$ monomers of each single strand now in the duplex or in the stem where previously in a random coil of size $a(n_s+n)^\nu\simeq a n^\nu$ for large enough $n$ [see figure~\ref{stem:flower}(b)]. Thus $n_s\simeq n^\nu$ and $\gamma(n)\simeq n^\nu$. From integration of Eq.~(\ref{friction}), it follows that 
\begin{equation}
n(t)\simeq \left(\frac{t}{\tau_f}\right)^{1/(1+\nu)}
\end{equation}
where $\tau_f=a\zeta_0/f$ is again the characteristic time of the junction [see Eq.~(\ref{tau00})]. The closure time $\tau_{\rm zip}$ defined by $n(\tau_{\rm zip})=N$ obeys 
\begin{equation}
\tau_{\rm cl}\simeq\tau_f N^{1+\nu}.
\end{equation}
Consequently $\alpha=1+\nu\simeq 1.59$, as observed numerically even for relatively short constructs~\cite{Frederickx2014}, because their length $N=20$ to 50 is already large as compared to the ssDNA persistence length, which validates the use of the Flory exponent $\nu$. In Ref.~\cite{Frederickx2014}, it was proposed that previously published experiments~\cite{Neupane2012} displayed closure times fully compatible with this scaling. However, in these experiments, the hairpin in the optical trap is under relatively large tension, larger than 10~pN~$>k_{\rm B}T/\ell_{p}^{\rm ss}$, so that the single strands are possibly not in a random coil configuration when the hairpin is open, which might question the use of the above stem-flower picture in this precise experimental context. To finish with, the authors of Ref.~\cite{Frederickx2014} do not expect hydrodynamic interactions to modify the exponent because the friction $\gamma(n)$ originates from the stretched region of the single strands. In reality, logarithmic corrections are expected in this context because of hydrodynamic interactions~\cite{DoiEdwards2004,Manghi2006}, however for the small range of values of $n$ under consideration, $\ln n$ is essentially a constant and the exponent $\alpha$ remains unchanged in practice.  

\begin{figure}[ht]
\begin{center}
\includegraphics*[width=6.5cm]{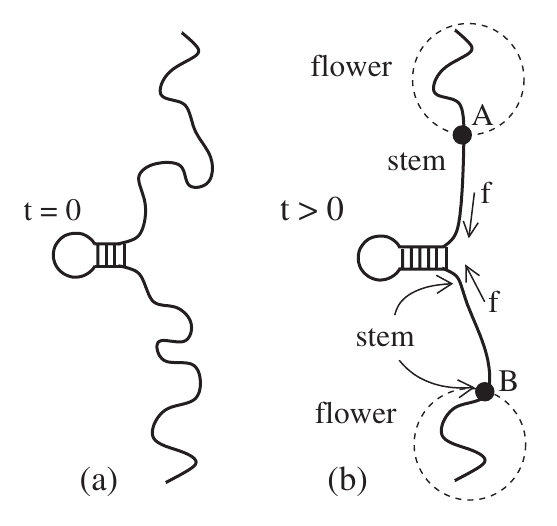}
\end{center}
\caption{The stem-and-flower picture of hairpin zipping. (a) When zipping starts at $t=0$ after a successful nucleation event close to the loop, both single strands are in a random coil configuration. (b) After $n<N$ base-pairs have been zipped, the ``flower'' base-pairs have not moved yet and remain close to their initial equilibrium configuration, whereas two stems of length $an_s$ link them to the already zipped region. Only the stems are under the influence of the driving force $f$. Since both stems have been substituted to $n+n_s$ monomers formerly in a random coil configuration, with end-to-end distance $\sim(n+n_s)^\nu\sim n^\nu$, the stem length also scales as $n^\nu$. The local driving force $f$ comes from the free energy gain $\Delta F_0$ consecutive to each monomer closure, $f \approx \Delta F_0/a$. Taken from~\cite{Frederickx2014}.
\label{stem:flower}}
\end{figure}

Dasanna and coworkers have tackled a similar issue by using more realistic mesoscopic numerical models of two interacting chains, where the difference in persistence lengths between ss and dsDNA~\cite{Dasanna2012}, and later the helical character of the duplex state~\cite{Dasanna2013} have successively been taken into account in a realistic way. In their case, zipping started from both extremities, as would occur when closing a large equilibrated bubble in the middle of a very long DNA molecule, or in a molecule with clamped extremities. Their findings were compatible with the previous ones in both the ``ladder model'', where the helical character of dsDNA is ignored~\cite{Dasanna2012}, and the fully helical model~\cite{Dasanna2013}. This indicates that, contrary to the unwinding regime explored in section~\ref{TgtrTm} below, the duplex helical character does not seem to play a pivotal,  limiting role during zipping. However, due to the increased complexity of the model, reachable system sizes where modest ($N \lesssim 100$) as compared to Carlon and coworkers' studies.

\subsection{Analogies with field-driven polymer translocation -- Different driving force regimes}
\label{transloc}

\begin{figure}[ht]
\begin{center}
\includegraphics*[width=3.8cm]{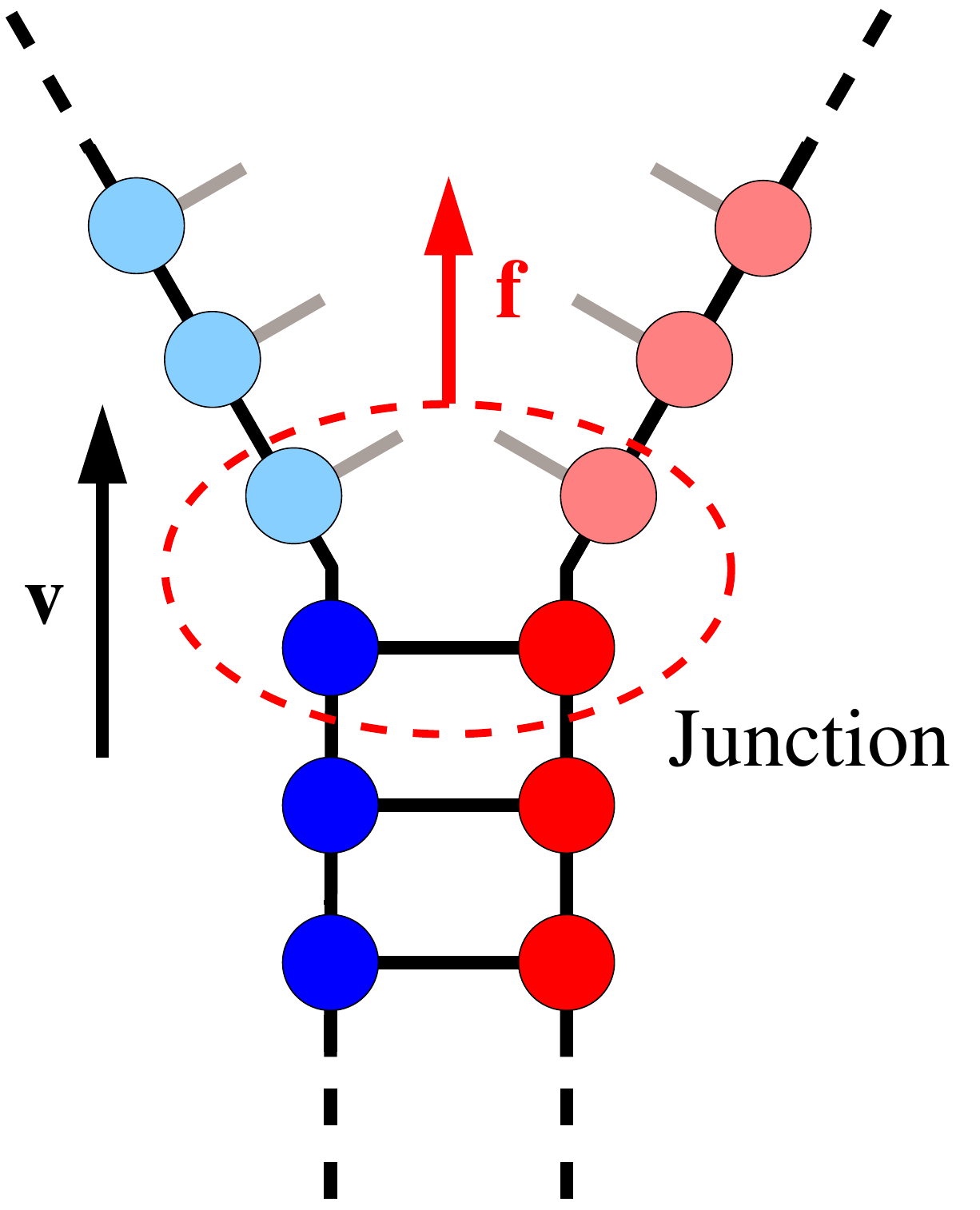}~~~~~~~\includegraphics*[width=6cm]{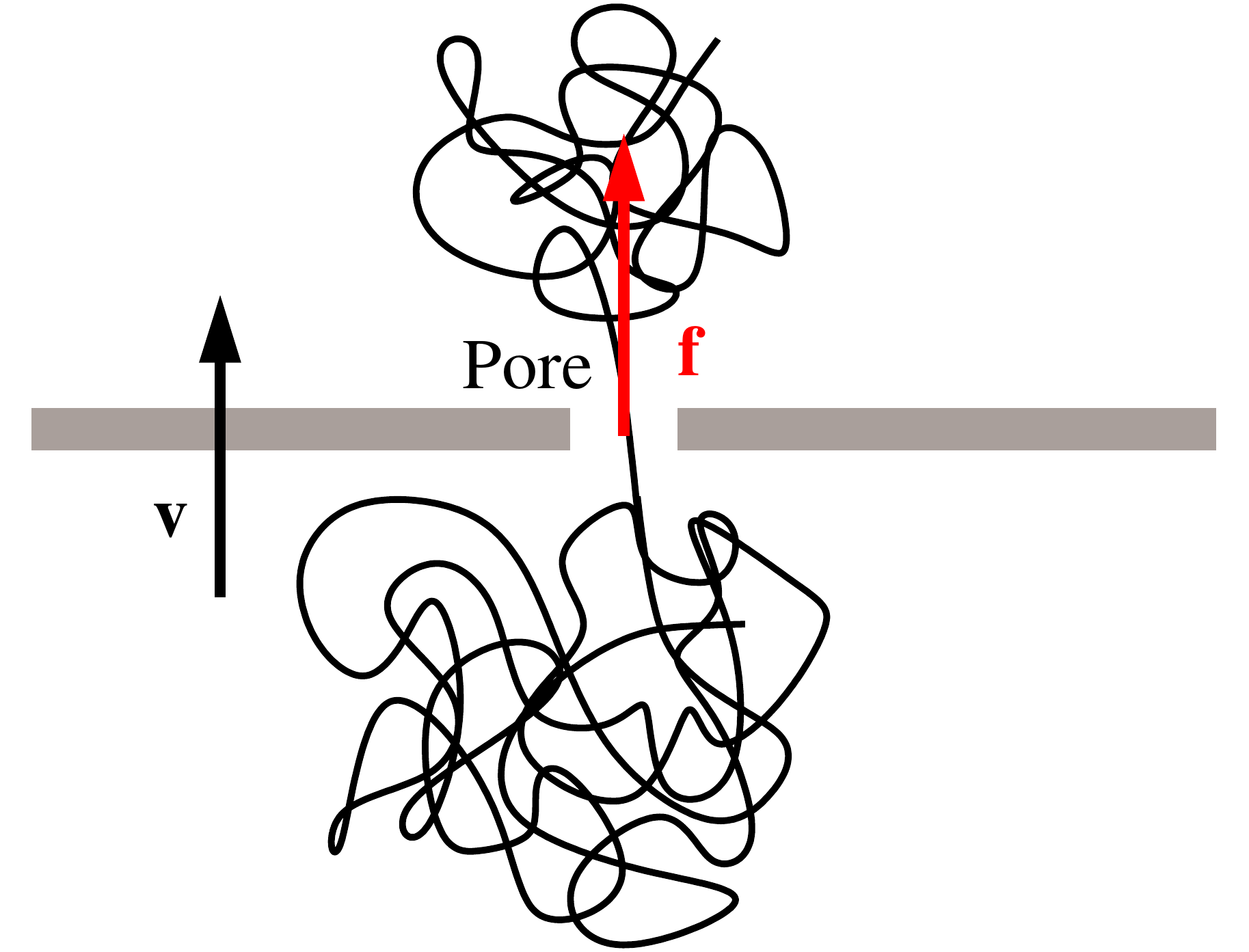}\\
\ \hspace{-0.9cm}{\bf (a)}\hspace{5.6cm}{\bf (b)} \
\caption{Analogy between DNA zipping at $T<T_m$ and DNA force-driven translocation. (a) A DNA molecule in the zipping regime, where $\mathbf{f}$ is the force acting at the Y-junction, driven by base-pair closure. (b) A polymer under the translocation process, where $\mathbf{f}$ is now the driving force inside the pore. In both cases, $\mathbf{v}(t)$ is the velocity of the interface between the two polymer states: ds-ss junction in (a) and before-after the pore in (b).
\label{analogy}}
\end{center}
\end{figure}

Polymer translocation across a nano-pore is an important and very active topic of investigation by itself (for a recent review focused on theoretical and numerical aspects of translocation, see Ref.~\cite{Palyulin2014}). Nucleic acid translocation across cell membranes is of interest in gene therapy and drug delivery, and has even been proposed as a technique of fast DNA-sequencing.  Experimentally, a membrane separates two compartments and a nanoscopic pore enables molecular transport between them (Figure~\ref{analogy}b). The pore size is supposed to be comparable to the thickness of the translocating molecules. The duration of the translocation event (also called the pore-blockade time) is of particular interest. A clear distinction must be made between unforced translocation (no force is applied), pulled translocation (a force is applied at a polymer end, see Ref.~\cite{Palyulin2014}), and field-driven translocation~\cite{Vocks2008}. In the former case, analogous to pairing dynamics at $T=T_m$ (see below), it was rapidly understood that a polymer cannot translocate through a succession of quasi-equilibrium states, because it would do so at a rate at which it has no time to equilibrate~\cite{Chuang2001,Walter2012}. In the latter case, a force $f$ is locally applied at the monomer inside the pore, either induced by a difference of chemical potential between both compartments, or, in the case of charged polymers such as DNA, by a DC voltage difference between the two sides of the membrane. We shall see below that the average translocation duration, $\tau_{\rm d}(N,f)$, essentially depends on the polymerization index $N$ and on the force $f$. An analogy between polymer translocation and zipping on the one hand, and polymer adsorption (physisorption) on a 2D substrate on the other hand, has also been proposed~\cite{Panja2009}.

The analogy between DNA zipping at $T<T_m$ and DNA force-driven translocation is as follows (see Figure~\ref{analogy} and Ref.~\cite{Ferrantini2011}): both are processive, a configuration being energetically favored over the other, and driven by a local force. In driven translocation, the driving force acts at the level of the pore and favors the presence of monomers in one compartment, whereas in the case of zipping, the driving force locally acts at the level of the ssDNA-dsDNA junction and favors the dsDNA state. 
Formally, strong similarities between partition functions {\em at equilibrium} can also be established~\cite{Walter2012}. There are however differences to keep in mind. In the case of DNA zipping, three polymers are involved at the junction, instead of two in translocation, on both sides of the wall. In addition, the zipped state is much more rigid than the single strands. Finally, when zipping progresses twist must be accumulated in the duplex and the ensuing rotational motion has to be considered properly. The analogy must be handled with care but it is certainly a good basis to tackle DNA zipping since translocation has been investigated in depth for  two  decades.

Findings on driven polymer translocation can be summarized as follows~\cite{Palyulin2014}: Saito and Sakaue~\cite{Saito2011,Saito2012} first proposed to classify dynamical regimes in function of the intensity of the driving force $f$. If we put aside very small forces (if $f < k_{\rm B}T/ (aN^\nu) $, the shape is the equilibrium one) and very large ones (if $f  > k_{\rm B}T \, N^\nu/a$, the polymer is stretched), there are two force regimes of interest: (a)~the intermediate force or ``trumpet'' regime for $k_{\rm B}T/ (aN^\nu) < f < k_{\rm B}T/a$; (b)~the strong force or stem-and-flower regime for $k_{\rm B}T/a < f < k_{\rm B}T \, N^\nu/a$. DNA zipping falls in this latter regime as discussed above, because $f\simeq 2.5 k_{\rm B}T/a$. This classification follows Brochard-Wyart's one in the context of tethered polymers in moderate or strong flow~\cite{BrochardWyart1993,BrochardWyart1995}. It can easily be extended to semi-flexible polymers by replacing the monomer length $a$ by the Kuhn length $\ell_{\rm K}$.

In both cases (a) and (b), following a scaling argument similar to the one developed in the previous section, Saito and Sakaue get $\alpha = 1+\nu $ in both regimes~\cite{Saito2012}, i.e. $\alpha  \simeq 1.59$ in 3D and 1.75 in 2D. At a given force $f$, whatever its value, owing to the bounds of the force intervals given above, the system asymptotically falls either in the trumpet regime or in the stem-and-flower one in the large $N$ limit, and one thus systematically recovers the exponent $\alpha=1+\nu$~[\footnote{Saito and Sakaue also studied the exponent, denoted by  $\beta$, characterizing the scaling of $\tau_{\rm d}$ with $f$: $\tau_{\rm d} \sim N^\alpha/f^\beta$. They found different values of $\beta$ in cases (a) and (b)~\cite{Saito2012}, but this issue is out of the scope of the present Report. See also Ref.~\cite{Palyulin2014}.}]. Concerning the variability of the exponents $\alpha$ found in the literature and inferred by numerical approaches (see Ref.~\cite{Palyulin2014} for an exhaustive review), by using a variant of Brownian dynamics simulations, Ikonen and coworkers arrived at the conclusion that it can be explained by finite-size effects~\cite{Ikonen2013}. 
Furthermore, as we have explained it above in the case of DNA zipping, hydrodynamic interactions are not expected to affect significantly the above scaling laws beyond finite-size effects~\cite{Izmitli2008,Ikonen2013}.
The correct scaling exponent likely equals $\alpha = 1+\nu $ in both the Rouse and Zimm (with logarithmic corrections) regimes, even though additional investigations will be necessary to ascertain it definitely.  

\subsection{Base-pairing dynamics close to the melting temperature}
\label{T:Tm}

When getting close to $T_m$, as in polymer translocation, it is expected to observe a crossover from driven renaturation with anomalous exponent $\alpha$, as reviewed above, to unbiased renaturation with a specific exponent $\alpha_m$ that we have not discussed so far. 

In Ref.~\cite{Walter2012}, in 2012, Walter and coworkers extended there previous investigations below (and above) the nucleic acid melting temperature~\cite{Ferrantini2011} by considering the dynamics at exactly $T=T_m$, again with the help of Rouse-Monte Carlo dynamics. In this case, there is no bias toward either the zipped or the unzipped state, since they have the same free energy in the large polymerization index $N$ limit. The dynamics is characterized by the time needed to reach either the zipped or the unzipped state. It is found numerically that the dynamics is anomalous, its characteristic time scaling as $\tau \propto N^{\alpha_m}$, with $\alpha_m=2.26\pm 0.02$. In this work, bubbles are again not allowed to form, because the complex is always supposed to adopt a Y-shape, as in the original zipper model. Note however that the Y-shape assumption is seriously questioned near $T_m$, since many denaturation bubbles are expected to proliferate in the duplex~\cite{Palmeri2008}.

The problem can be mapped on the 1D diffusion of the junction between single strands and the duplex. The fact than the exponent $\alpha_m$ is larger than 2 means that this process is sub-diffusive. Considering that pairing occurs through a succession of quasi-equilibrium states as in the zipper model would again lead to an exponent~2 as in the zipper model. For example, in Ref.~\cite{Calvo2008}, the authors assumed the validity of a pre-averaging procedure, amounting to consider base-pairing dynamics as a succession of quasi-equilibrium states. By solving an approximate adjoint Smoluchowski equation, they estimated the mean first passage time for the separation of the centers of mass of the two strands at $T \simeq T_m$ and indeed found $\alpha_m=2$ in the Rouse regime. This is again in contradiction with the fact that reaching quasi-equilibrium is slower than that. We recall that in the Rouse regime, quasi-equilibrium is reached in a time scaling like $N^{1+2\nu}>N^2$ [see \eq{Rouse2}]. Analogy with unbiased translocation would suggest that $\alpha_m = 1 + 2\nu$ or $2 + \nu \neq 2.26$ near $T_m$~\cite{Vocks2008}. At criticality, translocation and zipping thus possibly belong to two different dynamical universality classes, in case this notion of universality class were meaningful: in Ref.~\cite{Palyulin2014}, it is evoked that the high sensitivity of dynamical exponents in unbiased translocation in simulation details, notably the pore width, might rule out the relevance of universality.

\subsection{Thermal unzipping/unwinding above the melting temperature}
\label{TgtrTm}

We now address unzipping from the duplexed molecule above $T_m$, a problem already mentioned by Watson and Crick in 1953. They did not consider the nucleation of one or several denaturation bubbles above $T_m$, but assumed that the polymer unzips from one extremity. The duration $\tau_{\rm unzip}$ until complete strand separation is again expected to scales as $\tau_{\rm unzip} \propto N^{\alpha'}$, with a new unwinding exponent $\alpha'$. Melting of very short oligomers and hairpins, or energetic considerations at the base-pair scale~\cite{Wong2008} are not at stake here. At high temperature, far from $T_m$, this is expected to become an universal~\cite{Baiesi2010}, purely kinetically constrained problem, with no energy landscape that might for example arise from the sequence. Only entropic gain is  at play when the two strands unwind by rotating around each other, until they eventually get fully separated. At $T>T_m$ close to $T_m$, a free-energy landscape ensues from the intra-molecular heterogeneity arising from the difference in the stability of A-T and G-C base-pairs. We shall not consider the ensuing consequences here. We shall see that in any case the process is slowed because of the topological constraints. They imply that ``twist dissipation''  is allowed only at the two ends of the initial double helix.  

After an unsuccessful pioneering attempt by Kuhn in 1957 to estimate unwinding rates without taking a driving torque into account~\cite{Kuhn1957}, Longuet-Higgins and Zimm proposed in 1960 that $\alpha'=5/2$~\cite{Longuet1960} owing to the following argument\footnote{Note that at this time, the molecular machineries able to actively unwind DNA were unknown, and the authors were ``therefore led to consider the simplest possible scheme for the replication of DNA. In [their] view, (\ldots) the two strands of the old molecule separate spontaneously when the double helix becomes unstable as a result of some change in its environment''~\cite{Longuet1960} and this was considered to be the starting point of replication. This raised the interest of estimating unwinding rates.}: they assumed the rate-controlling process to be the unwinding of both ends. The equation governing the dynamics is simply the analogue of \eq{friction} for rotational motions (again neglecting Langevin random torque)
\be
\mathcal{T} -\zeta_r \omega=0
\ee
where the driving torque is $\mathcal{T}\simeq \Delta F_0/\phi_{\rm eq}$ ($\Delta F_0$ is the free energy gain when unwinding one bp, see Section~\ref{zipper:model}) and the friction torque is $-\zeta_r \omega$, where $\omega$ is the angular velocity of the rotating strands at the junction of the fork~\cite{Dasanna2012}.

\begin{figure}[ht]
\begin{center}
\includegraphics*[width=10.5cm]{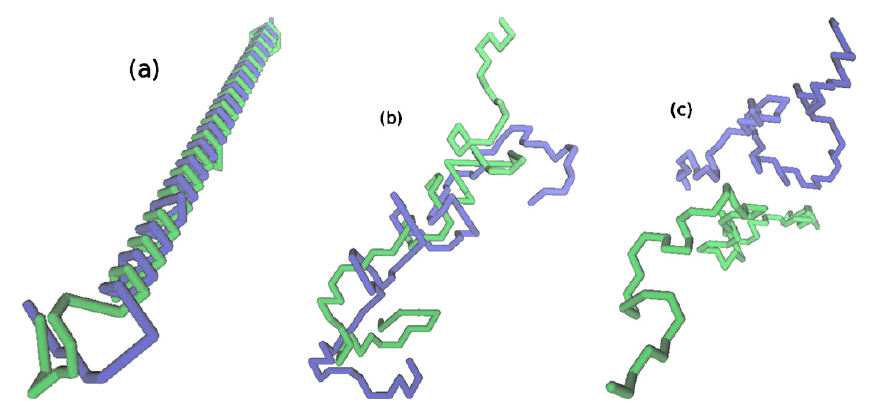}
\end{center}
\caption{Starting from two polymers tightly wound around each other that start unwinding from both ends (a), one ends with two fully separated single strands (c). An intermediate state is also shown in (b), where hydrogen-bonds are broken but the polymer are still entangled. Note that unwinding starts from both extremities ($N=100$ monomers). Taken from~\cite{Baiesi2010}.
\label{unwind:fig}}
\end{figure}

Longuet-Higgins and Zimm assumed that the  rotational friction coefficient $\zeta_r$ is essentially due to the hydrodynamic viscous drag on the ssDNA random coils. 
Assuming furthermore that unzipping starts from both molecule extremities, they understood that opposite and (on average) equal torques act at each end of the dsDNA helix, so that the helix itself does not move. Only the single strands that have already separated can rotate (see figure~\ref{unwind:fig}(a)) and dissipate twist. Hence $\zeta_r$ is the friction coefficient of each single strand of length $M$, that they assumed to be in a quasi-equilibrium random-coil configuration, or equivalently that the two strands in each free end are together equivalent to a random coil of length $2M$, in quasi-equilibrium, rotating about its midpoint. Taking hydrodynamic interactions into account, we have in the Zimm regime $\zeta_r \simeq a^2\zeta_0 M^{3\nu}$ (see \eq{Rouse2}). Finally, given that in absence of supercoiling, the Fuller-White theorem, \eq{Fuller}, leads to ${\rm d}M/{\rm d}t=\omega/\phi_{\rm eq}$ (with $\phi_{\rm eq}=2\pi/p$) and by integrating over the unzipping duration:
\begin{equation}
\tau_{\rm unzip} = \int_0^{\tau_{\rm unzip}} dt \simeq \frac{a^2\zeta_0}{\Delta F_0}\int_0^{N/2} M^{3\nu}dM \simeq \tau_f N^{1+3\nu}
\label{unzipLHZ}
\end{equation}
and thus $\alpha'=1+3\nu = 5/2$ in a $\Theta$-solvent ($\nu=1/2$) and $\alpha'\simeq 2.77$ for a self-avoiding chain ($\nu \simeq 0.59$).
 
In 1986, Baumg\"artner and Muthukumar performed the first numerical simulations of the unwinding of two initially intertwined chains on a cubic lattice~\cite{Baumgartner1986}. Their Monte Carlo algorithm allowed them to find an exponent $\alpha'=3.3 \pm 0.2$ for the disentangling process, but only small system could be addressed at this time (few helix turns, $N \leq 65$ monomers). The asymptotic regime was certainly not reached. 
 
Much more recently, Baiesi and collaborators have studied a similar simplified numerical model of double helix, on a face-centered cubic lattice, in the spirit of the model developed by this group in the zipping context discussed above. In particular, the double-helix pitch and the ssDNA  and dsDNA flexibilities are not required to meet the real ones. In their first work, in 2010~\cite{Baiesi2010}, they considered the unwinding of two lattice polymer strands initially tightly wound around each other in a double-helical conformation (no attractive energy between strands, formally $T_m=0$). Rouse dynamics were implemented and the system size ranged up to $N=10^3$ monomers. 
The simulations show that unwinding dynamics occurs in two stages, which was not anticipated by the analytical studies in the 1960s. A rather rapid stage, where hydrogen-bonds break and the entanglement loosens (Figure~\ref{unwind:fig}b), is followed by a slower complete disentanglement (Figure~\ref{unwind:fig}c). Firstly, these simulations showed that unwinding proceeds from the two ends of the initial double-stranded complex, and progresses inwards when time increases. However, this might me related to the model discreteness at the monomer level (see Section~\ref{torque:zip:unzip}). Moreover, the unwinding exponent was estimated to be $\alpha'=2.57 \pm0.03$. Using the same simple scaling argument as Longuet-Higgins and Zimm's one given above at quasi-equilibrium, but in the Rouse regime where $\zeta_r \propto M^{1+2\nu}$ (see \eq{Zimm}), they got $\alpha'=2+2\nu \simeq 3.18$, a larger exponent than the measured one. They concluded that the unwinding process that they observed was probably also a far-from-equilibrium process. It probably means that the number of monomers involved in the twist dissipation is less than the ones belonging to the full single strands, $M$ (as in the zipping case, Section~\ref{far:from:eq:zipping}).

Another promising and somewhat simpler numerical approach has recently been proposed to decipher unwinding dynamics, namely a single polymer strand attached to an end to an infinite straight rod and tightly wrapped around it at $t=0$. Such a systems has been considered  for the first time in Ref.~\cite{Walter2011} by Walter, Barkema and Carlon, and considered as a ``good surrogate of a double helical structure", because it has an unambiguously defined reaction coordinate, the winding angle $\theta(t)$ of the last monomer at the polymer free end: $\theta(t)$ measures the angle accumulated around the rod from the first (attached) monomer to the last (free) one. The unwinding dynamics of this model (without hydrodynamics) has been studied in two consecutive works~\cite{Walter2013,Walter2014}. It presumably displays logarithmic corrections to a simple power law: its characteristic unwinding time is demonstrated to scale like $N^{1+2\nu} (\ln N)^{2\gamma}$ where $\gamma \approx 0.75$ is another exponent. As proposed by the authors, the exponent $\alpha' \simeq 2.57$ discussed above in the case of double-stranded complex unwinding might be simply due to the logarithmic corrections to the exponent $1+2\nu \simeq 2.18$. 

\medskip

In conclusion, effective 1D models of Sections~\ref{bpb} and~\ref{1Damod} are again in difficulty to tackle this dynamical system because they ignore both out-of-equilibrium issues and chain entanglement~\cite{vanErp2012}, even if specifically adapted to take into account that twist can only be dissipated at the molecule extremities~\cite{Baiesi2009}. Double helix topology can be not fundamental when one focuses on the equilibrium properties of DNA only, but the fact that strands cannot cross each other and rotate around each other is a pivotal ingredient of their unwinding dynamics. In this respect, the reader can also refer to Ref.~\cite{Baiesi2013} for a relatively recent short review on theoretical approaches to DNA denaturation dynamics. In this work, Baiesi and Carlon also conclude that ``models of three-dimensional polymers are the most suitable for describing full denaturation of very long DNA" above $T_m$. They also stress that in order to measure disentangling times experimentally, an indicator of polymers separation will be needed beyond the traditional measure of hydrogen-bond breaking between both strands such as UV absorbance. Indeed we have seen that they presumably break on timescales much faster than the disentangling ones [Figures~\ref{unwind:fig}(b) and~(c)].

\subsection{Bubble final closure and metastable bubble opening}
\label{wooed}

Unzipping and zipping of a long dsDNA are far-from-equilibrium processes. However there are situations where the opening/closure kinetics of a segment of few base-pairs can be studied using the tools of the statistical equilibrium mechanics in the framework of the quasi-static approximation. Indeed, when the system is characterized by two states, usually one corresponding to the equilibrium state and the other to a metastable state, the dynamics can sometimes be studied as a diffusion process in a low dimensional equilibrium free energy landscape. This approach has been fruitful when studying the folding/unfolding of small DNA (or RNA) hairpins within the two-state approximation. In this Section, we present another situation where it is useful, the final closure of  denaturation bubble in long or clamped DNA where the far-from equilibrium zipping has stopped in a metastable bubble state of $\simeq 10$~bp.

\begin{figure}[ht]
\centerline{\includegraphics[width=7cm]{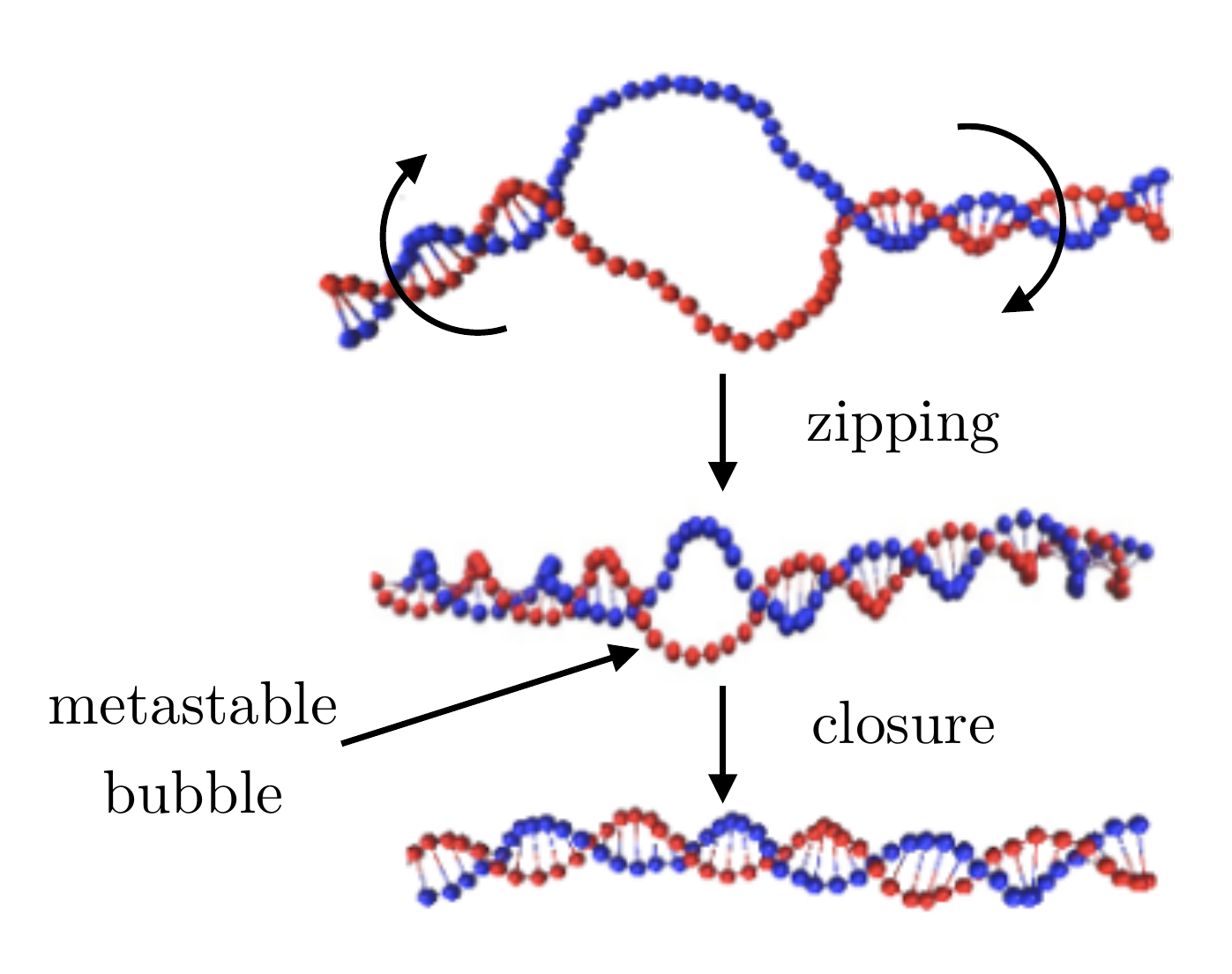}\hspace{2cm}}
\caption{When zipping starts from the two extremities of a large bubble (top), the two dsDNA arms rotate in the opposite direction to increase the twist of the chain and thus to close base-pairs (following the law $\Delta\mathrm{Tw}=\Delta\mathrm{Lk}$ in the absence of supercoiling). One eventually ends with a metastable denaturation bubble of about 10~bp (middle). A free-energy barrier related to twist must then be overcome to close this bubble (bottom).
\label{bubble:fig}}
\end{figure}

In Section~\ref{far:from:eq:zipping}, we have considered renaturation of DNA molecules starting at any place along the molecule and ending at one of its extremities, assuming that nucleation is a rare event and occurs only once per molecule at room temperature. Once the extremity has been reached, closure is achieved instantaneously. However, it is possible that zipping starts at two distant sites, in particular if one starts with a only partially denaturated duplex, in which case closure ends at a site situated in the middle of the chain (Figure~\ref{bubble:fig}). Double helix opening for transcription or duplication is also responsible for bubbles localized in the middle of the molecule.

This problem of the closure of a large (say $>7$~bp) bubble at room temperature ($T<T_m$) has recently been tackled through Brownian simulations and (biased) metadynamics~\cite{Tiwary2013} on the helical mesoscopic model evoked in section~\ref{far:from:eq:zipping}, together with scaling arguments~\cite{Sicard2015,Dasanna2013}. The primary goal was to account for Altan-Bonnet, Libchaber, and Krichevsky's (ALK) experiments~\cite{Altan2003}, already presented in Section~\ref{hairpins}.

It has been demonstrated that once the far-from equilibrium zipping is almost achieved, as the bubble size during zipping reaches the order of the persistence length of ssDNA, the bending and torsion forces start competing with the force driving closure~\cite{Dasanna2013,Dasanna2012}, due to the connectivity (or geometrical) constraint on both bubble sides. When the bubble size becomes $\bar n\approx10$~bps, the bubble stops closing and the resulting metastable bubble survives for long before ultimate closure occurs. A simple estimation of $\bar n$ is obtained by balancing the energy gain of closing one base-pair $\Delta F_0$ with the variation of the bubble bending energy $E_{\rm bend}^{\rm ss loop}(\bar n)-E_{\rm bend}^{\rm ss loop}(\bar n-1)$. The bending energy of the loop is $E_{\rm bend}^{\rm ss loop}(\bar n)=\bar n \kappa_{ss}/R^2 = \pi^2\kappa_{ss} /\bar n$ (where $R= \bar n/\pi$ is the loop radius in units of $a$ when approximated by a circle) which yields $\bar n\simeq \pi \sqrt{\kappa_{ss}/\Delta F_0}$, i.e. a few base-pairs.
Note that the facts that the bending and torsional moduli are very different for dsDNA and ssDNA and that the base-pairing is coupled to them~\cite{Palmeri2008,Manghi2009,Palmeri2007} (see the Introduction) is a key ingredient in this picture.

\begin{figure}[t]
\begin{center}
\includegraphics*[width=6cm]{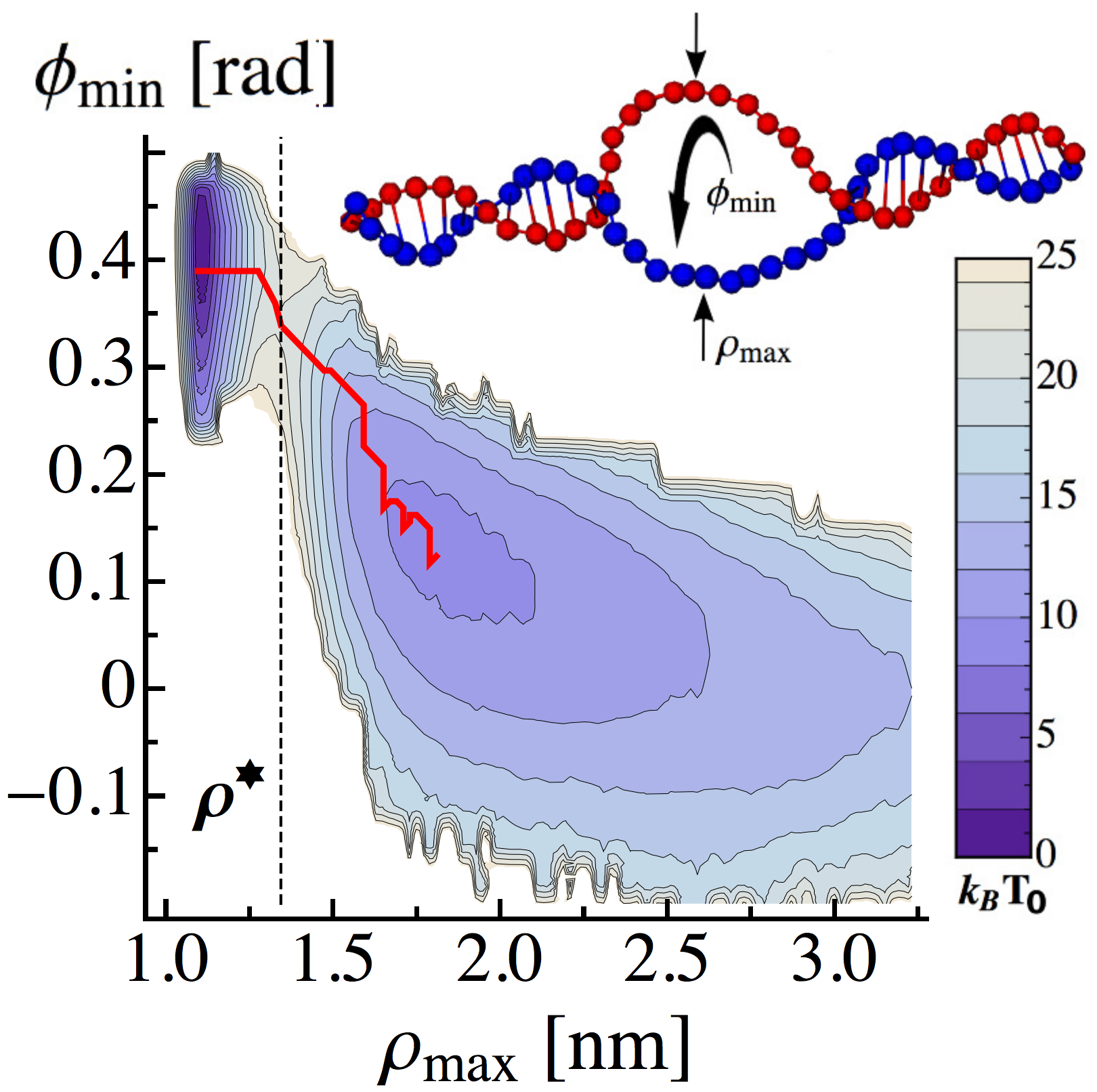}\hspace{1cm}
\includegraphics*[width=7cm]{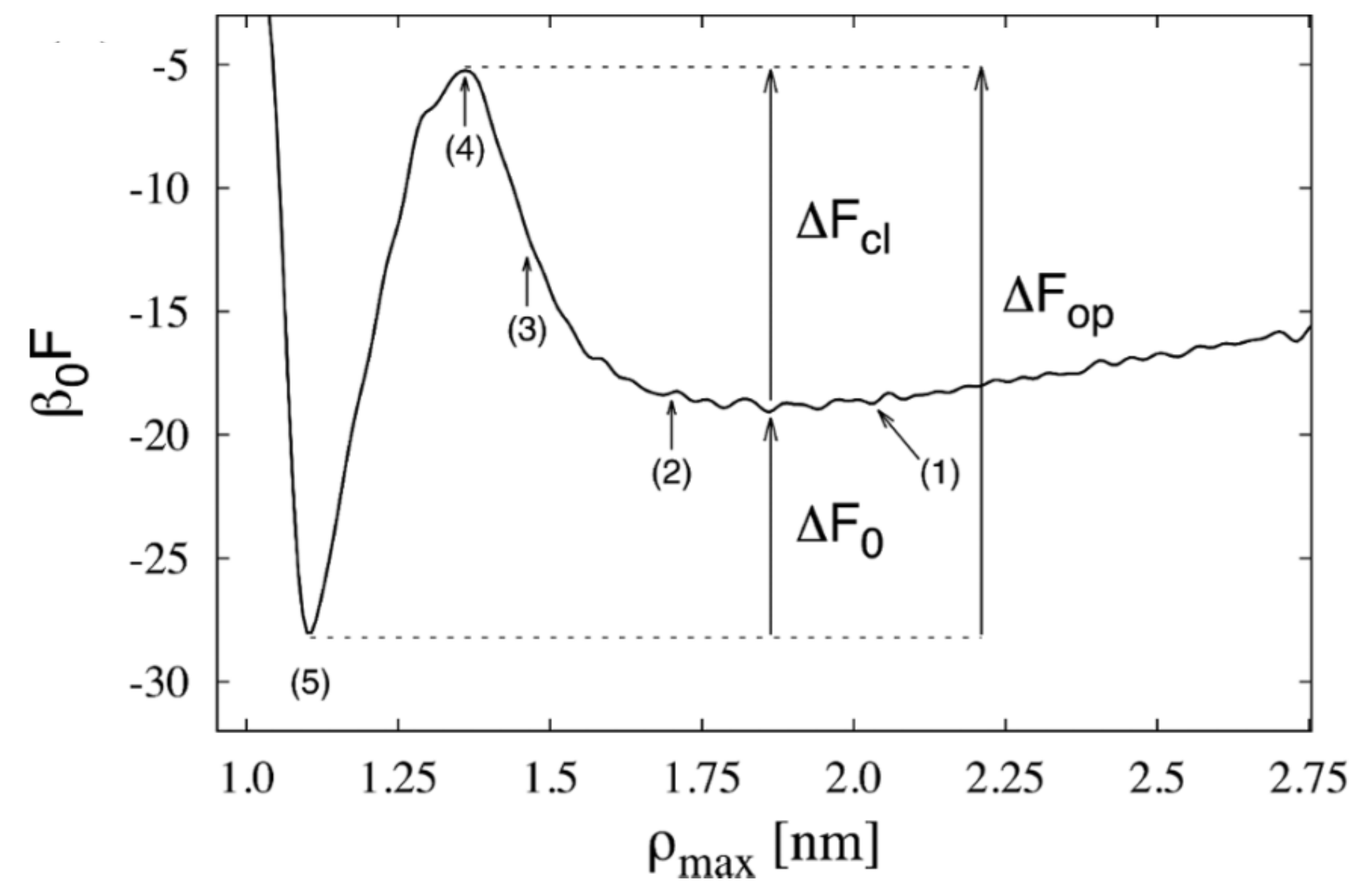}\\
\ \hspace{-.6cm}{\bf (a)}\hspace{8cm} {\bf (b)} \
\end{center}
\caption{(a)~Free-energy surface associated with the bubble closure/nucleation mechanism projected along two observables: the maximal distance between paired bases $\rho_{\max}$ and the minimal twist angle between successive base-pairs, $\phi_{\min}$ (see inset). The saddle point is located at $\rho^*= 1.35$~nm. The typical minimal free-energy path is shown in red color, and the contour lines are every $2 k_{\rm B}T_0$, $T_0$ being the room temperature. (b)~Free-energy profile associated with the opening/closure mechanism corresponding to the projection of the free-energy landscape on the $x$-axis. The closure and opening free-energy barriers are denoted by $\Delta F_{\rm cl}$ and $\Delta F_{\rm op}$. The difference free-energy between the closed and open bubble states is $\Delta F_0$ and therefore corresponds to $\simeq \bar n\Delta F_0$ with the previous definition of $\Delta F_0$. Taken from~\cite{Sicard2015}.
\label{Sicard:fig}}
\end{figure}
Once in the metastable regime, the simulations show that the bubble closure time $\tau_{\rm cl}$ grows exponentially with the torsional modulus. Closure of the metastable bubble is therefore a collective mechanism where the bubble twists as long as base-pairs close~\cite{Sicard2015}. Indeed, the closure time is related to the energy barrier $\Delta F_{\rm cl}$ \textit{along the transition path} following the Kramers theory given in \eq{Kramers}  (see Figure~\ref{Sicard:fig}). This energy barrier, $\Delta F_{\rm cl}$, ensues essentially from the penalty associated with the torsional energy. Closure is a thermally activated process, as expected from experiments. 

To understand the origin of the energy barrier, one can use the following simplified picture: the two dsDNA arms, free to rotate around their own axis, are not free to rotate relatively to one another. The small bubble forms a ``single joint'', the base-pairs of which twist collectively, with a torsional modulus that depend on the maximal  interstrand distance in the bubble $\rho_{\max}$, $\kappa_\phi(\rho_{\max})$. Numerically it has been shown that $\kappa_\phi\sim\rho_{\max}^{-\delta}$ in the range of interest (where $\delta\simeq 2.2$) and interpolates between $k_{\rm B}T$ for the metastable bubble and $\simeq 500 k_{\rm B}T$ for the dsDNA case. Hence the bubble torsional energy changes from a large potential well $\kappa_\phi(\rho)\Phi^2/2\simeq k_{\rm B}T\Phi^2/2$ in the metastable state, to the much sharper potential well $\kappa_\phi(\rho)(\Phi-\Phi_{\rm eq})^2/2\simeq \kappa_\phi^{\rm ds}(\Phi-\Phi_{\rm eq})^2/2$ in the ds state where $\Phi$ is the twist angle measured consecutively between the base-pairs defining each extremity of the bubble of size $\bar n$ ($\Phi_{\rm eq}=\bar n \phi_{\rm eq}$ is the dsDNA value). These two potential wells are clearly visible in Figure~\ref{Sicard:fig}a showing the free-energy landscape in the $(\rho_{\max},\phi_{\min})$ plane, reconstructed using metadynamics simulations. In particular, they are separated by a barrier of around $\Delta F_{\rm cl}=13~k_{\rm B}T$ at the saddle point.

The closure times measured in these metadynamics simulations~\cite{Sicard2015} are on the order of 50~$\mu s$, comparable to the experimental result of Altan-Bonnet \textit{et al.}~\cite{Altan2003}.
The metadynamics allows one to reconstruct the whole free energy landscape and therefore to deduce the bubble \textit{opening }time as well, which follows the same law as \eq{Kramers} where $\omega_{\rm met}$ and $\Delta F_{\rm cl}$ should be replaced by $\omega_{\rm ds}$ and  $\Delta F_{\rm op}$. It is defined as the time for escaping of the closed state basin of Figure~\ref{Sicard:fig}b and corresponds to the nucleation of a bubble of 3 to 4 bps with well separated strands. This opening time is $\tau_{\rm op} \simeq 15$~ms. Accordingly, opening times measured in NMR experiments in a similar context~\cite{Gueron1987} are also on the 10~ms timescale.

In this picture, the breather modes of Section~\ref{bpb} correspond to diffusion in the closed state basin. Indeed the spring constant in the closed state basin is given by $V_{\rm Morse}''=2 U_0\alpha_1^2=1.6$~N/m where $\beta U_0\simeq 8$ is the depth and $\alpha_1^{-1}=0.2$~nm the width of the Morse potential at its global minimum in the closed state. It exactly corresponds to the spring constant values measured from the sound velocity as explained in Section~\ref{bpb} and leads to the $\simeq 10$~ps timescale as given by \eq{diffusion_potential}.

Note that Molecular Dynamics simulations by Zeida \textit{et al.} ~\cite{Zeida2012} were performed on the ALK sequence. In their 250~$\mu$s long simulation they observed one large bubble event of lifetime of 3~$\mu$s which is still smaller by one order of magnitude than the observed one. The PBD model was also applied to fit the experimental temporal autocorrelation function~\cite{Srivastava2009}. The friction coefficient value had to be adjusted to match the experimental time scale, leading to the unphysical value of $\zeta\simeq10^{-4}$~kg/s (the same friction coefficient as a bead of radius 5~cm in water). Indeed, as recognized later by Peyrard himself and co-workers~\cite{Peyrard2009}, physical timescales are attainable only when adding and \textit{ad hoc} energy barrier of $6k_{\rm B}T$ to the base-pairing potential (see Section~\ref{bpb}). 
This is another instance where it is difficult to correctly tackle base-pairing dynamics by considering an effective 1D model only. 


\section{Prospects and open questions}
\label{poq}

\subsection{Summary}

We have surveyed how physics is able to throw light on base-pairing dynamics of the DNA macromolecule, in situations of experimental interest where the polymer length varies from few to thousands of base-pairs. This includes zipping/unzipping of long molecules and closure/opening of short hairpins or oligomers. These structural modifications can be induced either by a temperature (or $p$H) change or by an external force or torque. Thermally activated fluctuations have also been explored, and we have been led to propose a distinction between (i) small  (a few base-pairs) and very short-lived ($< 1$~ns) breathers where the single strands remain essentially frozen while base-pairs are weakly distorted; and (ii) large ($> 7$~bp) and long-lived ($> 1$~$\mu$s), metastable openings of the double helix. 

This distinction led us to categorize the reviewed works in three major groups, depending on the relative roles played by internal (base-pairing) and external (chain conformation) degrees of freedom. The models were systematically compared to \textit{in vitro} biophysical experimental results when available. In Section~\ref{bpb}, we have first focused on the mesoscopic models describing breathing bubbles dynamics, principally the celebrated Yakushevich and Peyrard-Bishop-Dauxois models. Even though they do not focus on the same base-pair degrees of freedom (base rotation in the first case and hydrogen-bonds elongation in the second one), the underlying physics are very close and appeal to the same concepts of non linear-physics such as solitons or breathers. The models tackled in the following Section~\ref{1Damod} remain essentially 1D in nature, as the previous ones. However instead of considering chain degrees of freedom as frozen variables, they are pre-averaged as fast variables as compared to the supposedly slower base-pairing variables. This concerns the original Ising-like zipper model developed in the early 1960s, as well as its more recent developments such as the kinetic one proposed by Metzler \textit{et al.} based on the Poland-Scheraga model. The two-state approximation used in the description of small hairpin folding/unfolding also belongs to this category. However, in circumstances such as fast zipping/unzipping of long molecules provoked by a temperature jump or an applied force, this quasi-static approximation consisting in pre-averaging all chain degrees of freedom is no more valid and tools from polymer dynamics have to be used. In particular hydrodynamic friction and hydrodynamic interactions are then at play, the role of the latter requiring further investigations in the future. Reviewing these models was the object of Section~\ref{ffeq}. 

From a biological perspective, the different mechanisms explored in this Report cannot systematically be considered as definitive answers to questions arising from the exploration of molecular processes in real cells. However, base-pairing dynamics in general are at the core of keystone mechanisms of Life, such as transcription, replication, recombination or DNA repair, generically assisted by active enzymes. \textit{In vivo}, active molecular machineries exert constraints on the biopolymer, which motivates the studies on force- and torque-induced denaturation reviewed in this Report. Supercoiling is also implicated in several mechanism, concerning e.g. naturally supercoiled plasmids or DNA segments attached to the protein network in the chromatin. The next step would be to consider base-pairing dynamics at this larger and much more complex scale of chromatin. From a physical point of view, this long-time project has been undertaken but chromatin dynamics is a topic \textit{per se} where additional mechanisms play pivotal roles \cite{Koch2002,Smith2006,Kouzarides2007,Cairns2007,Hajkova2008,Hall2009,Lavelle2010,Lugert2012}. It has not reached the same level of quantitative understanding as the one at the molecular scale presented in the present work and might require future paradigmatic shifts to be deciphered. 

In spite of the undeniable numerous successes of the physical approaches presented in this Report, we have noted at several places that important open questions remain. We now recall them in a synthetic way in the following sections, and we propose strategies to tackle them in a near future. 

\subsection{On the biological relevance of breathers}
\label{Breather:vs:bubble}

It has been argued~\cite{Alexandrov2009,Alexandrov2009c} that there are situations of experimental interest where chain degrees of freedom do not have time to equilibrate and where the mechanism of the models which describes the small bubbles as breathers reported in Section~\ref{bpb} are at play.

In the context of the ``pre-melting'' transition, the localization of these DNA breathers has been studied by inserting different parameter values to mimic sequence effects observed in promoters, possibly related to biological function. These studies where done using the PBD model without~\cite{Kalosakas2006,Alexandrov2009,Ares2005} or with a very small viscous damping coefficient ($\gamma=0.005$ or 0.05~ps$^{-1}$)~\cite{Alexandrov2009b,Alexandrov2006,Alexandrov2009c,Kalosakas2004,Blagoev2006}. Numerical studies were done on promoter sequences by suggesting that they exhibit a propensity for spontaneous thermal strand fluctuation. In particular, they suggest that ``functionally relevant structural variation in genomic DNA occurs at the level of fast motions not readily observed by traditional molecular structure analysis''~\cite{Alexandrov2009c}.
These models have also been applied to study the effect of a terahertz radiation on DNA which can induces localized breathers (of a few ps with an amplitude smaller than 0.1~nm) which can evolve during 100~ps for a matched frequency~\cite{Alexandrov2010}.
Moreover, Peyrard and coworkers showed, by simulating the PBD model using a faster method, that the large bubbles appear only seldom so that their location cannot be directly correlated to DNA transcription start sites~\cite{vanErp2005,vanErp2006b}. 

Another important biological issue is whether these small breathing bubbles are accessible to small reactants and relevant to DNA-protein interaction mechanism. Von Hippel and coworkers showed that, at physiological temperature, these breathing bubbles are only accessible to hydrogen exchange probes~\cite{vonHippel2013} as exploited in NMR experiments~\cite{Gueron1995,Warmlander2000}. However, for small reactants such as formaldehyde (HCHO) which attacks the AT base-pair, the access to DNA is undetectable at temperatures well below $T_m$ ($\simeq53^\circ$C for low formaldehyde concentrations and DNA with 50\% of AT)  and is only possible for large HCHO concentrations and temperatures close to $T_m$. Concerning larger proteins such as the gene 32 protein (gp32) which has a high affinity with ssDNA (and binds to a site of seven nucleotide residues), it has been experimentally shown that even at high concentrations and at temperatures more than a few degrees below $T_m$ their binding was completely kinetically blocked.

Therefore these breathing fluctuations cannot provide a pathway for large enzymes to access the DNA interior, and Nature consequently had to develop active proteins such as RNA polymerases and DNA helicases~\cite{Alberts2002} to melt duplexes and give molecular machineries a pathway to the single strands.

\subsection{Free-energy barriers slowing down the closure at the base-pair level}
\label{F:barrier}

In Section~\ref{zipper:model}, we have estimated both the closure time $\tau_{\rm zip}$ of a single base-pair predicted by the zipper model at $T<T_m$, notably at room temperature, and we found $\tau_{\rm zip}\sim 1$~ns. In the same way, the characteristic diffusion time at $T=T_m$ was estimated to be $\tau_0 \sim 0.1$~nm$^2$/ns. It corresponds to the time needed for a domain wall between a double-stranded DNA region and a denaturation bubble to diffuse one base-pair away. We have also discussed the values of the same quantities extracted from either experiments or atomistic numerical simulations. In Section~\ref{exp:motiv} we have given several such values of  $\tau_{\rm zip}$ in the range $0.1$ to 5~$\mu$s, and Crothers proposed values of $\tau_0$ about $10^3$ times longer than our previous estimate~\cite{Crothers1964}. These values where estimated in the context of small hairpins (or oligomers), where the slowing down due to hydrodynamic friction on the single strands is supposedly small. In the worst case, the single strands are $\sim 10$~bp long, and the friction coefficient $\zeta_0$ of a single base-pair used in our estimates should be replaced by $\sim 10\zeta_0$. This can explain a difference of one order of magnitude but not the observed three orders of magnitude. Two orders of magnitude are still lacking. Is it possible to reconcile the zipper model and experimental values? 

A beginning of answer might come from the all-atom simulations of Qi and collaborators~\cite{Qi2011}, which computed free-energy barriers that might slow down the closure of each base-pair. They are associated with the water molecules and their hydrogen-bonding capability with base-pairs as explained in Section~\ref{all_atom} (see Fig.~\ref{ThreeBeads}a). The potential barrier to be overcome to close the base-pair is then measured in the simulations to be $\Delta F_{\rm cl} \approx 3 k_{\rm B}T$ at room temperature. Another source of a free-energy barrier is associated with the counter-ions dynamics because they interact with the negatively charged phosphate groups. On the experimental side, Chen and collaborators measured a barrier of height between 1 and $2.5 k_{\rm B}T$ by analyzing FCS experiments on hairpins~\cite{Chen2008}. They also attributed it to the adjustment of the solvation environment and charge distribution of the approaching bases.

\begin{figure}[ht]
\centerline{\includegraphics[height=4.5cm]{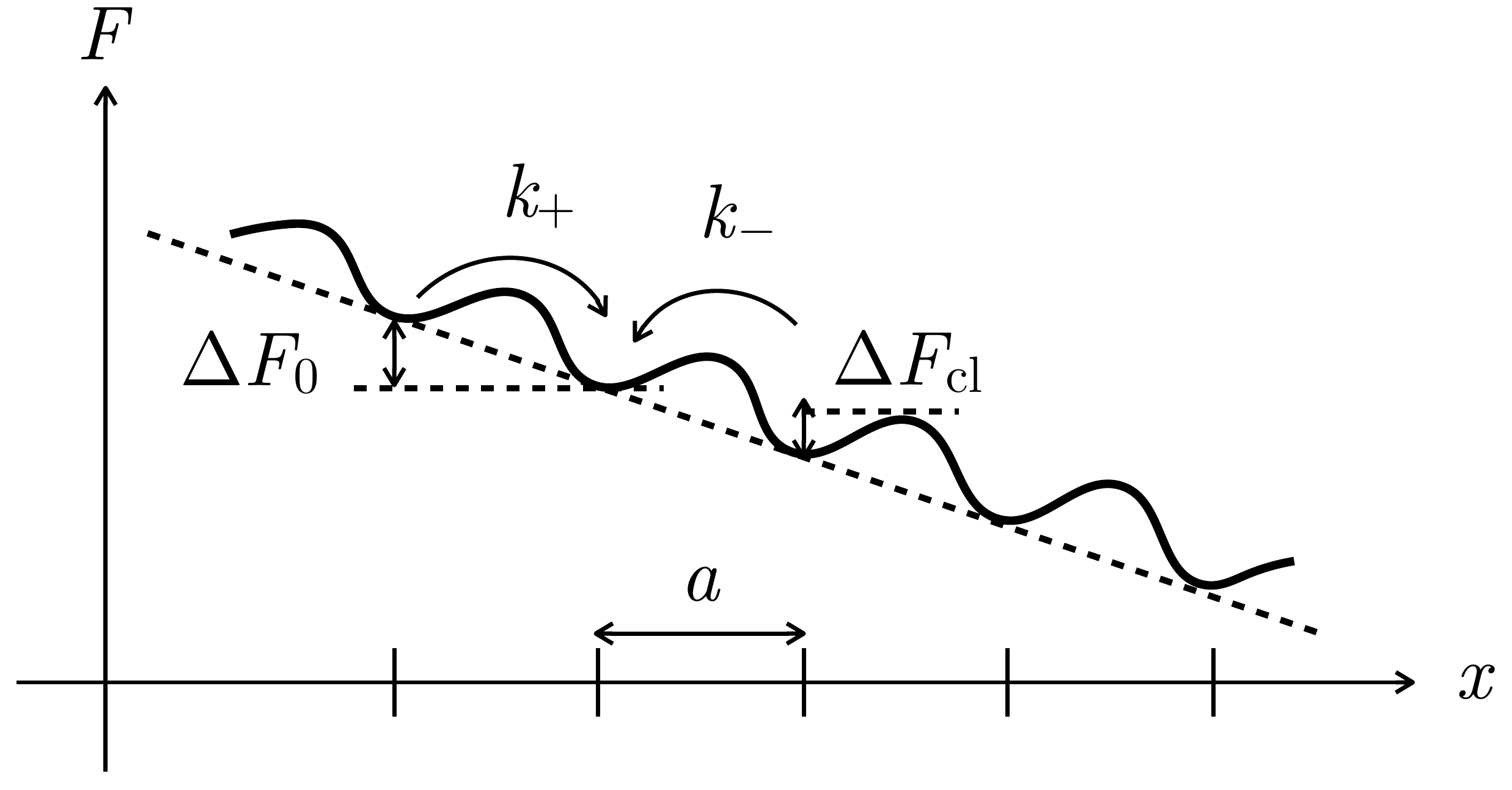}}
\caption{Schematic free-energy landscape of the zipper model at $T<T_m$ (and zero external force), in the hypothesis where the chain degrees of freedom can be integrated out (quasi-static approximation)  and where an
energy barrier $\Delta F_{\rm cl}$ of few $k_{\rm B}T$ slows down successive base-pairs closure. The closure and opening rates are still denoted by $k_+$ and $k_-$ respectively but their values are increased by a factor $\approx e^{\beta \Delta F_{\rm cl}}$. Note that increasing values of $x$ corresponds to DNA closure. \label{zipper:with:barrier}}
\end{figure}
These nontrivial barriers likely play a role in slowing-down the hybridization dynamics, as schematized in Figure~\ref{zipper:with:barrier}b. Indeed, the simple relation $v_{\rm zip}=f/\zeta_0= \Delta F_0/(a\zeta_0)$ becomes inappropriate because $f$ is no more constant on the $x$ interval of length $a$. Even though they still respect the detailed balance in \eq{balance}, $k_+$ and $k_-$ are now greater by a factor $\approx e^{\beta \Delta F_{\rm cl}} \simeq 20$. This increases the zipping time $\tau_{\rm zip}$ and the domain wall diffusion time $\tau_0$ by roughly the same amount, owing to the relations given in Section~\ref{zipper:model}, $v_{\rm zip}=a(k_+-k_-)$ and $D=a^2(k_++k_-)/2$. This rough argument will need to be improved, notably because the relevant free-energy landscape dimension might be larger than 1~\cite{Qi2011}, but it is a promising idea to investigate further. If its existence is confirmed this barrier ought to be included in future mesoscopic models.

\subsection{Importance of the friction torque on far-from-equilibrium zipping and unzipping}
\label{torque:zip:unzip}

Two types of model have been reviewed in Section~\ref{ffeq} for the zipping and unzipping of a long DNA: the ladder model in which the helicoidal structure of the DNA is neglected and led to the analogy with DNA translocation, and the more realistic double-helical model (two chains are wrapped around each other) where any base-pair closure or opening is associated with a rotational motion of one part of the DNA following \eq{Fuller}.

For the zipping of a DNA assumed to occur in a Y-shape configuration, both the ladder model, using the stem-and-flower argument, and the helical model~\cite{Dasanna2013T} (note that in the second case the exponent was calculated for $N\leq 100$) lead to the same scaling law $\tau_{\rm zip}\propto N^{1+\nu}$. In a different geometry where the zipping occurs from both ends (see the first step in Figure~\ref{bubble:fig}) the scaling exponent is also close to $\alpha =1+\nu$. It suggests that when the short dsDNA part rotates in the helical model, the associated hydrodynamic torque, which is absent in the ladder model, does not contribute too much to the zipping time and the drag is dominated by the translation of the ss strands. What happens if the bubble is flanked by two very long dsDNA arms or at the late stage of the renaturation in the Y-shape configuration? If the dsDNA is assumed to be rigid (which does not hold for $N>\ell_p^{\rm ds}/a\simeq150$~bp), one can estimate the dissipated power $P$ in the two following extreme cases: i)~if the ds strand of length $N-n$ rotates at an angular velocity $\omega$ and the ss strands are assumed to stay fixed, $P=\mathcal{T}_{\rm fric}\omega\simeq \eta a^3(N-n) \omega^2$; ii) in the reverse case, $P=f_{\rm fric}v\simeq \eta an v^2\simeq \eta a^3 n^2 \omega^2$ (with $v\simeq \omega an^{1/2}$ the velocity of the single strands center of mass in the quasi-static approximation). Hence, if we assume that the zipping is kinetically controlled, the first situation occurs while $n(t)\geq N^{1/2}$, that is most of the renaturation time. Only at the very end of zipping do the single-strand extremities rotate while the wound dsDNA part stops rotating. If one goes beyond the quasi-static approximation, the situation is even more complex because of ``thinning'' of the rotating single-strands, as studied very recently  in Ref.~\cite{Laleman2016} in a related context. Clearly, more simulations are needed, using for instance the model of Refs.~\cite{Sicard2015,Dasanna2013}, for longer DNAs with long ss strands at the initial configuration.

In the case of unwinding, this issue has been raised for a long time. In the 1960s, several authors have questioned a pivotal hypothesis of the Longuet-Higgins and Zimm approach leading to $\tau_{\rm unzip}\propto N^{1+3\nu}$, namely the fact that unwinding starts from both ends~\cite{Crothers1964,Fixman1963}. It was argued that if unwinding starts from the middle, it is the wound double-helix parts that rotate in the solvent~\cite{Levinthal1956,Fong1964}. The viscous drag on the unwound parts becomes quickly large enough such that the wound arms actually rotate around their principal axis. This might modify \eq{unzipLHZ} by replacing $3\nu$ by 1 yielding $\tau_{\rm unzip}\propto N^2$. This is indeed what has been proposed by Fixman in 1963~\cite{Fixman1963} and later by Cocco \textit{et al.} in the context of quasi-static rezipping as discussed at the end of Section~\ref{force_unzip}. The scaling found by Walter and coworkers $\tau_{\rm unzip}\propto N^{1+ 2\nu}(\ln N)^{1.5}$ relies on the assumptions that the unwinding starts form the end, the dsDNA part cannot rotate, and that the driven unwinding torque is not constant but of entropic nature only (formally $T_m=0$) equal to $k_{\rm B}T \theta$ where $\theta$ is the winding angle~\cite{Walter2013}.

All these arguments ignored the fact that when the molecule length exceeds the dsDNA persistence length (typically $N \geq 10 \ell_{\rm K}\simeq 3000$~bp), or when it becomes partially denaturated, the rigid rod image is not valid and twist strain can be relieved by the dsDNA writhe (see \eq{Fuller})~\cite{Moroz1997,Kamien1998,Powers2010} with the possible formation of supercoiled structures~\cite{Marko1994}. This whirling instability has been studied theoretically~\cite{Wolgemuth2000,Wada2006}. The equilibrium between the twist strain and the unwinding (or winding) driving torque  $\mathcal{T}$ at the fork junction yields $C{\rm d}\phi/{\rm d}s = \mathcal{T}/a$ and the straight DNA becomes instable when $\mathcal{T}/a$ is equal to the characteristic bending moment of the rod of length $L$, $\kappa_{\rm ds}/L$. Hence the critical dsDNA length for which the whirling instability occurs is given by $L_{\rm crit}=8.9 a\kappa_{\rm ds}/ \mathcal{T}$. One estimates $\mathcal{T}\simeq\Delta F_0/\phi_{\rm eq}\simeq 2 k_{\rm B}T$ (for $T=T_m+10$~K) which yields $N_{\rm crit}=L_{\rm crit}/a\simeq 660$~bp. Hence for dsDNA wound parts larger than $\sim 1$~kbp, the DNA writhes and the drag increases dramatically. The process is finally slower than the one proposed by Longuet-Higgins and Zimm. Since the current simulations on this system are done for $N\leq 1$~kbp, more theoretical work is needed to fully understand this complicated issue.

\subsection{Some propositions of experiments}

To conclude this section devoted to prospects, it is worth returning to experiments because they will be the ultimate referee of the theoretical developments reviewed in this Report. We mention some propositions of experiments related to the questions raised in this Report. 

\begin{figure}[ht]
\begin{center}
\includegraphics[height=4.5cm]{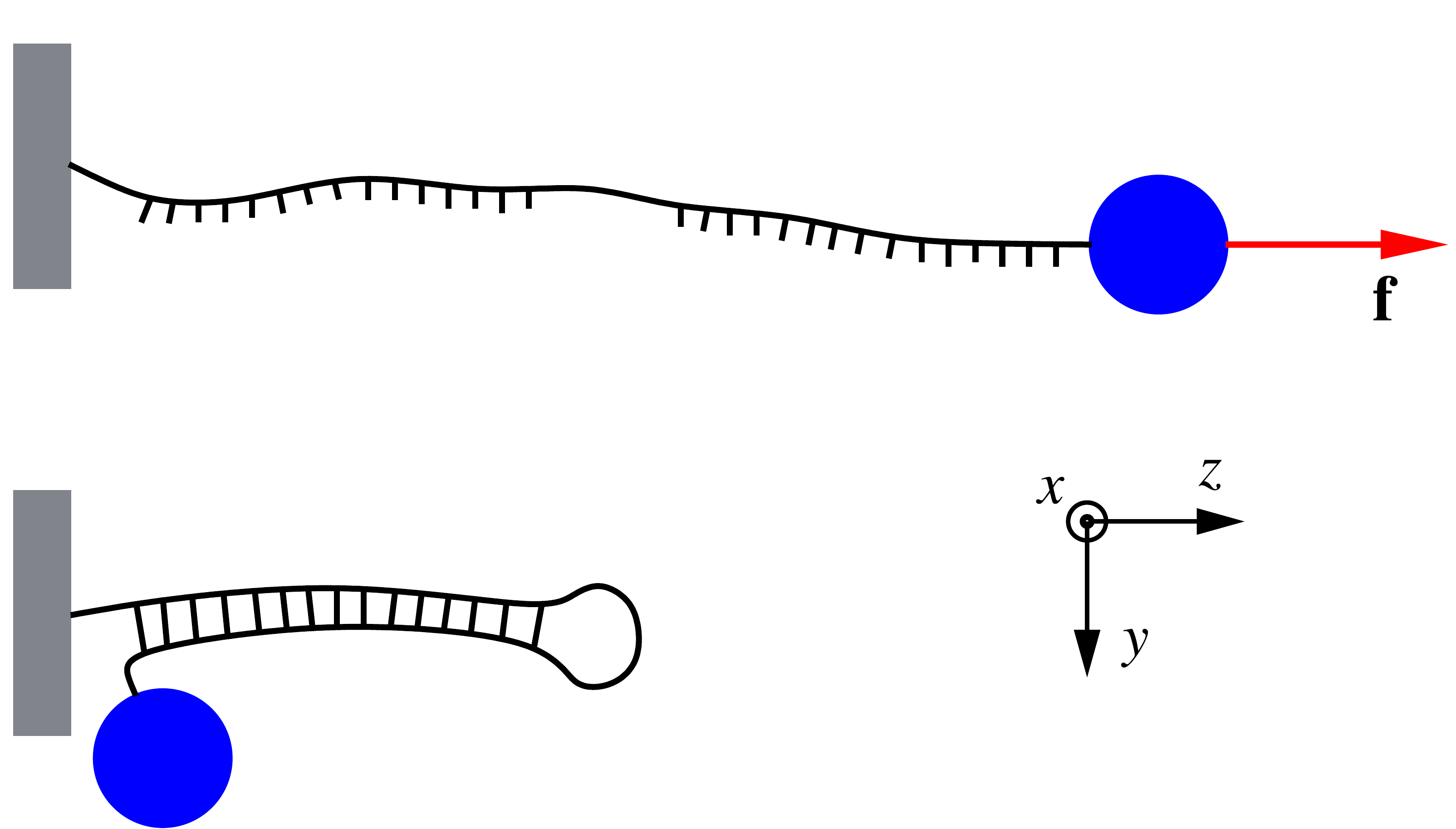}
\end{center}
\caption{Principle of Tethered Particle Motion with an applied magnetic field exerting a force $f$ on the paramagnetic particle (in blue). When the field is switched on (top), the force opens the duplex and elongates the ssDNA along the $z$-axis. The particle is far from the substrate (in gray). When the field is switched off (bottom), the molecule closes and the particle gets close to the substrate. Measuring $z(t)$ should give information about the zipping velocity provided that the particle is small enough and does not slow down zipping.\label{TPM}}
\end{figure}

There appear to be few experimental data on the zipping velocity in the Y-shape geometry at $T<T_m$ described in Section~\ref{far:from:eq:zipping}. Bulk experiments are not readily adapted to accurately measure them because temperature jumps or any modification of the solution properties (ionic strength, $p$H) are not instantaneous~\cite{Nagapriya2010} and the starting point ($t=0$) of zipping is not well defined (see however Ref.~\cite{Zadegan2012} for advances in this issue). Single-particle force experiments are a promising solution, as discussed in Section~\ref{force_unzip} (Figure~\ref{force:unzip}a), but acquiring sufficient data is fastidious. A possibility might arise from multiplexing the acquisition on a biochip as the one developed in Salom\'e's group~\cite{Plenat2012}. The idea would be to replace the fluorescent particles used in Tethered Particle Motion (TPM) by paramagnetic particles and to use magnetic tweezers~\cite{Strick1996,Vilfan2009} to unzip/rezip hundreds of hairpins in parallel (Figure~\ref{TPM}).  A sufficient magnetic field inducing a force $f > 15$~pN would unzip all hairpins. Once equilibrium has been reached, switching the field off would set the starting point of zipping. Measuring the evolution rate of the bead amplitude of movement would then give an indication of the  zipping velocity. The position along the $z$-axis can also be measured with a $\sim10$~nm precision~\cite{Vilfan2009}. This is another way to assert how fast zipping progresses, since at $t=0$ the bead is far away from the glass substrate while at zipping end it is very close to it. 

However, the particle radius $R$ will have to be significantly reduced, at least down to $100$~nm, so that its viscous drag $6\pi\eta R$ remains negligible as compared to the ssDNAs being zipped ($\approx 6 \pi \eta N a_{\rm ss}$, if neglecting hydrodynamic interactions). With $N\sim 10^3$ and $a_{\rm ss} \approx 0.5$~nm (the ssDNA base radius~\cite{Manghi2012}), this yields $R \ll 500$~nm. 

The temperature can be finely controlled in this setup, and measuring velocities when getting close to $T_m$ is also of broad interest. Having measures on an extended range of hairpin lengths will provide tests of the scaling laws discussed in Section~\ref{far:from:eq:zipping} as well as those of Section~\ref{force_unzip} concerning force-induced unzipping and will enable one to clarify the questions raised in Section~\ref{torque:zip:unzip} on an experimental basis. Thermal unzipping could also be explored at $T>T_m$, with or without applied force (Section~\ref{TgtrTm}). A small applied force would interestingly favor the $Y$-shape geometry. 

When modeling ALK experiment~\cite{Altan2003} in Section~\ref{wooed}, we have seen that a prediction of the analysis by Sicard and collaborators~\cite{Sicard2015} is that closure of the open AT-rich core of the hairpin is a temperature-activated process of local nature. Its duration should be weakly sensitive to the hairpin length. This means that the closure mechanism of a bubble in the middle of the molecule is different from the zipper-like one when renaturation started from one single end (compare Figures~\ref{Ferrantini:fig} and \ref{bubble:fig}). Since crossing the twist-related energy barrier is the limiting step, an experimentally testable prediction of these works is that the closure times determined by FCS should not depend on the molecule length.

Finally, we have proposed in Section~\ref{F:barrier} that a free-energy barrier might slow-down zipping because the base-pair is trapped in a metastable state where a water molecule establishes a bridge between both bases (Figure~\ref{zipper:with:barrier}a). Even though getting access to this information at the experimental level seems difficult (even though NMR experiments might be appropriate), all-atom numerical experiments should be pushed forward to ascertain the validity of our proposition, in particular from a quantitative perspective.


\section*{Acknowledgments}
\addcontentsline{toc}{section}{Acknowledgments}

We warmly thank our colleagues and collaborators John Palmeri, Laurence Salom\'e, Catherine Tardin, Philippe Rousseau, Fran\c cois Sicard, Anil Kumar Dasanna, Anna\"el Brunet, Jean-Charles Walter, and Enrico Carlon for fruitful discussions and sound advices during the writing of this Report.
We are tributary to the Universit\'e Toulouse III-Paul Sabatier and the Centre National de la Recherche Scientifique (CNRS).

\newpage

\section*{Appendix: Main experimental values}
\addcontentsline{toc}{section}{Appendix: Main experimental values}

\begin{table}[h]
\begin{center}

\begin{tabular}{|c|p{8cm}|c|c|}
\hline
Notation & Quantity & Consensual value & Refs. \\
\hline
\hline
$a$ & {Monomer (base-pair) length in the duplex (B-DNA)} & 0.34~nm & \cite{Alberts2002} \\
$p$ & {Duplex pitch (in B-DNA)} & 10.5~bp & \cite{Alberts2002} \\
$\ell_p$ & {Persistence length of ds/ssDNA in physiological conditions ($\ell_p=\kappa/(k_{\rm B} T)$)} & $\simeq 50$~nm/$\simeq 1$~nm & \cite{FrankKamenetskii1997,Hagerman1998} \\
$\ell_p^{\rm torsion}$ & {Torsional persistence length of dsDNA/bubble in physiological conditions ($\ell_p^{\rm torsion}=C/(k_{\rm B} T)$)} & $\simeq 110$~nm/$\simeq 1$~nm & \cite{Kahn1994,Bryant2003,Lipfert2010} \\
$T_m$ & {Denaturation temperature in physiological salt conditions (sequence-dependent)} & 50 $\leq T_m \leq$ 90$^\circ$C & \cite{Gotoh1983,Blake1998} \\
$\Delta F_0$ &  {Average free-energy gained when closing one base-pair at 25$^\circ$C/37$^\circ$C} & $\simeq 2.3 k_{\rm B} T$/$2.8 k_{\rm B} T$ &  \cite{SantaLucia1998} \\
$\Delta S_0$ &  {Average entropy loss when closing one base-pair at physiological or room temperature} & $\simeq 10 k_{\rm B}$ & \cite{SantaLucia1998} \\
$\varphi_{\rm B}$ & {Fraction of open base-pairs in physiological conditions (sequence-dependent)} & $10^{-7} \leq \varphi_{\rm B} \leq 10^{-5}$ & \cite{FrankKamenetskii2014} \\
\hline \hline
$\tau_{\rm cl,transient}$ & {Transient denaturation bubble lifetime} & $\sim1$~ns & \cite{vonHippel2013,Warmlander2000} \\
$\tau_{\rm cl,equil}$ & {Equilibrated denaturation bubble lifetime} & $10 < \tau_{\rm cl} <$ 200~$\mu$s & \cite{Warmlander2000,Altan2003,Phelps2013} \\
$\tau_{\rm op,bubble}$ & {Typical bubble opening time at room temperature} & $\sim 10$~ms & \cite{Gueron1987}\\
$\tau_{\rm op,hairpin}$ & {Typical opening time of short hairpins at zero external force and at room temperature} & $10 < \tau_{\rm op} <100~\mu$s & Sec.~\ref{hairpins}\\
$\tau_{c}$ & {Base-pair coasting time (timescale above which inertial effects are irrelevant)} & $\sim 0.1$~ps & Sec.~\ref{dissipation} \\
$\tau_0$ & {Base-pair relaxation time in the zipper model} & $\simeq 0.5$~ns & Sec.~\ref{1Damod} \\
$\tau'_0$ & {Relaxation time of a dsDNA segment of length $\ell_{\rm K}=2\ell_p$} & $\simeq 80~\mu$s & Sec.~\ref{chain_dynamics} \\
$f_{\rm ext}$ & {Minimal force needed to unwind DNA in single-molecule experiments (sequence-dependent)}  &  $9< f_{\rm ext}< 20$~pN & \cite{Rief1999} \\
$v_{\rm zip}$ & {Zipping velocity of short oligomers (few base-pairs)} & $1 < v_{\rm zip} < 10$~bp/$\mu$s & \cite{Porschke1974,Craig1971,Chen2008} \\
$v_{\rm zip}$ & {Average zipping velocity of long duplexes ($\lambda$-phage $\simeq 48.5$~kbp)} & $\simeq 10$~bp/ms &  \cite{Thomen2002} \\
$v_{\rm unzip}$ & Average unzipping velocity of long duplexes (T2-phage $\simeq 20$~kbp) at $T>T_m$ & $\simeq 1$~bp/ms &  \cite{Record1972} \\
$v$ & {Sound velocities in DNA} & $100< v < 2000$~m/s & \cite{Yakushevich2004,Yakushevich2002} \\
\hline
\end{tabular}

\caption{The typical experimental values of the main quantities involved in DNA dynamics as discussed in this Report. The first part of the table lists structural or equilibrium quantities, whereas the second part addresses dynamical ones.
\label{tab2}}
\end{center}
\end{table}

\newpage

\end{document}